\documentclass[]{aa}
\usepackage{graphicx}
\usepackage{color}
\usepackage{colortbl}
\usepackage{soul}
\usepackage{natbib}
\bibpunct{(}{)}{;}{a}{}{,}
\graphicspath{{figures/}}
\usepackage{subfigure}
\usepackage{algorithm,algorithmicx}

\begin{document}
%%%%%%%%%%%%%%%%%%%%%%%%%%%%%%%%%%%%%%%%%%%%%%%%%%%%%%%%%%%%%%
%
% Title
%
%%%%%%%%%%%%%%%%%%%%%%%%%%%%%%%%%%%%%%%%%%%%%%%%%%%%%%%%%%%%%%
\title{The effects of numerical resolution, heating timescales
and background heating on thermal non-equilibrium in coronal loops}
\author{
C. D. Johnston\inst{1}
\and P. J. Cargill\inst{1, 2}
\and P. Antolin\inst{1} 
\and A. W. Hood\inst{1} 
\and I. De Moortel\inst{1} 
\and S. J. Bradshaw\inst{3}
}
\institute{School of Mathematics and Statistics, University of St Andrews, St Andrews, Fife, KY16 9SS, UK.
\and
Space and Atmospheric Physics, The Blackett Laboratory, Imperial College, London, SW7 2BW, UK.
\and
Department of Physics and Astronomy, Rice University, Houston, TX 77005, USA.
\\
\email{cdj3@st-andrews.ac.uk}
}

%%%%%%%%%%%%%%%%%%%%%%%%%%%%%%%%%%%%%%%%%%%%%%%%%%%%%%%%%%%%%%
%
% Abstract
%
%%%%%%%%%%%%%%%%%%%%%%%%%%%%%%%%%%%%%%%%%%%%%%%%%%%%%%%%%%%%%%
  \abstract
  {
  Thermal non-equilibrium (TNE) is believed to be a potentially 
  important process in understanding some properties of the 
  magnetically closed solar corona. Through one-dimensional 
  hydrodynamic models, this paper addresses the importance of 
  the numerical spatial resolution, footpoint heating 
  timescales and background heating on TNE. 
  Inadequate transition 
  region (TR) resolution can lead to significant discrepancies 
  in TNE cycle behaviour, with TNE being suppressed in 
  under-resolved loops. 
  A convergence on the periodicity and plasma 
  properties associated with TNE required spatial resolutions 
  of less than 2 km for a loop of length 180 Mm. 
  These numerical 
  problems can be resolved using an approximate method that 
  models the TR as a discontinuity using a jump condition, as 
  proposed by Johnston et al. (2017a,b). The resolution 
  requirements (and so computational cost) are greatly 
  reduced while retaining good agreement with fully resolved 
  results. Using this approximate method we (i) identify
  different regimes for the response of coronal loops to 
  time-dependent footpoint heating including 
  one where TNE does not
  arise and (ii) demonstrate that TNE in a loop with footpoint 
  heating is suppressed unless the background heating is 
  sufficiently small. The implications for the generality of 
  TNE are discussed.
  }
  \titlerunning{Thermal non-equilibrium in coronal loops}
  \maketitle
  
%%%%%%%%%%%%%%%%%%%%%%%%%%%%%%%%%%%%%%%%%%%%%%%%%%%%%%%%%%%%%%
%
% Introduction
%
%%%%%%%%%%%%%%%%%%%%%%%%%%%%%%%%%%%%%%%%%%%%%%%%%%%%%%%%%%%%%%
\section{Introduction}
  \label{Sect:intro}
  \indent
  The numerical modelling of energy release in the solar corona 
  has a long history, yet remains computationally challenging. 
  In a multi-dimensional magnetohydrodynamic (MHD) approach the 
  difficulty concerns the very small values of diffusion 
  coefficients that are necessary for the correct modelling of, 
  for example, shocks and magnetic reconnection. If the 
  observational consequences of energy release are to be 
  assessed, the difficulty is compounded by the very severe 
  restriction on the time step imposed by the need to model 
  thermal conduction accurately through the narrow transition 
  region (TR).
  \\
  \indent
  One approach has been to decouple the MHD from the plasma 
  response by solving the one-dimensional (1D) hydrodynamic 
  equations along a field line, or collection of field lines, 
  in response to a prescribed heating function. Here the 
  numerical problems are at least tractable with adaptive 
  re-gridding
  \citep[][]{paper:Bettaetal1997,
  paper:Antiochosetal1999,
  paper:Bradshaw&Mason2003,
  paper:Bradshaw&Cargill2013}. 
  Translating this to 3D remains challenging 
  due to (a) the requirement for many more grid points and 
  consequent increase in computing requirements and (b) the 
  competition for where any adaptive re-gridding 
  is carried out (i.e. 
  whether to prioritise getting the TR or current sheet
  behaviour correct).
  \\
  \indent
  The consequences of under-resolving the TR were fully 
  documented by 
  \citet[][hereafter BC13]{paper:Bradshaw&Cargill2013}
  for impulsive heating where the amplitude 
  of the heating covered a range between nanoflares and small 
  flares. Without adequate resolution, the coronal density 
  increase in response to the heating could be far too small. 
  We note that in 1D this \lq brute force\rq\
  approach of ultra-high 
  resolution is feasible, but not in 3D. Thus there is 
  considerable interest in approximate methods for handling 
  this problem that avoid the severe time step limitations of 
  solving the full equations.
  \\
  \indent
  In two recent papers
  \citep{paper:Johnstonetal2017a,paper:Johnstonetal2017b}, 
  we have proposed an 
  approximate method that addresses this problem
  for 1D hydrodynamic models. 
  [\cite{paper:Mikicetal2013} 
  have proposed an alternative method that will be 
  discussed fully in subsequent papers.] Below a certain 
  temperature the 
  TR is treated as an unresolved discontinuity across which 
  energy is conserved (we call this the unresolved transition 
  region (UTR) approach). 
  A closure relation for the 
  radiation in the unresolved TR is 
  used to permit a simple jump relation between the
  chromosphere 
  and upper TR. 
  The method was tested against 
  the HYDRAD code
  \citep[][BC13]{paper:Bradshaw&Mason2003,
  paper:Bradshaw&Cargill2006}
  and was found to give good 
  agreement.
  \\
  \indent
  In these papers we focussed on impulsive heating that was 
  either uniform across the loop or concentrated near the 
  footpoints (such as might arise from the precipitation of 
  energetic particles). 
  In this third and final paper on
  1D UTR modelling, we 
  have applied 
  the method to a different computationally 
  challenging problem, 
  namely thermal non-equilibrium (TNE) in coronal loops. 
  \\
  \indent
  TNE is a phenomenon that can occur
  in coronal loops when the heating is concentrated towards the
  footpoints 
  \citep[e.g.][]
  {
  paper:Antiochosetal2000,
  paper:Karpenetal2001,
  paper:Mulleretal2003,
  paper:Mulleretal2004,
  paper:Mulleretal2005, 
  Mendozabriceno_2005ApJ...624.1080M,
  paper:Moketal2008, 
  paper:Antolinetal2010, 
  paper:Susinoetal2010,
  paper:Lionelloetal2013,
  paper:Mikicetal2013,
  paper:Susinoetal2013, 
  paper:Moketal2016}.  
  This
  localised energy deposition 
  drives evaporative upflows that
  fill the loop with hot dense plasma, increasing
  the coronal density and radiative losses. The loop 
  evolution is then determined primarily by an enthalpy flux 
  injection from the footpoints to sustain radiative and 
  conductive losses 
  \citep{Serio_1981ApJ...243..288S,paper:Antiochosetal2000}.
  Eventually,
  when the coronal radiative losses overcome the
  heating source(s) at the top of the loop,
  the thermal
  instability is triggered locally in the corona 
  \citep[e.g.][]
  {paper:Parker1953,paper:Field1965, paper:Hildner1974}. 
  The subsequent runaway cooling leads to the formation of
  coronal
  condensations 
  in the region around the loop apex
  \citep
  {paper:Moketal1990, paper:Antiochos&Klimchuk1991, 
  paper:Antiochosetal1999}. These condensations then
  fall back down to the TR and chromosphere  due to gas 
  pressure or gravitational forces, with the loop
  draining along the magnetic
  field. These cool and dense condensations are 
  thought to manifest as coronal rain, observed in 
  chromospheric and transition region lines 
  \citep{Kawaguchi_1970PASJ...22..405K,
  Leroy_1972SoPh...25..413L,
  Levine_Withbroe_1977SoPh...51...83L,
  Kjeldseth_Brekke_1998SoPh..182...73K,
  Schrijver_2001SoPh..198..325S,DeGroof_2004AA...415.1141D,
  DeGroof05,Oshea_etal_2007AA...475L..25O,
  Tripathi_etal_2009ApJ...694.1256T,
  Kamio_etal_2011AA...532A..96K,
  Antolin_Rouppe_2012ApJ...745..152A,
  Antolin_etal_2012SoPh..280..457A}.
  \\
  \indent
  Furthermore, 
  if the heating frequency is high and sustained for a 
  relatively long time in comparison to the
  characteristic cooling time of the loop then this evolution 
  of evaporation followed by condensation can become cyclic 
  \citep{Mendozabriceno_2005ApJ...624.1080M,
  paper:Antolinetal2010,
  paper:Susinoetal2010}. 
  The response of a loop to such quasi-steady heating is to 
  undergo evaporation and condensation 
  cycles with a period on the timescale of hours 
  independent of the characteristic timescale of the 
  heating events \citep{paper:Mulleretal2003,
  paper:Mulleretal2004}. This highly 
  nonlinear and unstable behaviour has been termed TNE
  \citep
  {paper:Antiochosetal2000,
  paper:Karpenetal2001,
  paper:Mikicetal2013}
  and 
  we refer to these evaporation and condensation cycles as TNE 
  cycles
  \citep{paper:Kuin&Martens1982}. 
  \\
  \indent
  Debate exists on whether TNE, as a coronal response to 
  footpoint heating 
  theory matches long standing observational constraints on 
  coronal loops 
  \citep{paper:Moketal2008,Klimchuk_2010ApJ...714.1239K,
  Klimchuk_2015RSPTA.37340256K,
  Peter_2012AA...548A...1P,
  Lionello_2013ApJ...773..134L,
  Lionello_2016ApJ...818..129L,paper:Moketal2016,
  Winebarger_2016ApJ...831..172W,
  Winebarger_2018ApJ...865..111W}. 
  Recently, TNE has further gained considerable interest as a 
  mechanism for explaining the discovery of 
  long period intensity pulsations, particularly those in 
  active region loops
  \citep{
  paper:Auchereetal2014,paper:Fromentetal2015,
  paper:Fromentetal2017,paper:Fromentetal2018}, 
  observed to be accompanied by periodic coronal rain   
  \citep[][]{paper:Antolinetal2015,
  paper:Auchereetal2018}.
  \\
  \indent
  Modelling
  TNE in coronal loops 
  is a computationally
  challenging problem because (a) the heating is 
  applied to a region where numerical resolution is likely to 
  be poor (especially in 3D), (b) the presence or 
  absence of coronal 
  condensations, and their precise characteristics 
  (i.e. densities, temperatures,  periodicity, etc.), requires 
  the correct evaporative response to the heating 
  injection, and (c) the presence of such condensations further 
  requires correct modelling of a second hot-cold interface in 
  the
  corona. This constitutes an excellent challenge for the UTR 
  method and we demonstrate its use on a series of TNE 
  problems. 
  \\
  \indent
  We describe the key features of the numerical methods
  in Section \ref{Sect:methods}.
  A second aspect of the paper is to extend the analysis of 
  BC13 to TNE heating profiles. That is done in 
  Section \ref{Sect:Resolution} and 
  it is shown that the same problems arise as with impulsive 
  heating. Indeed TNE does not occur in under-resolved loops. 
  In 
  Section \ref{Sect:Resolution_LareJ} 
  we demonstrate that the UTR method performs well 
  on these problems. 
  Sections \ref{Sect:Time_dependent_fp_heating} -- 
  \ref{Sect:Q_bg} 
  further demonstrate the 
  method on other problems of interest to TNE, including 
  a clear demonstration of different TNE regimes 
  obtained with unsteady footpoint heating. 
  Our conclusions are stated in Section 
  \ref{Sect:dis_concl}. 
  %
  %
%%%%%%%%%%%%%%%%%%%%%%%%%%%%%%%%%%%%%%%%%%%%%%%%%%%%%%%%%%%%%%
%
% Results
%
%%%%%%%%%%%%%%%%%%%%%%%%%%%%%%%%%%%%%%%%%%%%%%%%%%%%%%%%%%%%%%
\section{Results
  \label{Sect:res}}
  %
  %
%%%%%%%%%%%%%%%%%%%%%%%%%%%%%%%%%%%%%%%%%%%%%%%%%%%%%%%%%%%%%%
%
% Numerical methods
%
%%%%%%%%%%%%%%%%%%%%%%%%%%%%%%%%%%%%%%%%%%%%%%%%%%%%%%%%%%%%%%
\subsection{Numerical methods
  \label{Sect:methods}}
  \indent
  To study TNE, we solve the one-dimensional field-aligned 
  time-dependent hydrodynamic equations 
  (see
  \cite{paper:Johnstonetal2017a}) 
  using two methods. The HYDRAD code 
  \citep[][BC13]{paper:Bradshaw&Mason2003,
  paper:Bradshaw&Cargill2006} 
  uses adaptive re-gridding to ensure adequate spatial 
  resolution in the TR, with the grid being 
  refined such that cell-to-cell 
  changes in the temperature and density are kept between 
  5\% and 10\% where possible. This is achieved by each 
  successive refinement splitting a cell into two, and a 
  refinement level of RL leads to cell sizes decreased by 
  $1/2^{\textrm{RL}}$. 
  The maximum value of RL is taken as 13. This can 
  lead to very small cells, with a commensurate decrease in the 
  time step required for numerical stability. However 
  individual 
  loops can be simulated for reasonable real times (a few 
  hours). 
  Further details of the HYDRAD numerical method,
  including the finite difference schemes used,
  can be found in Appendix A2 of BC13 and references therein.
  \\
  \indent
  Such a \lq brute force\rq\
  method is unlikely to be a viable way 
  of running multi-dimensional codes. To this end we have 
  developed an alternative approach, tested on one-dimensional 
  problems 
  \citep{paper:Johnstonetal2017a,paper:Johnstonetal2017b}, 
  that treats the lower TR as an 
  unresolved layer. By integrating the energy equation across 
  this layer, and imposing a closure condition, we are able to 
  provide rapid solutions to 1D problems with an accuracy that 
  compares well with HYDRAD results. The details are found in 
  \cite{paper:Johnstonetal2017a}, 
  and we refer to this as the 
  UTR method. It is incorporated into a one 
  dimensional version of the Lagrangian remap (Lare) code
  whose computational details are discussed in 
  \cite{paper:Arber2001}, referred to as
  \lq LareJ\rq .	
  
%%%%%%%%%%%%%%%%%%%%%%%%%%%%%%%%%%%%%%%%%%%%%%%%%%%%%%%%%%%%%%
%
% Influence of numerical resolution
%
%%%%%%%%%%%%%%%%%%%%%%%%%%%%%%%%%%%%%%%%%%%%%%%%%%%%%%%%%%%%%%
\subsection{Influence of numerical resolution
 \label{Sect:Resolution}}
  \indent
  We start by exploring the effect of numerical   
  resolution
  on TNE cycles in coronal loop models.
  We model a coronal loop of total length 180 Mm with a small 
  chromosphere attached to each end and select
  the largest grid cells in our calculations
  to have a width of 1 Mm. Thus, at the highest refinement 
  level, the minimum cell width is 122 m.
  %
  %
  %%%%%%%%%%%%%%%%%%%%%%%%%%%%%%%%%%%%%%%%%%%%%%%%%%%%%%%%%%%%%    
  %
  % Fig:tne_T_Q_180Mm_steady
  %
  \begin{figure}
  \vspace*{-1.4cm}
  \hspace*{-0.8cm}
  \subfigure{\includegraphics[width=1.15\linewidth]
  {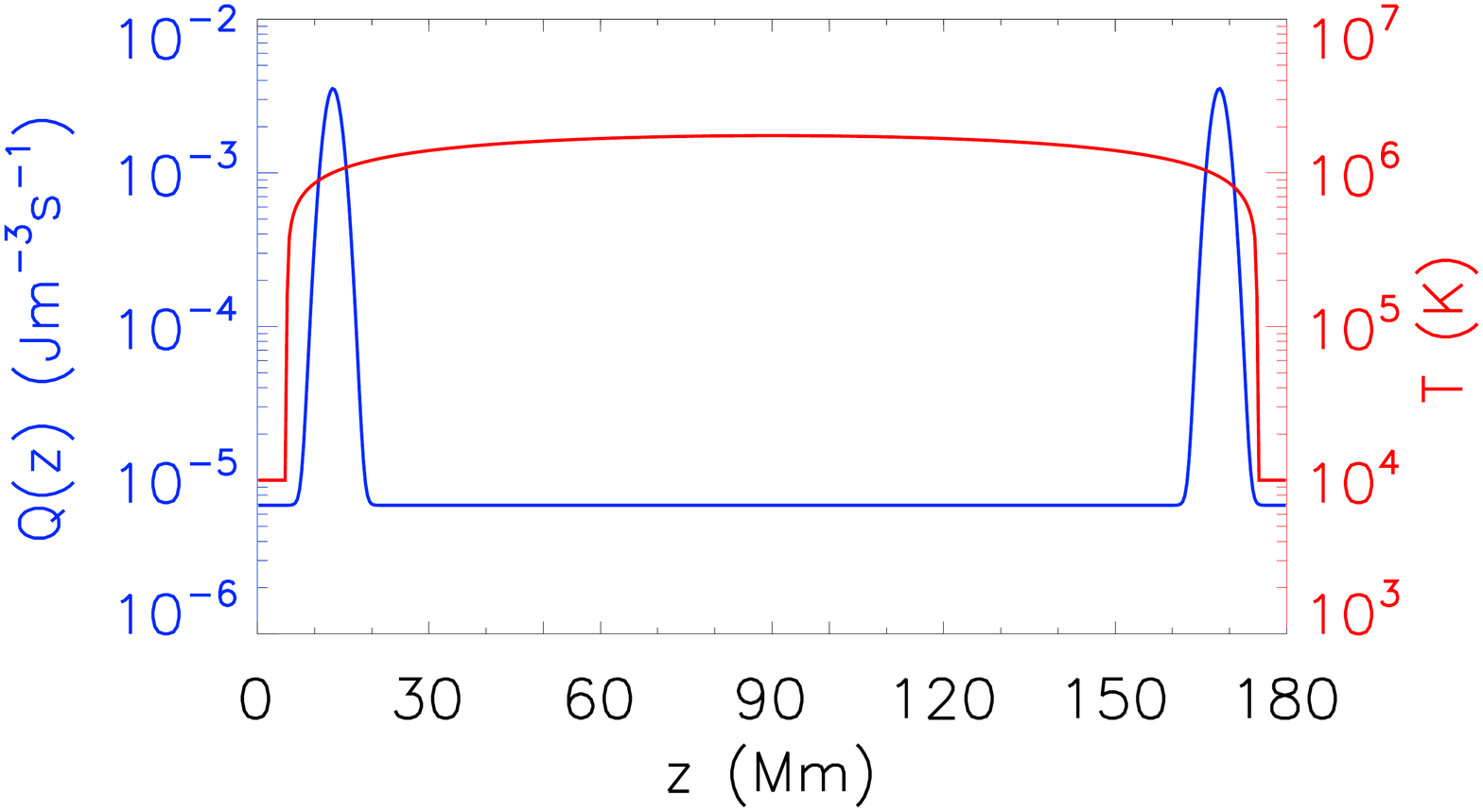}}
  \vspace*{-0.8cm}
  \caption{
    Steady footpoint heating profile $Q(z)$
    (blue line, left-hand axis)
    used in Sections
    \ref{Sect:Resolution} and 
    \ref{Sect:Resolution_LareJ},
    imposed on top of the temperature initial condition
    (red line, right-hand axis). 
    The temperature is determined by the imposed background 
    heating.
    \label{tne_T_Q_180Mm_steady}
    }
\end{figure}
  %
  %
  %%%%%%%%%%%%%%%%%%%%%%%%%%%%%%%%%%%%%%%%%%%%%%%%%%%%%%%%%%%%%    
  %
  % Fig:IoR_HYDRAD
  %
\begin{figure*}
  \hspace*{0.04\linewidth}
  \subfigure{\includegraphics[width=0.35\linewidth]
  {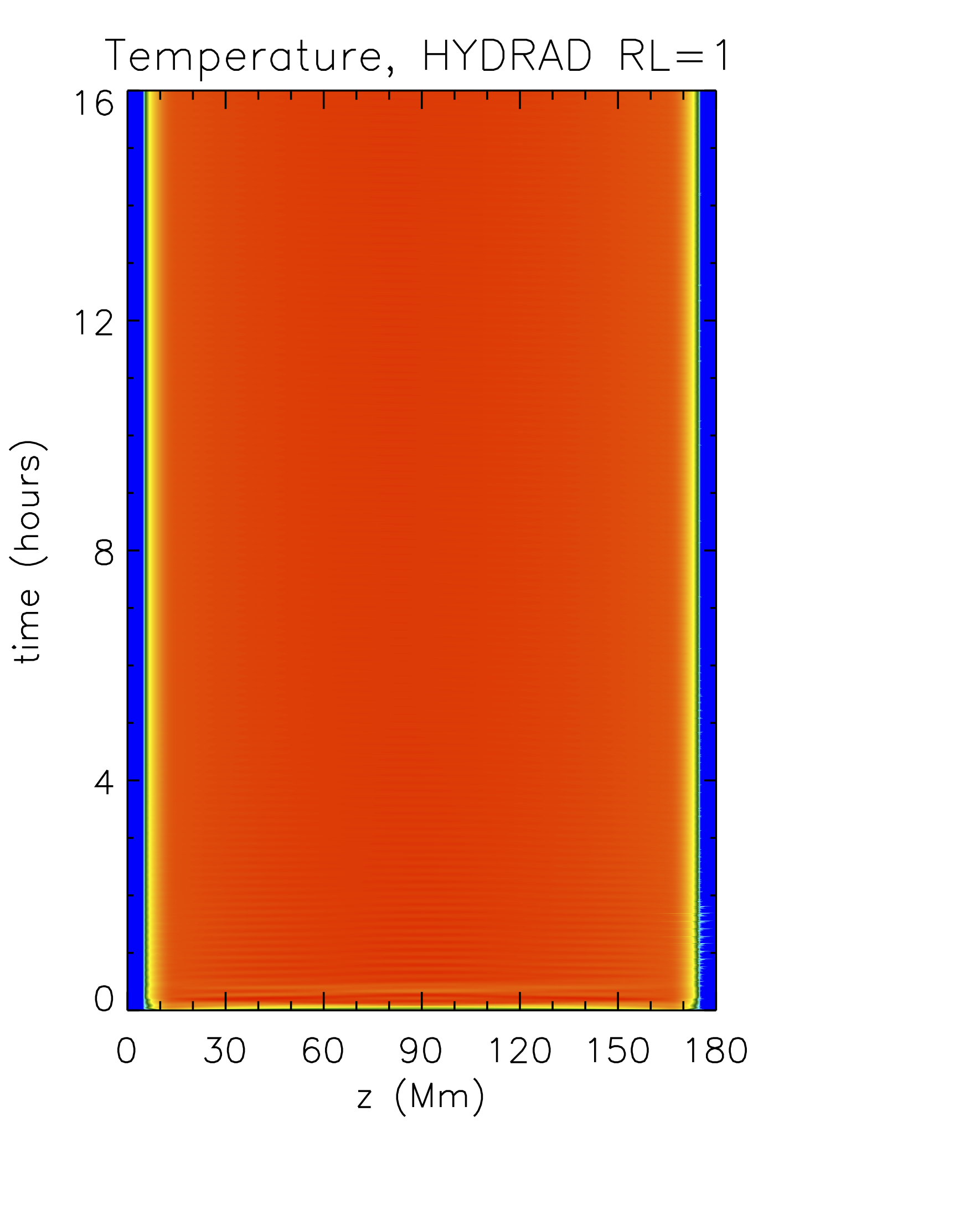}}
  \hspace*{-0.08\linewidth}
  \subfigure{\includegraphics[width=0.35\linewidth]
  {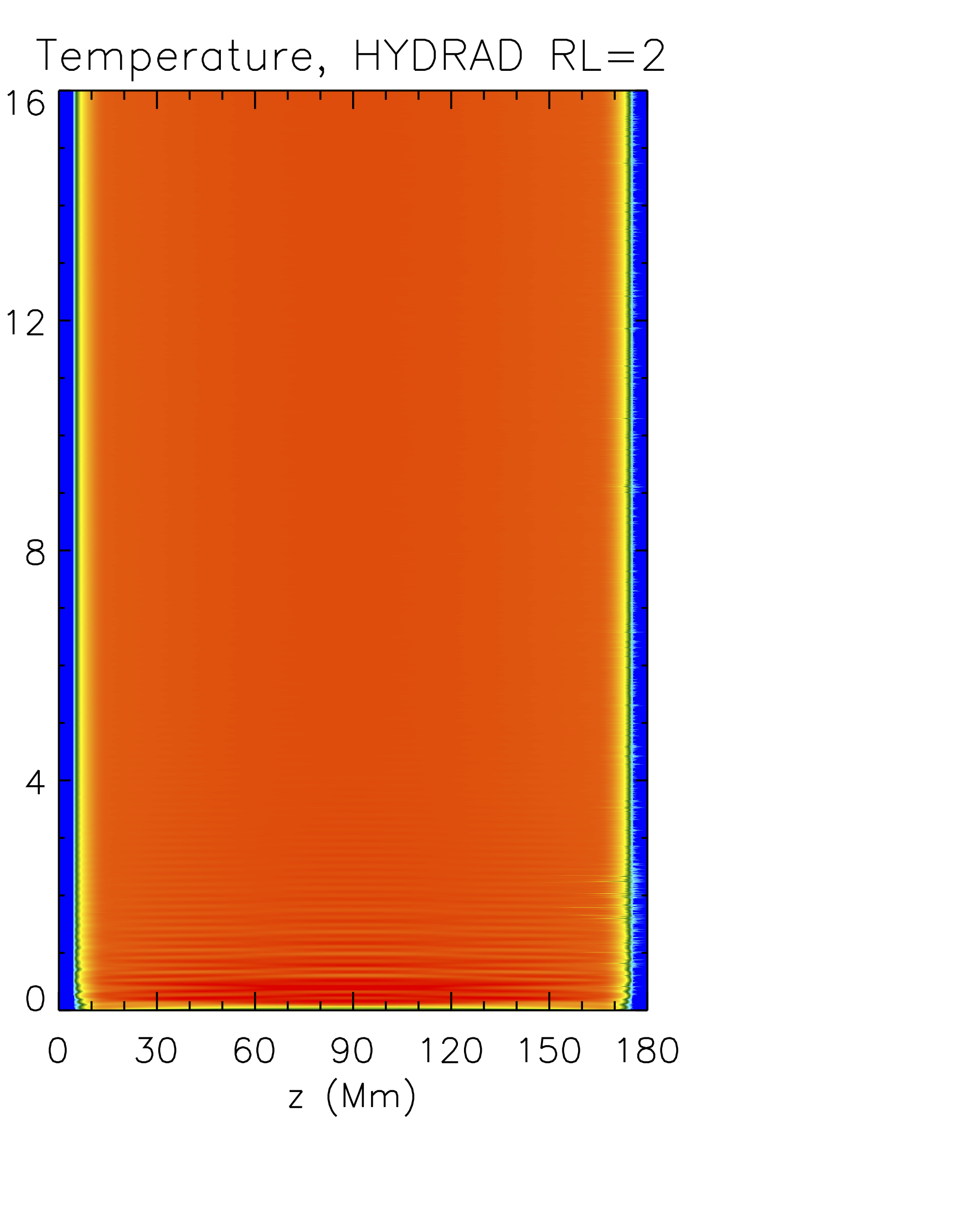}}
  \hspace*{-0.1\linewidth}
  \subfigure{\includegraphics[width=0.35\linewidth]
  {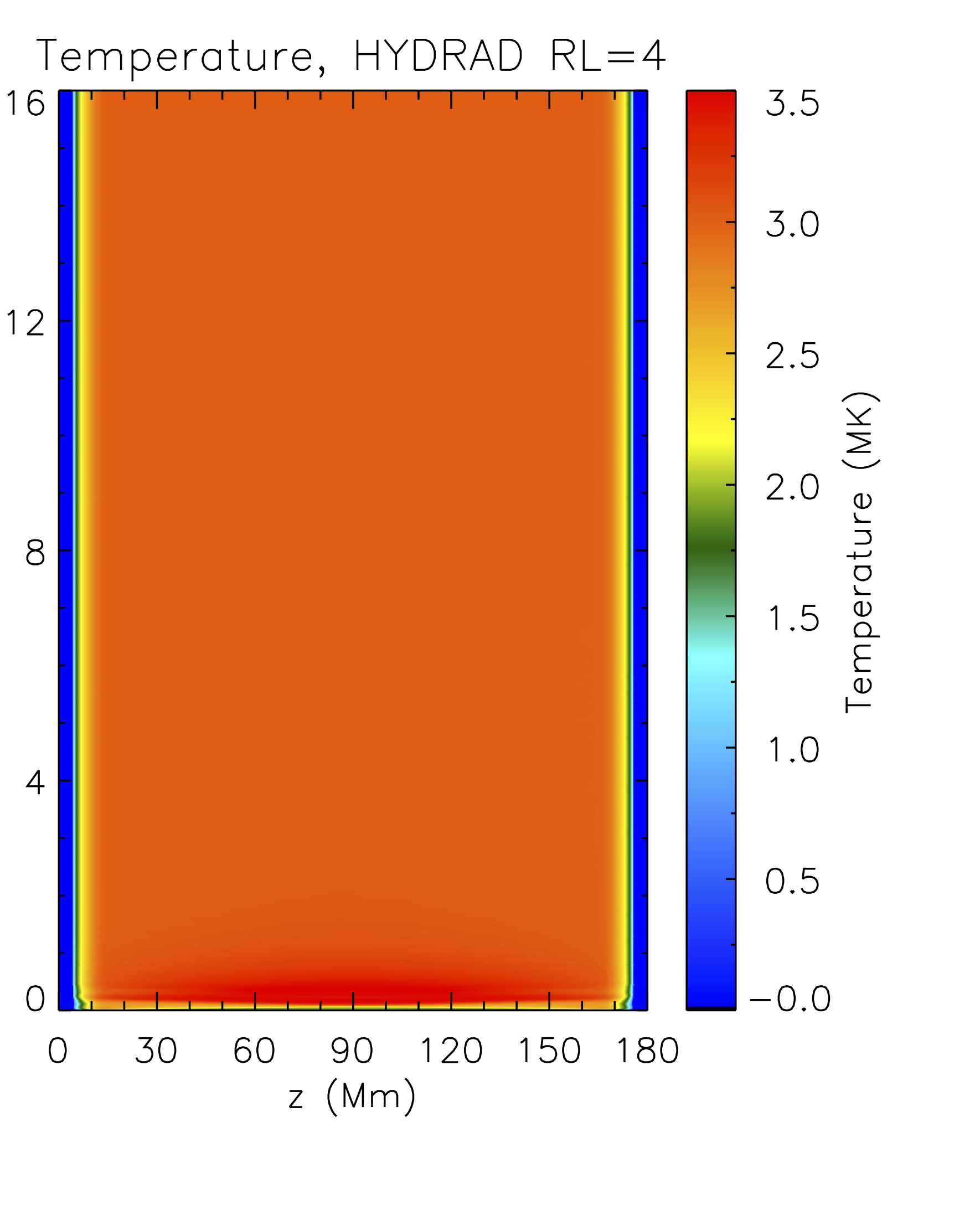}}
  \\[-0.75cm]
  \hspace*{0.04\linewidth}
  \subfigure{\includegraphics[width=0.35\linewidth]
  {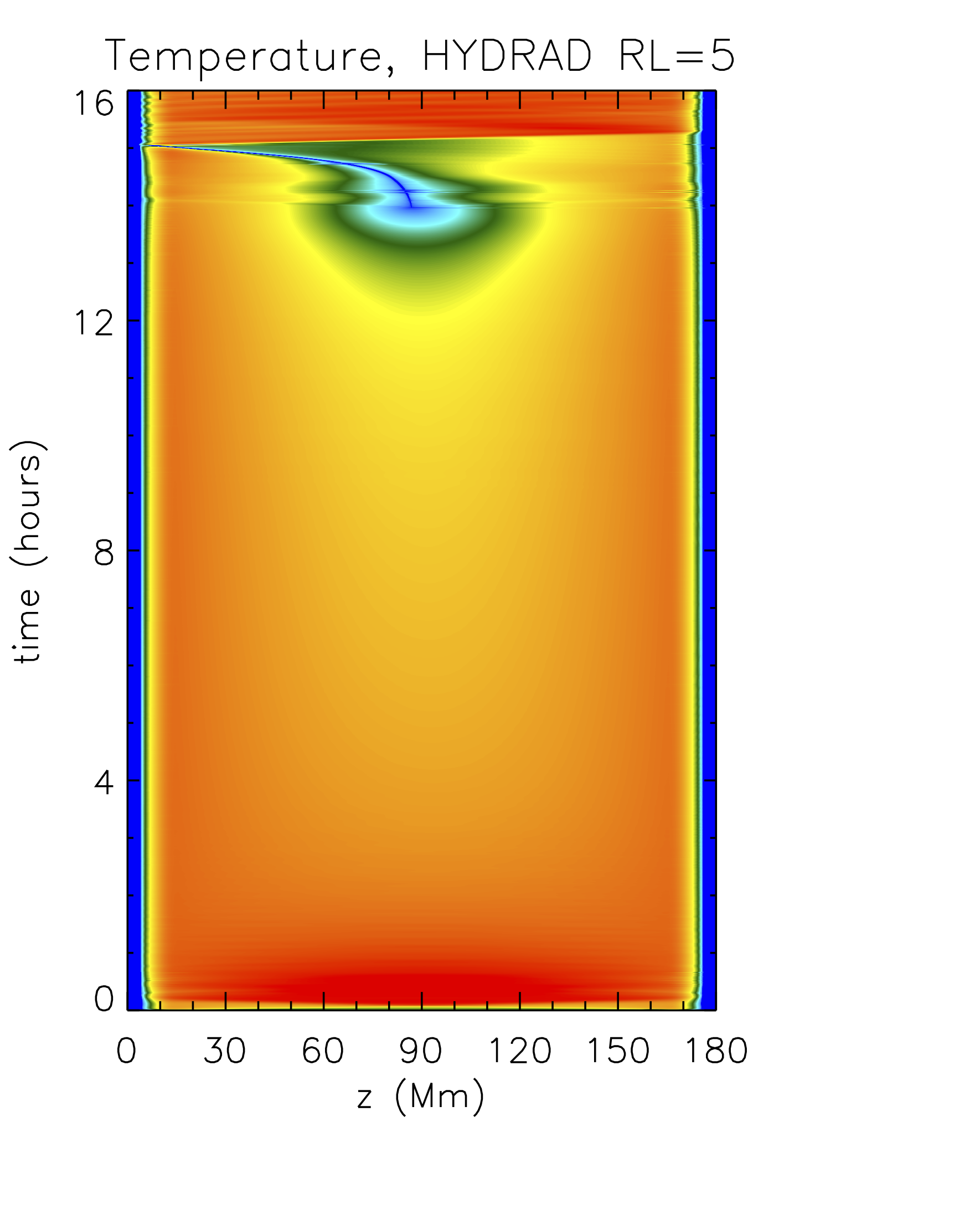}}
  \hspace*{-0.08\linewidth}
  \subfigure{\includegraphics[width=0.35\linewidth]
  {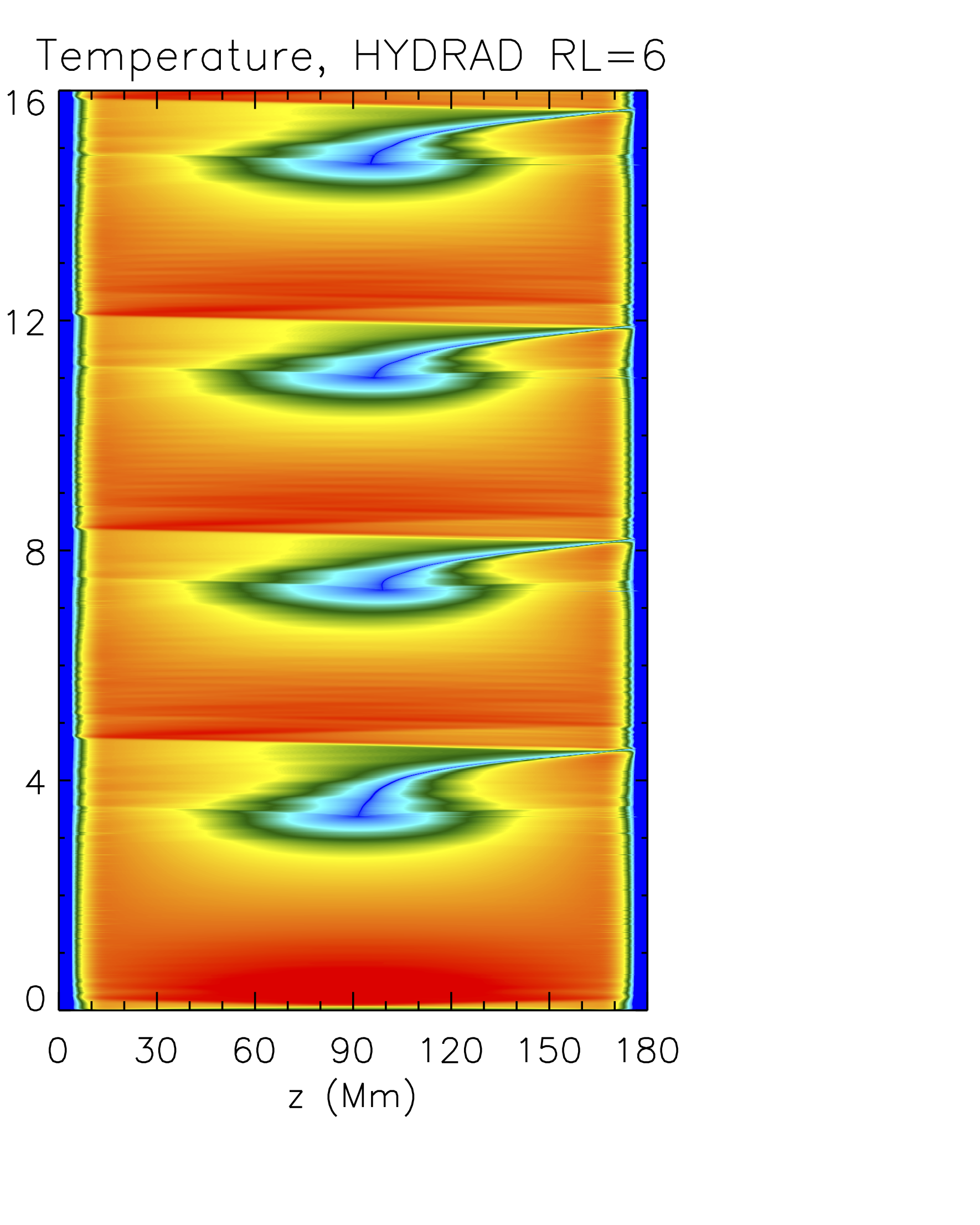}}
  \hspace*{-0.1\linewidth}
  \subfigure{\includegraphics[width=0.35\linewidth]
  {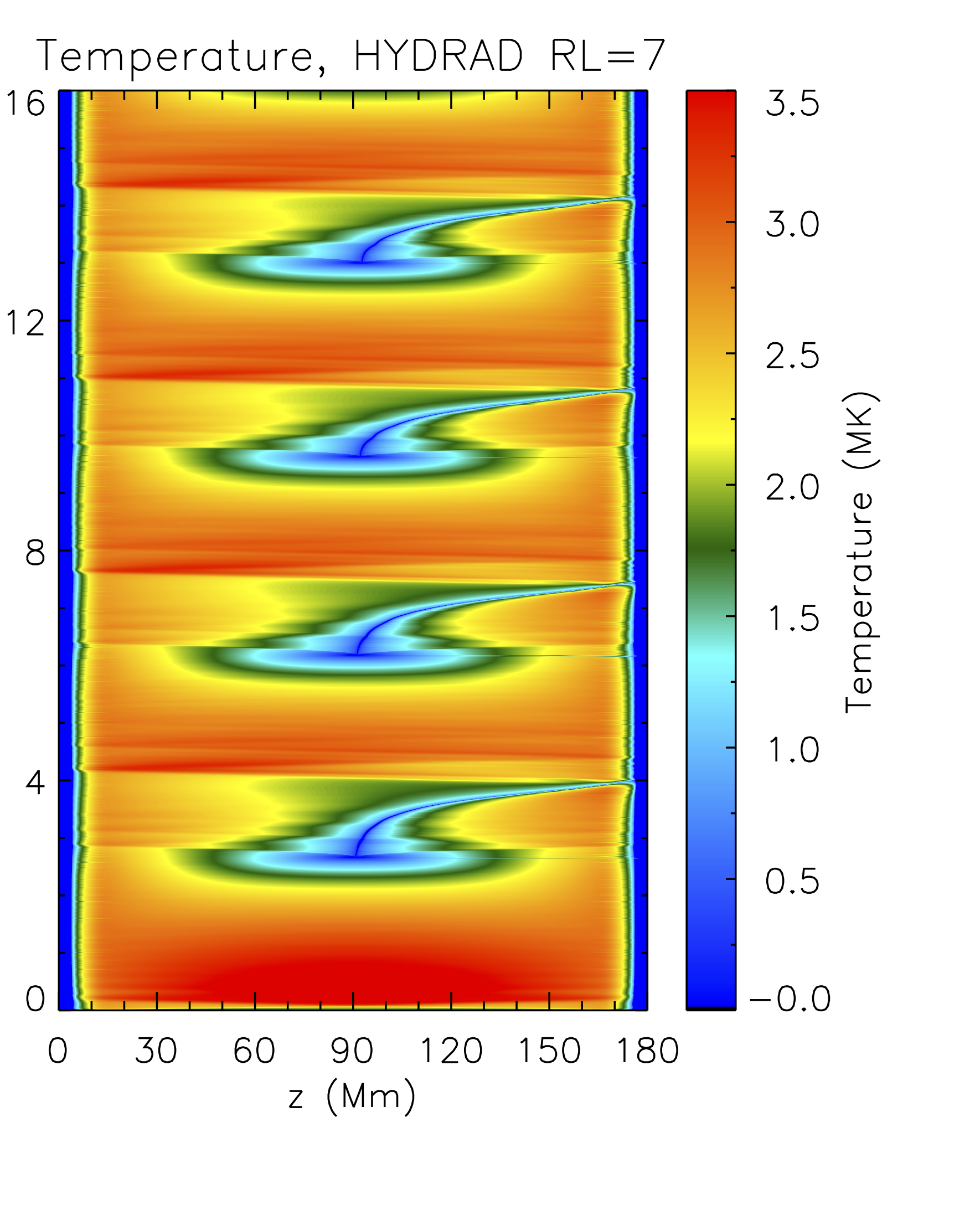}}
  \\[-0.75cm]
  \hspace*{0.04\linewidth}
  \subfigure{\includegraphics[width=0.35\linewidth]
  {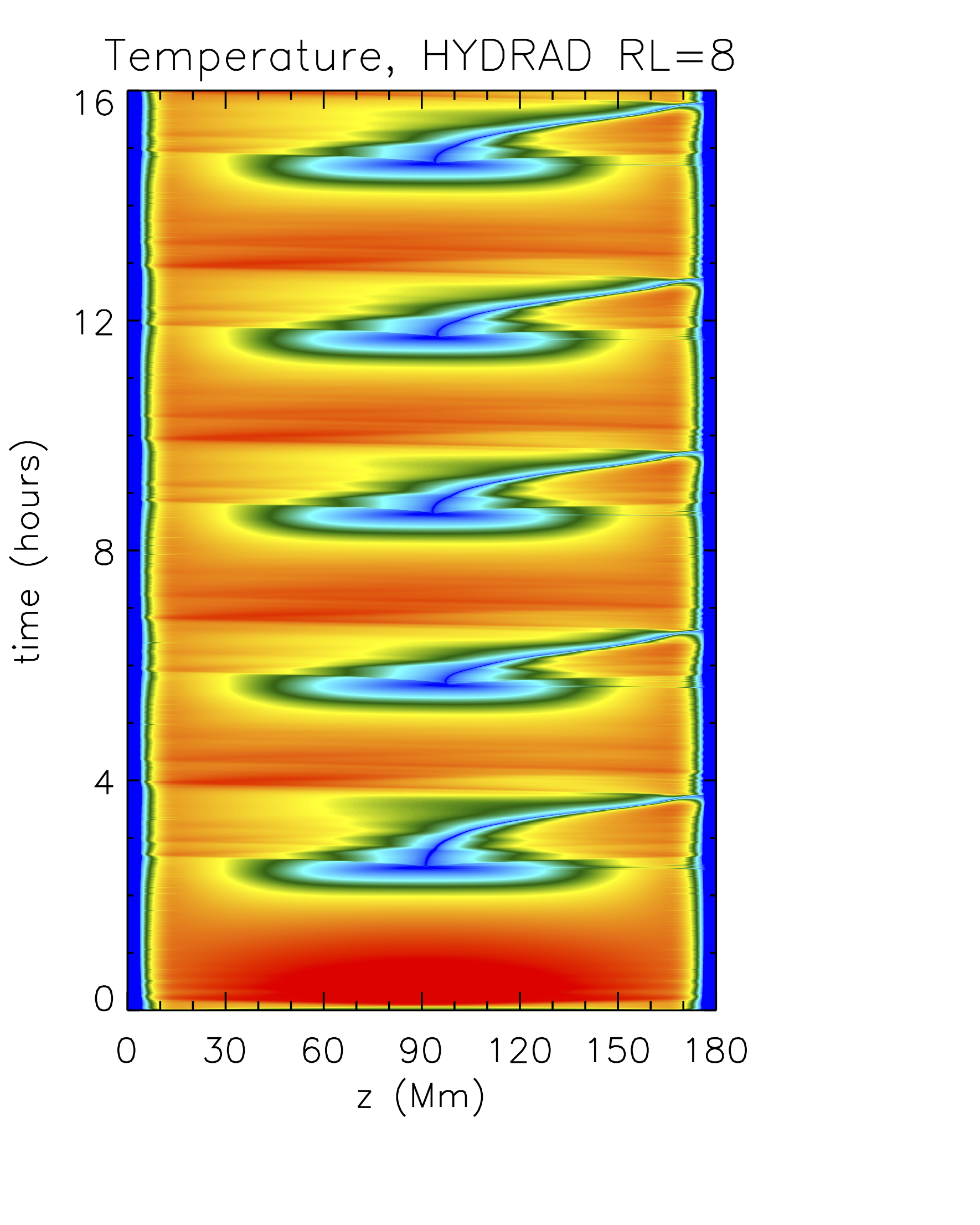}}
  \hspace*{-0.08\linewidth}
  \subfigure{\includegraphics[width=0.35\linewidth]
  {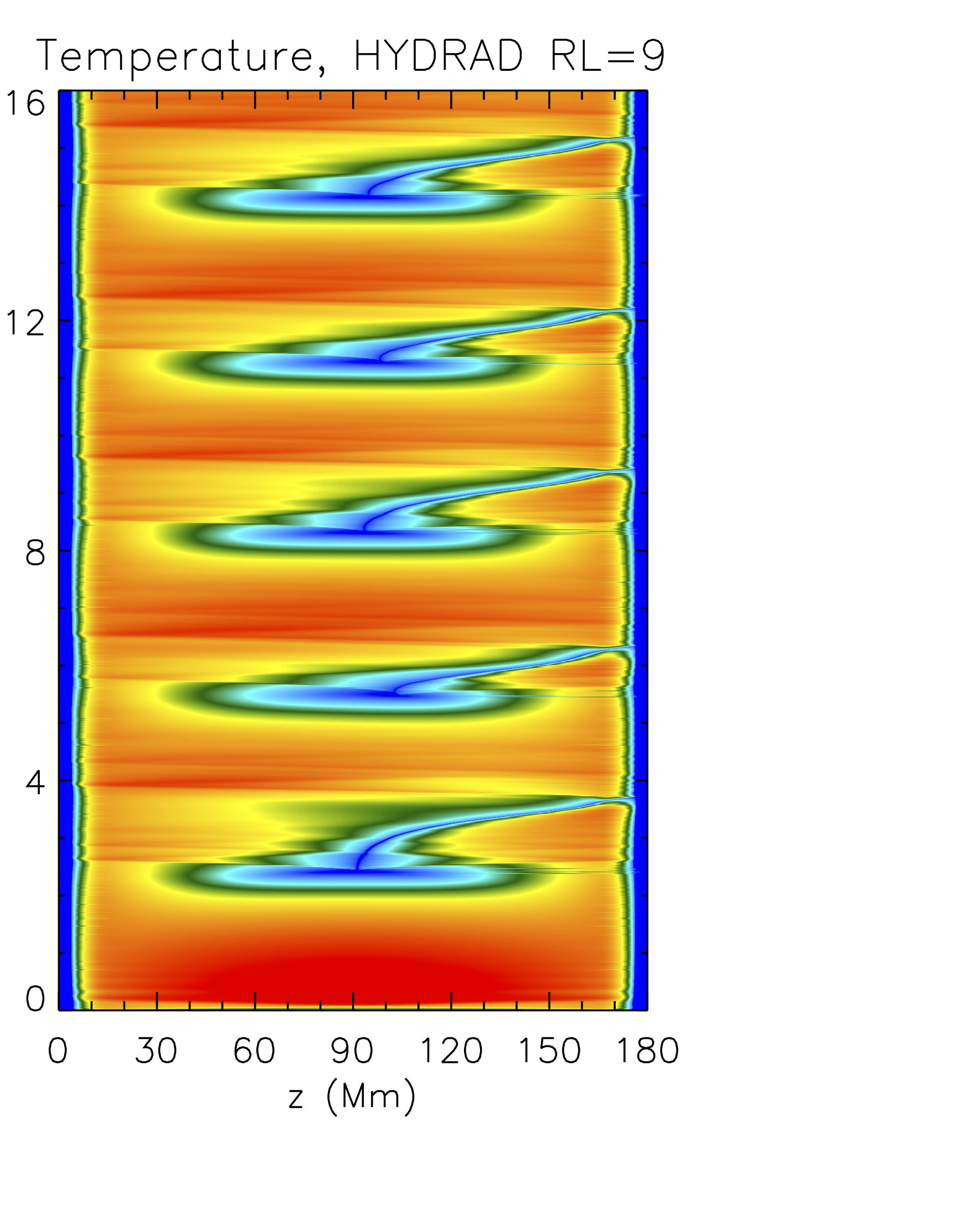}}
  \hspace*{-0.1\linewidth}
  \subfigure{\includegraphics[width=0.35\linewidth]
  {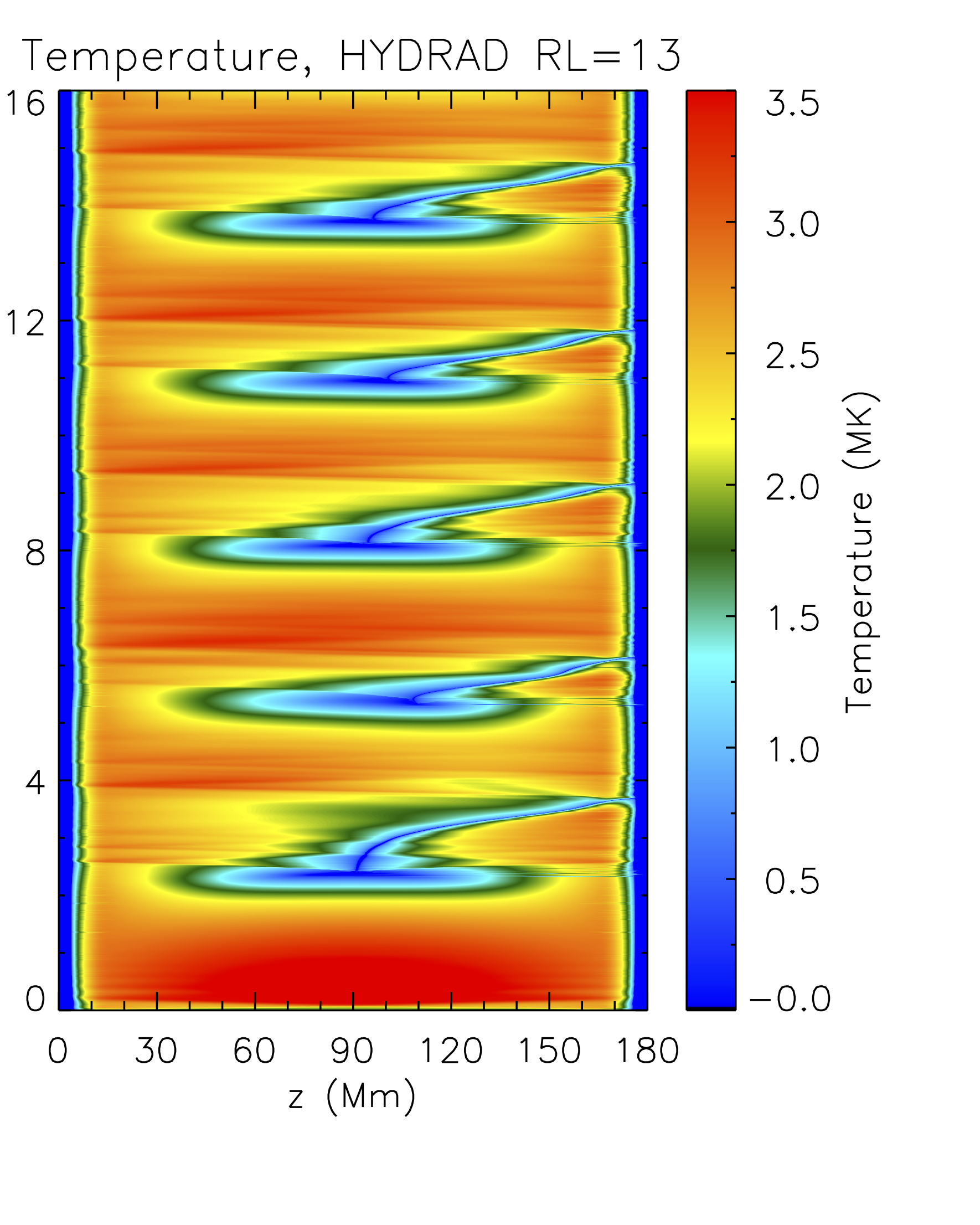}}
  \vspace*{-0.75cm}
  \caption{Influence of numerical resolution
    on the loop temperature and TNE cycle evolution for steady 
    footpoint heating using HYDRAD simulations. 
    Each plot shows the spatial dependence of temperature 
    (horizontal axis) and the temporal evolution (vertical 
    axis). 
    The various panels represent different values of RL, as 
    indicated above the panel. 
  \label{Fig:IoR_HYDRAD}
  }
\end{figure*}
  %
  %
  %%%%%%%%%%%%%%%%%%%%%%%%%%%%%%%%%%%%%%%%%%%%%%%%%%%%%%%%%%%%%    
  %
  % Fig:IoR_HYDRAD_RL_ca
  %
\begin{figure*}
  \vspace*{-0.3cm}
  \hspace*{1cm}
  \subfigure{\includegraphics[width=0.825\linewidth]
  {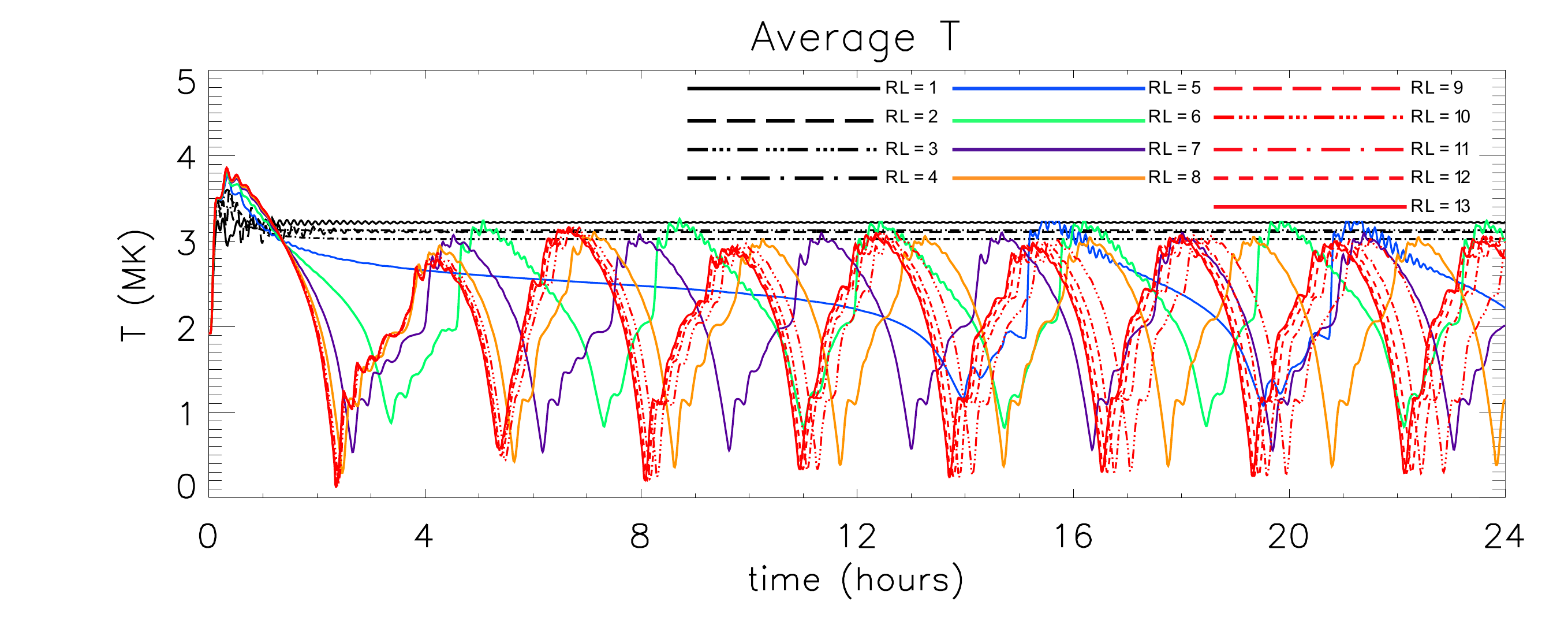}}
  \\[-0.5cm]
  \hspace*{1cm}
  \subfigure{\includegraphics[width=0.825\linewidth]
  {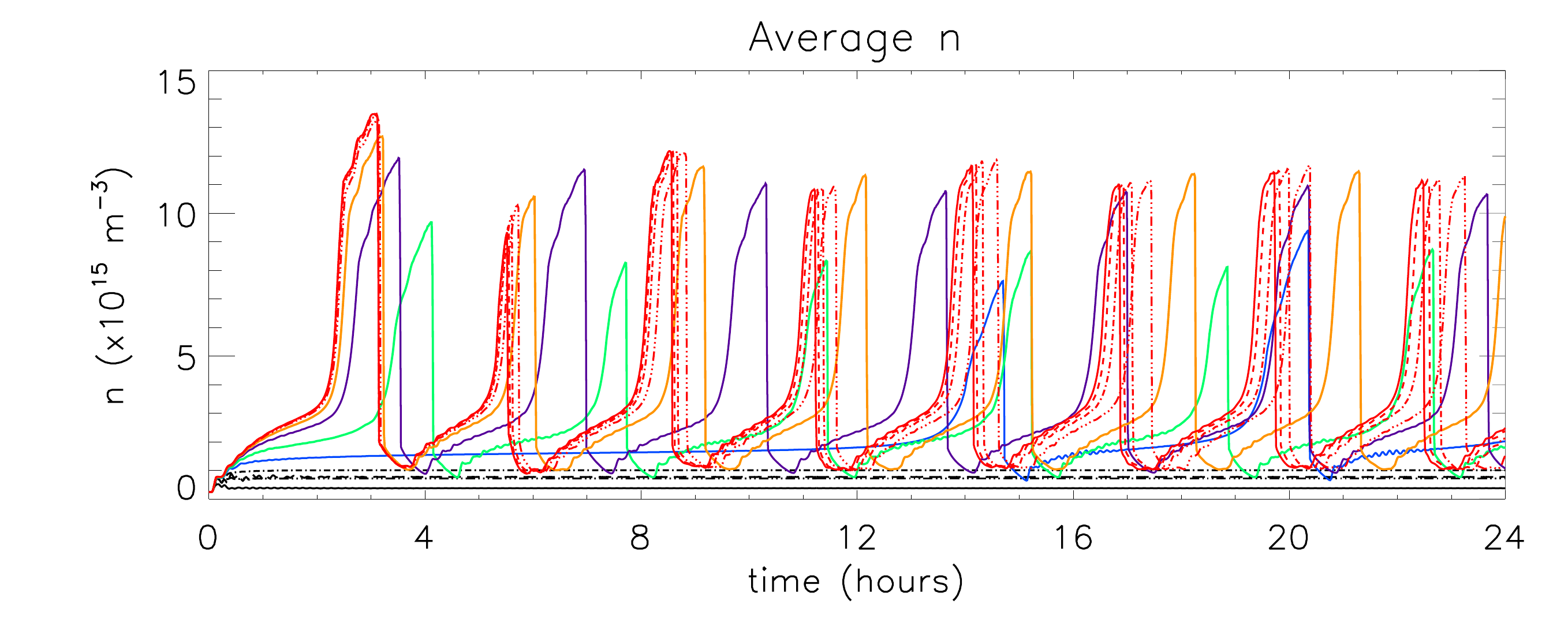}}
  \\[-0.5cm]
  \hspace*{1cm}
  \subfigure{\includegraphics[width=0.825\linewidth]
  {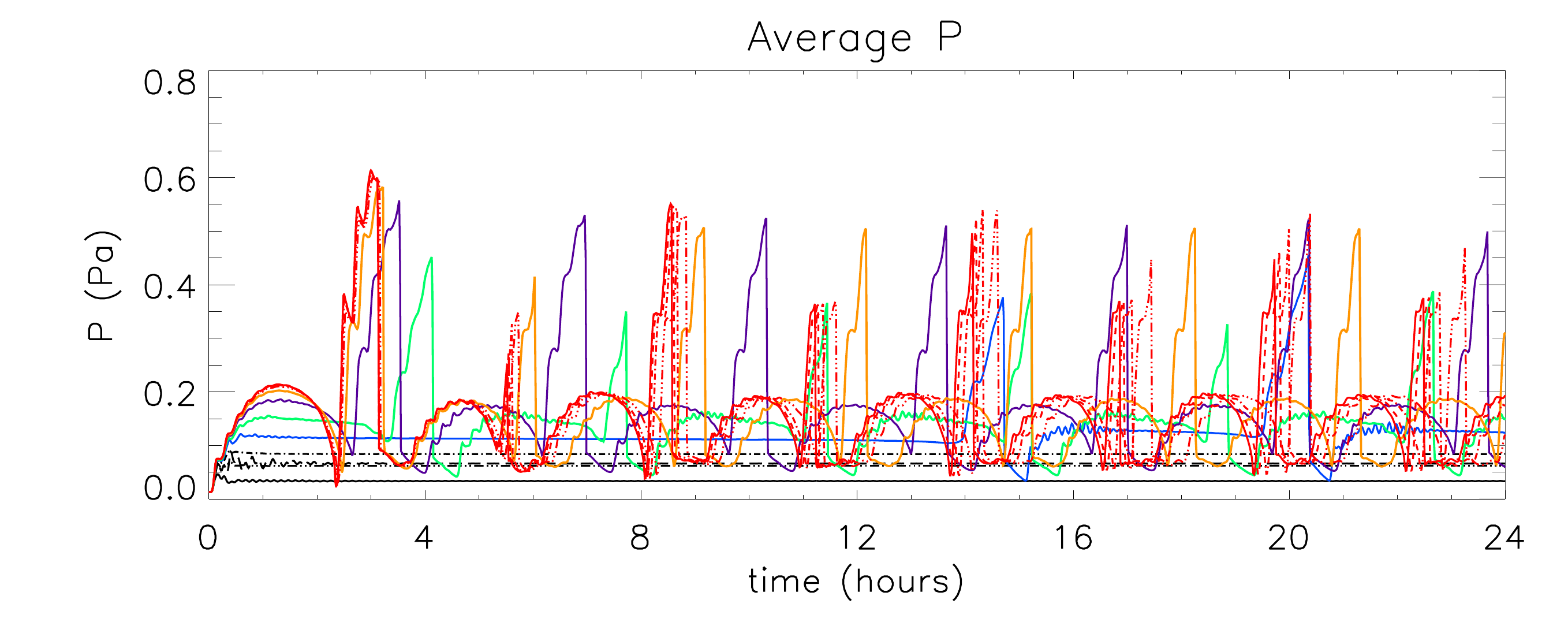}}
  \\[-0.5cm]
  \hspace*{1cm}
  \subfigure{\includegraphics[width=0.825\linewidth]
  {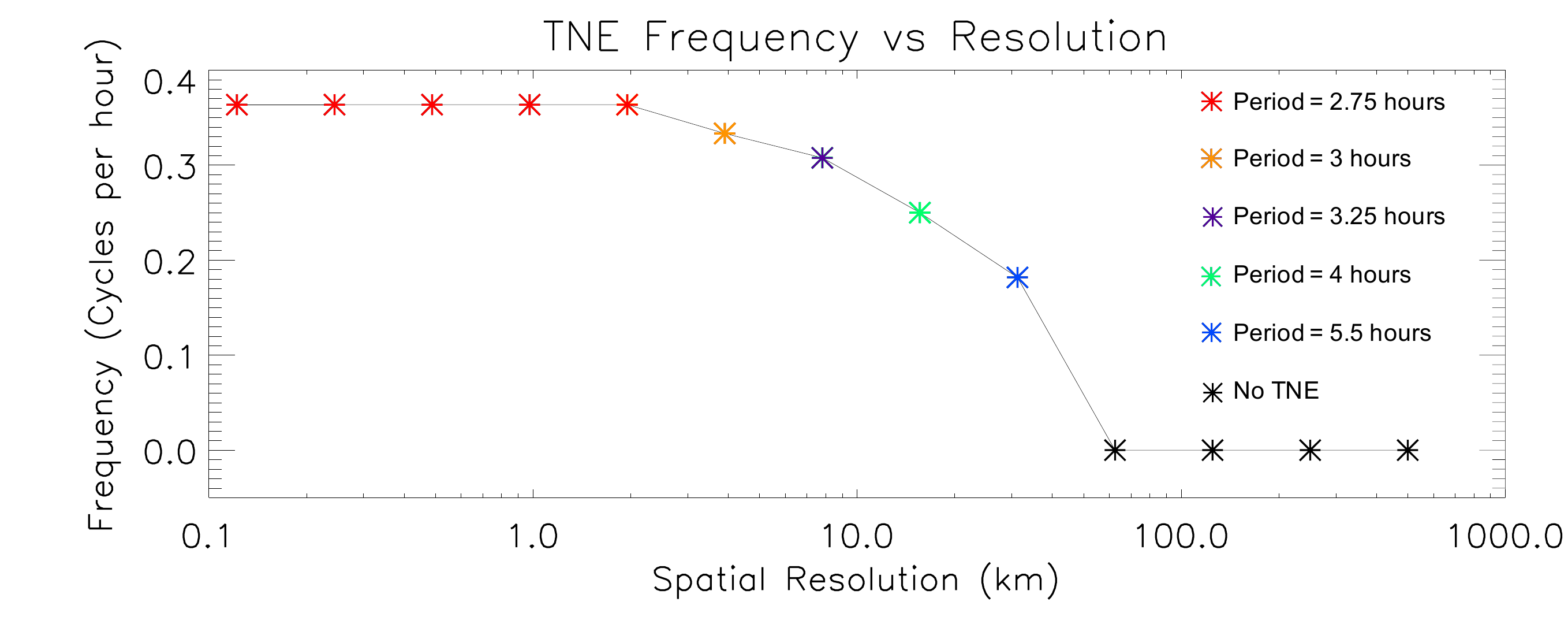}}
  \vspace*{-0.5cm}
  \caption{The top three panels show 
    the coronal averaged temperature, density and pressure as a
    function of time, for thirteen 
    values of RL. The lowest panel 
    shows how the
    TNE cycle frequency depends on the minimum permitted 
    spatial resolution (coarser resolution is associated with 
    smaller RL). The lines in the upper three panels are 
    colour-coded in a way that reflects the period of the TNE 
    cycle.
    \label{Fig:IoR_HYDRAD_RL_ca}
  }
\end{figure*}
  %
  %
  %%%%%%%%%%%%%%%%%%%%%%%%%%%%%%%%%%%%%%%%%%%%%%%%%%%%%%%%%%%%%
  %
  % Fig:IoR_Lare1d_vs_LareJ
  %
\begin{figure*}
  \subfigure{\includegraphics[width=0.5\linewidth]
  {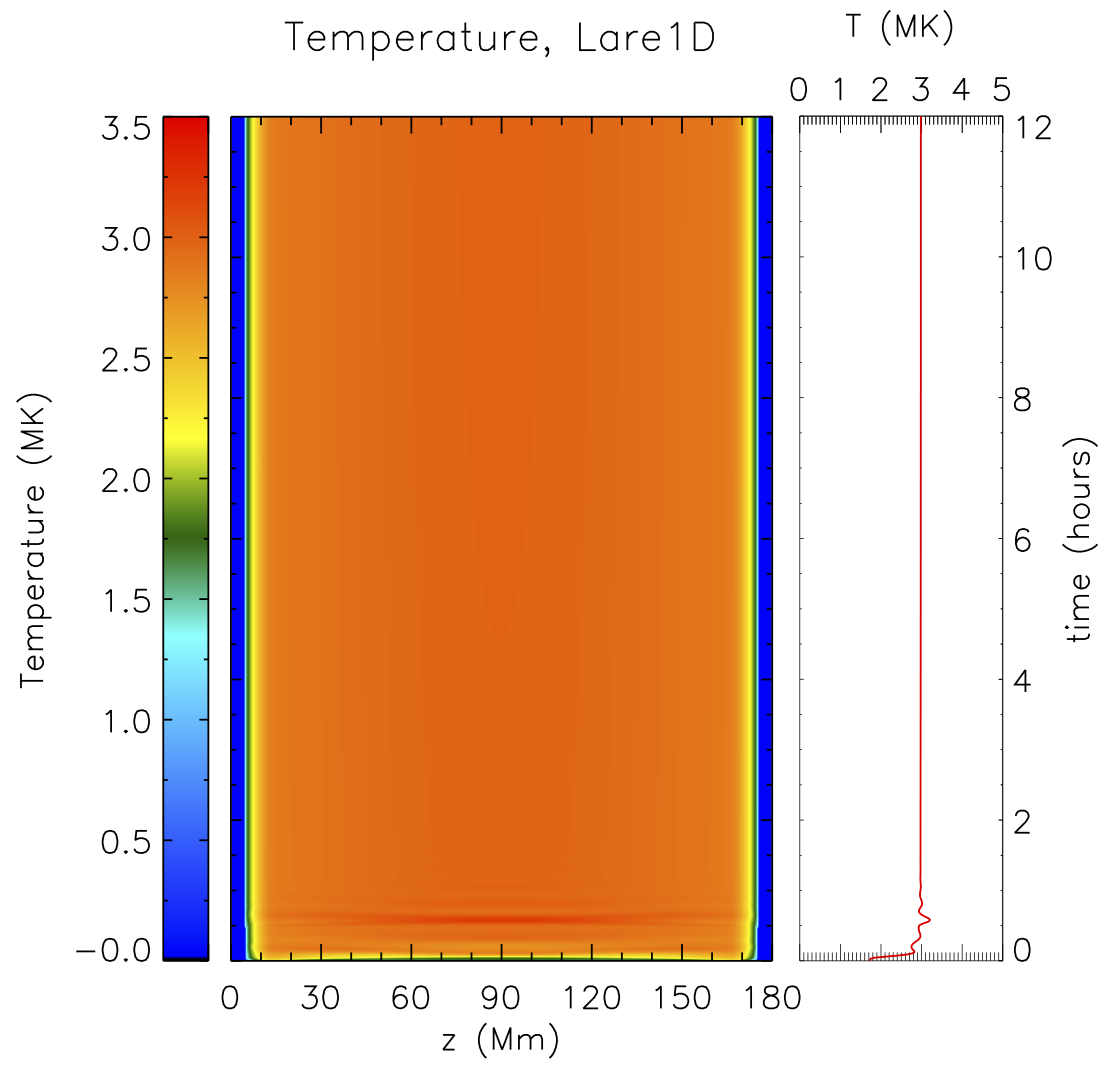}}
  \subfigure{\includegraphics[width=0.5\linewidth]
  {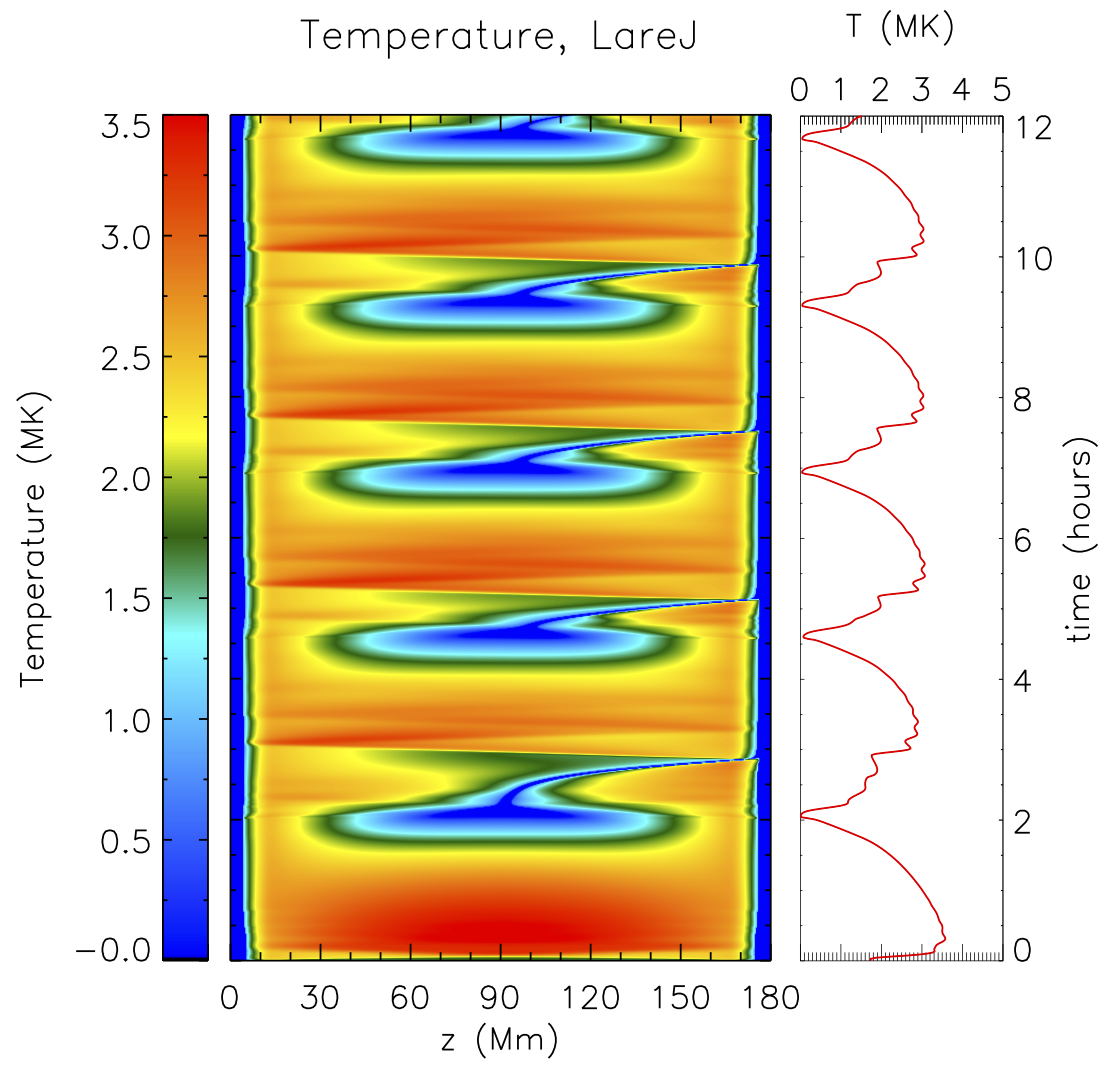}}
  \vspace*{-0.5cm}
  \caption{Effect of obtaining the correct evaporative 
    response on TNE cycles. Results for steady 
    footpoint heating. 
    The panels show the time evolution of
    the temperature as a function of position along the loop
    obtained in
    two different simulations, each run with 
    the same spatial
    resolution of  500 grid points
    along the length 
    of the loop (coarse resolution - 360 km).
    The left and right-hand panels correspond to the
    Lare1D  and LareJ 
    (Lare1D with the UTR
    jump condition method) solutions, 
    respectively. On the right of the 2D plots, we display the
    evolution of the coronal averaged temperature (computed by
    spatially averaging over the uppermost 25\% of the loop).
    \label{Fig:IoR_Lare1d_vs_LareJ}
    }
\end{figure*}
  %
  %
  %%%%%%%%%%%%%%%%%%%%%%%%%%%%%%%%%%%%%%%%%%%%%%%%%%%%%%%%%%%%%    
  %
  % Fig:IoR_HYDRAD_ct
  %
  \begin{figure*}
  \hspace*{1cm}
  \subfigure{\includegraphics[width=0.85\linewidth]  
    {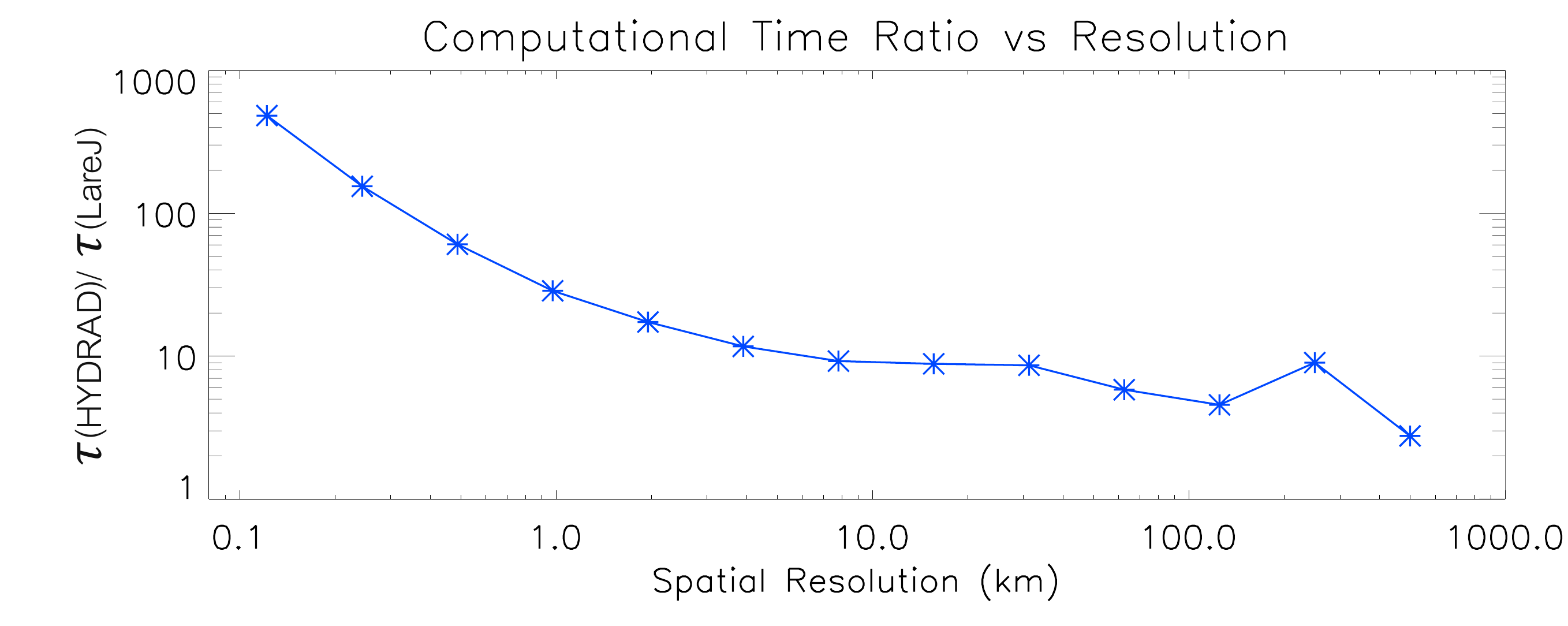}}
  \vspace*{-0.25cm}
  \caption{
    Numerical simulation computation times (run on 
    a single processor)
    for steady footpoint heating. 
    The panel shows the computational 
    time ratio between HYDRAD and LareJ 
    (Lare1D with 500 grid points and
    the UTR
    jump condition method) as a function of
    the HYDRAD spatial resolution for the different
    RL runs.
    \label{Fig:IoR_HYDRAD_ct}
    }
\end{figure*}
  %
  %
  %%%%%%%%%%%%%%%%%%%%%%%%%%%%%%%%%%%%%%%%%%%%%%%%%%%%%%%%%%%%%
  %
  %
  \subsubsection{Steady footpoint heating: HYDRAD simulations
  \label{Sect:Resolution_HYDRAD}}
  \indent
  Representative of the conditions necessary to induce TNE
  in coronal loop models 
  \citep[e.g.][]
  {paper:Mulleretal2003,paper:Antolinetal2010,
  paper:Peteretal2012,paper:Mikicetal2013,
  paper:Fromentetal2018},
  we first consider the case of steady 
  footpoint 
  heating where
  the spatial profile is
  given by the sum of two Gaussian peaks
  (one at each loop leg), each defined as,
  \begin{align}
    Q_H(z)
    =
    Q_{H_0}
    \exp
    \left(
    \dfrac{-(z-z_0)^2}{2z_H^2}
    \right).
  \end{align}
  These peaks are
  localised between the base of the corona and
  base of the TR 
  with a maximal value
  at $z=12.5$ Mm and we take
  $z_H=1.5$ Mm as the length scale of heat deposition.
  This is
  shown in 
  Figure \ref{tne_T_Q_180Mm_steady} as the blue curve. 
  We note that in this part of the paper a 
  small spatially uniform
  background heating term is always present so that
  $Q(z) = Q_{bg} + Q_{H}(z)$. This is commonly done in 1D loop 
  models to ensure that $T$ and $n$ remain positive: here 
  $Q_{bg}=6.8682 \times 10^{-6}$ Jm$^{-3}$s$^{-1}$
  and the effect of including a range of 
  values of $Q_{bg}$ will be examined in
  Section \ref{Sect:Q_bg}. 
  Moreover,
  in order to avoid unrealistic pile up of condensations 
  at the loop apex due to the symmetry of the model, 
  the spatial symmetry of the heating profile
  is perturbed by adding a small enhancement
  of 0.4\% to the Gaussian peak at the
  left-hand leg of the loop. The initial state of the loop is 
  determined using just $Q_{bg}$, leading to a temperature of 
  order 1 MK. The footpoint heating is then ramped up linearly 
  over 30 s to a constant value with a peak of $Q_{bg} + 
  Q_{H_0}$ with $Q_{H_0}=3.5\times10^{-3}$Jm$^{-3}$s$^{-1}$. 
  This
  gives a maximum temperature of
  approximately $3.5$ MK
  \citep[a similar energy input and loop length as  
  Models 1 and 2 in][]{paper:Mikicetal2013}.
  \\
  \indent 
  We run the HYDRAD code in single fluid mode
  and perform the steady footpoint heating simulations for a 
  sequence of refinement levels: 
  RL = [1, 2, 3, 4, 5, 6, 7, 8, 9, 10, 11, 12, 13].
  This 
  corresponds 
  to grid sizes that range from 
  500 km for one level of
  refinement (RL=1) 
  down to 122 m in the case of maximum refinement (RL=13).
  \\
  \indent
  The results are shown in Figure \ref{Fig:IoR_HYDRAD}. Each 
  panel shows the temperature as a function of position 
  (horizontal axis) at a sequence of times (vertical axis). The 
  temporal snapshots are shown every 54 seconds. The 
  refinement levels are indicated above each panel, and 
  increase going from upper left to lower right. The 
  simulations are identical in all respects except for the 
  value of RL. [We note that RL = 3, 10, 11 and 
  12 are not shown.]
  There are major differences in the evolution as RL 
  increases.
  The cases with RL=[1, 2, 4]
  settle to  
  static equilibria while for
  RL=[5, 6, 7, 8] there is 
  TNE with condensations, but in each case the
  cycles have a 
  different period ranging from 5.5 to 3 hours.
  Convergence of the TNE cycle period and 
  thermodynamic evolution 
  (i.e. the same temperature and density extrema) 
  is seen only for $\textrm{RL}\geq 9$, thus requiring a TR 
  grid resolution of
  1.95 km or better.
  \\
  \indent
  Figure \ref{Fig:IoR_HYDRAD_RL_ca} shows 
  the temporal evolution of the coronal averaged 
  temperature ($T$), density ($n$) and pressure ($P$) 
  for all the values of RL (upper three panels)
  and the dependence of TNE cycle frequency  
  on the minimum 
  spatial resolution (lowest panel).
  The coronal averages are calculated by spatially
  averaging over the uppermost 25\% of the loop.
  These quantities are particularly useful for 
  demonstrating the range of coronal responses obtained
  in the thirteen simulations run with different values
  of RL 
  and the periods of the TNE cycles are 
  estimated 
  from the troughs in the coronal averaged temperatures.
  In the upper three plots each value of RL is associated with 
  a specific colour
  that we associate with a particular cycle period: these 
  colours can also be seen in the star symbols in the lower 
  panel.  
  For example, the red lines and stars correspond to 
  simulations where the TNE cycle evolution has
  a period of 2.75 hours
  (and the various RL values within this group are 
  separated by different 
  line styles). This Figure confirms the earlier conclusion of 
  the importance of adequate resolution on obtaining the 
  correct TNE cycle behaviour. 
  Even if computationally one can achieve a TR resolution 
  of 10 km, then an error in the cycle period of order 20\% 
  is still to be expected.
  \\
  \indent
  We now turn our attention to understanding 
  why the loops computed with different levels of 
  spatial resolution show such significant inconsistencies in 
  their temporal evolution.
  We start by considering the
  first TNE cycle of the RL=13 loop.
  For the first 30 minutes, the temperature
  and density in the corona both
  increase in response to the ramped up footpoint heating, the 
  density by the usual evaporation process 
  \citep{paper:Antiochos&Sturrock1978,
  paper:Klimchuketal2008,
  paper:Cargilletal2012a}. 
  The subsequent evolution follows the familiar TNE pattern 
  with the temperature falling quite rapidly from 
  3.9 MK to $10^4$K  
  between 30 and approximately 150 minutes, during which time
  the coronal density continues to increase.
  The rapid cooling is
  driven locally by the thermal instability
  and leads to the
  formation of the 
  condensation at the loop apex, as found by others 
  \citep[e.g.][]
  {
  paper:Mulleretal2003,
  paper:Mulleretal2004,
  paper:Mulleretal2005, 
  paper:Moketal2008, 
  paper:Antolinetal2010, 
  paper:Susinoetal2010,
  paper:Peteretal2012,
  paper:Lionelloetal2013,
  paper:Mikicetal2013,
  paper:Susinoetal2013, 
  paper:Moketal2016,
  paper:Fromentetal2018}.
  The condensation has a peak density of around 
  $14 \times 10^{15}$ $\rm{m}^{-3}$ at 195 minutes 
  but then 
  quickly falls down the right-hand leg of the loop.
  After 195 minutes, the coronal density decreases 
  due to the draining of the \lq condensed\rq\
  plasma back into the TR and chromosphere.
  After this stage, the coronal plasma
  is reheated and coronal temperatures re-reached.
  The TNE  cycle then repeats with a period of
  about 2.75 hours.
  We note though that the coronal temperature and density 
  oscillate
  throughout the evolution (upper panels of Figure 2)
  due to  the shock waves that are generated
  during the formation of the condensation
  and when the mass associated with the condensation falls
  down the loop leg
  \citep{paper:Mulleretal2003,paper:Mulleretal2004}.
  \\
  \indent The 
  examples with RL=[9, 10, 11, 12] all 
  behave in an identical way, 
  while though RL = [7,8] show differences in
  the cycle period, the error in the 
  density and 
  temperature when averaged over a cycle are just 7\% and 3\% 
  respectively.
  \\
  \indent
  In contrast, the behaviour of the loop computed with 
  one level of refinement (RL=1, 500 km resolution) is 
  completely different. 
  Initially, the temperature in the corona increases 
  but the evaporative
  response is significantly underestimated.
  Rather than passing through the TR continuously in a series 
  of steps, the heat flux jumps across the TR.
  The incoming energy is then strongly radiated 
  (BC13), leaving
  little residual heat flux to drive the upflow.
  Therefore, the lack of spatial resolution leads to an 
  enthalpy flux and
  coronal
  density that are artificially low for the 
  prescribed 
  heating profile.
  This ensures that the loop remains thermally stable.
  The outcome is that after a transient phase of 
  around 1 hour,  
  the loop settles to a static
  equilibrium with a coronal temperature and density of
  3.2 MK and $0.4\times 10^{15}$ $\rm{m}^{-3}$,
  respectively. The loops calculated with RL=[2, 3, 4]
  all show broadly similar behaviour while RL = 5 and 6 are
  transition cases.
  %
  %
  %%%%%%%%%%%%%%%%%%%%%%%%%%%%%%%%%%%%%%%%%%%%%%%%%%%%%%%%%%%%%     
  %
  % Table:LareJ_td_fph_simulations
  %
  \begin{table*}
    \caption{
    \label{Table:LareJ_td_fph_simulations}
    A summary of the parameter space used and results from
    the time dependent footpoint heating cases
    computed with LareJ.}
    \centering
    \resizebox{\hsize}{!}
    {
\begin{tabular}{lcccccc}
    \hline\hline
    \\
    Case & $t_d$ & $t_w$ &
    $Q_{H_0}$ &
    Behaviour & TNE Cycle & Heating Cycle
    \\
    & (s) & (s) & factor  & 
    &  Period (hrs) & Period (s)
    \\
    \hline
    1  & 62.5 & 62.5  & 2 &
    TNE with condensations &
    2.25 & 125                        
    \\
    2  & 125 & 125  & 2 &
    TNE with condensations &
    2.5 & 250             
    \\
    3  & 250 & 250  & 2 &
    TNE with condensations &
    2.75 & 500          
    \\
    4  & 500 & 500  & 2 &
    TNE with condensations &
    2.75 & 1000               
    \\
    5  & 1000 & 1000  & 2 &
    TNE with condensations &
    3.0 & 2000             
    \\
    6  & 2000 & 2000  & 2 &
    TNE with condensations, 
    global cooling \& draining &
    3.5 & 4000              
    \\
    7  & 4000 & 4000  & 2 &
    Global cooling \& draining &
    - & 8000             
    \\
    8  & 8000 & 8000  & 2 &
    Catastrophic cooling with 
    global cooling \& draining &
    - & 16000                  
    \\
    9  & 125 & 375  & 4 &
    TNE with condensations &
    3.25 & 500          
    \\
    10  & 250 & 750  & 4 &
    TNE with condensations &
    3.25 & 1000               
    \\
    11  & 500 & 1500  & 4 &
    TNE with condensations, 
    global cooling \& draining &
    3.5 & 2000               
    \\
    13  & 1000 & 3000  & 4 &
    Global cooling \& draining &
    - & 4000          
    \\
    13   & 2000 & 6000  & 4 &
    Global cooling \& draining &
    - & 8000 
    \\
    14  & 125 & 875  & 8 &
    TNE with condensations &
    3.75 & 1000        
    \\
    15  & 250 & 1750  & 8 &
    Global cooling \& draining &
    - & 2000  
    \\
    16  & 500 & 3500  & 8 &
    Global cooling \& draining &
    - & 4000   
    \\    
    \hline
    \end{tabular}
    }
    \tablefoot{
      From left to right
      the columns show the case number,
      the heating duration and waiting times 
      that comprise a single
      heating cycle, 
      the amplification factor 
      for the 
      peak heating rate ($Q_{H_0}$) 
      that is required to ensure
      the average total energy released is equivalent 
      to the steady footpoint heating simulation,
      the characteristic simulation behaviour 
      and the periods of the TNE and heating cycles,
      respectively.
    }
  \end{table*}
  %
  %
  %
  %
  %%%%%%%%%%%%%%%%%%%%%%%%%%%%%%%%%%%%%%%%%%%%%%%%%%%%%%%%%%%%%    
  %
  % Fig:td_fph_LareJ_T
  %
  \begin{figure*}
  \hspace*{0.05\linewidth}
  \subfigure{
  \includegraphics[width=0.35\linewidth]
  {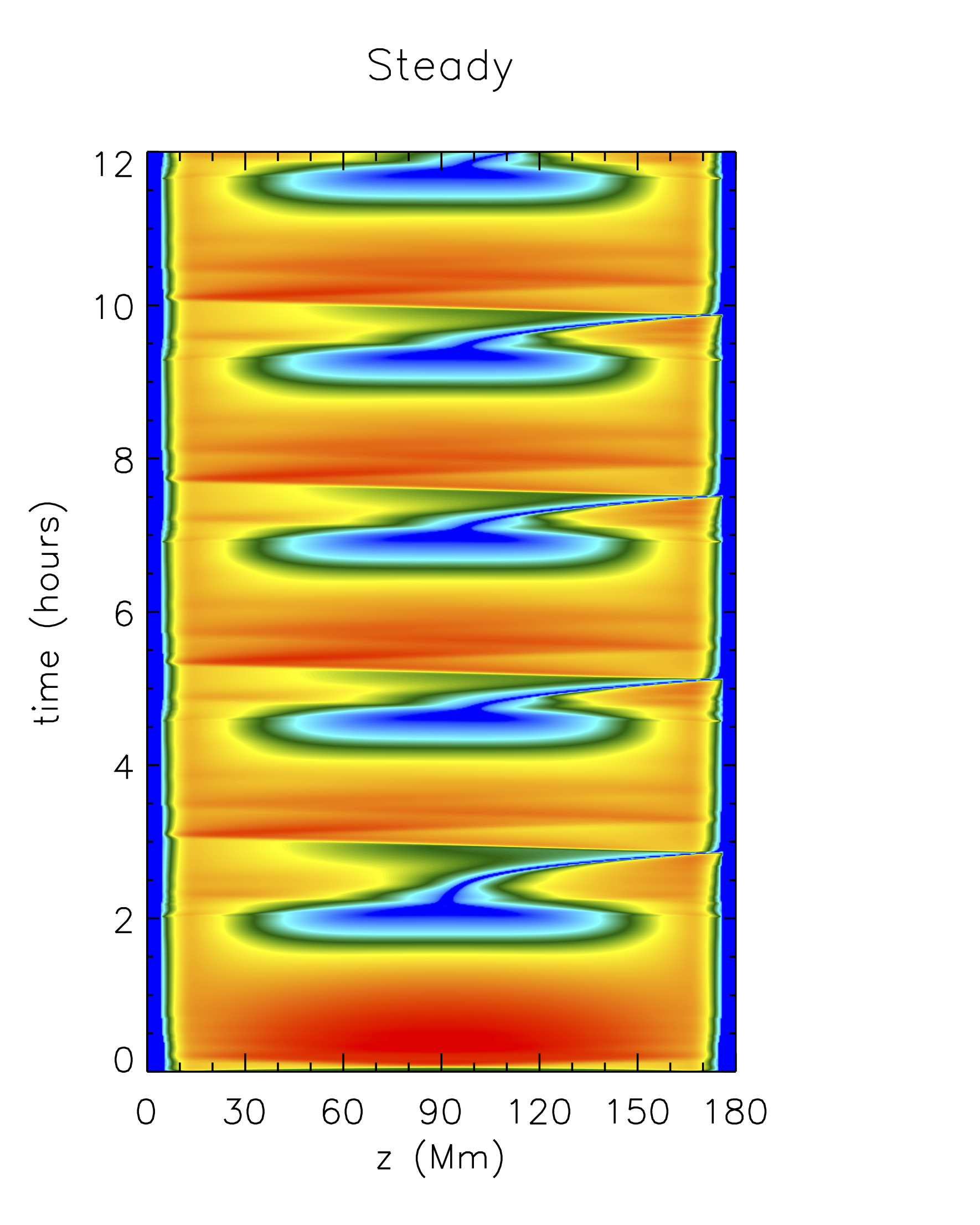}}
  \hspace*{-0.08\linewidth}
  \subfigure{  
  \includegraphics[width=0.35\linewidth]
  {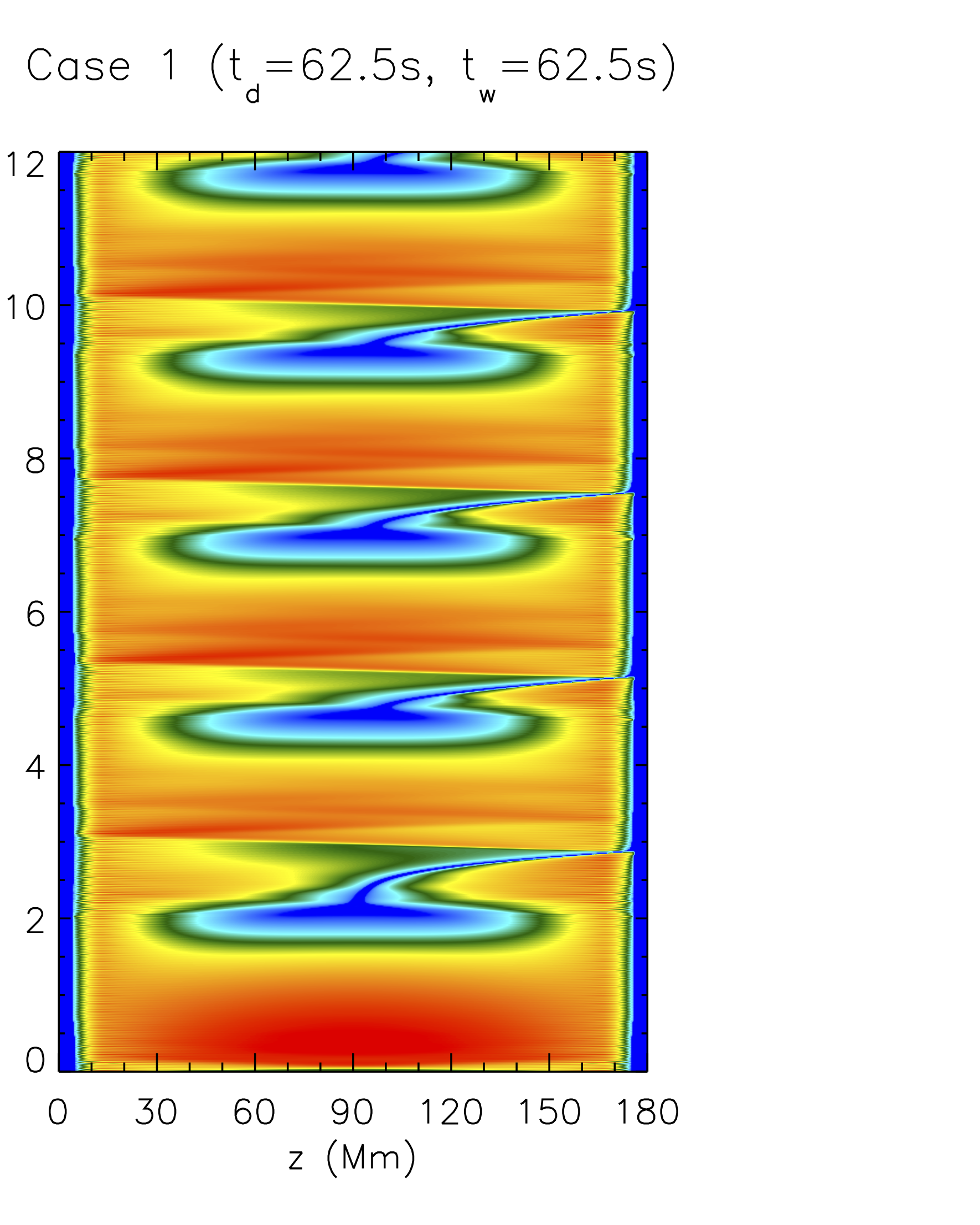}}
  \hspace*{-0.11\linewidth}
  \subfigure{  
  \includegraphics[width=0.35\linewidth]
  {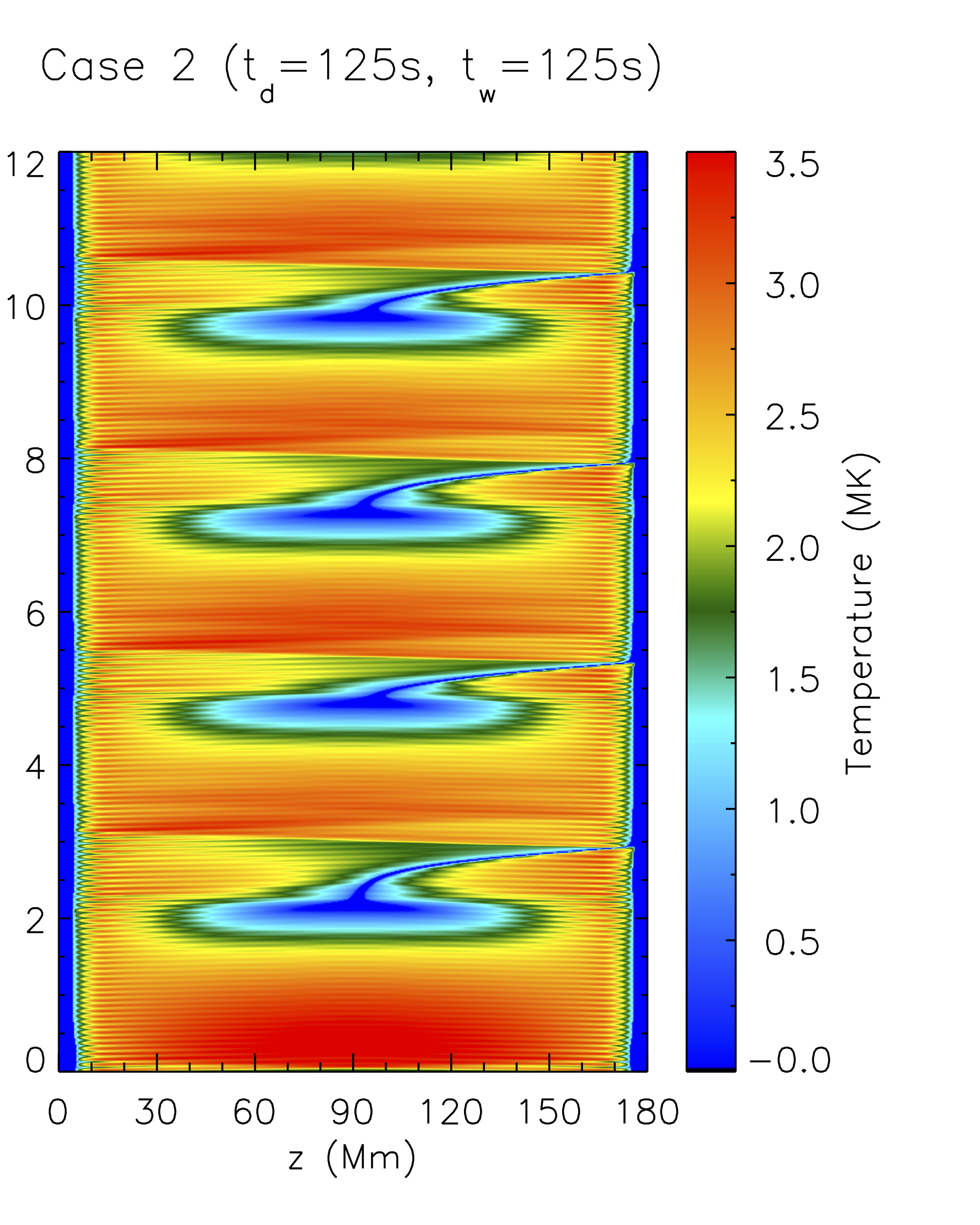}}
  \\[-0.5cm]
  \hspace*{0.05\linewidth}
  \subfigure{
  \includegraphics[width=0.35\linewidth]
  {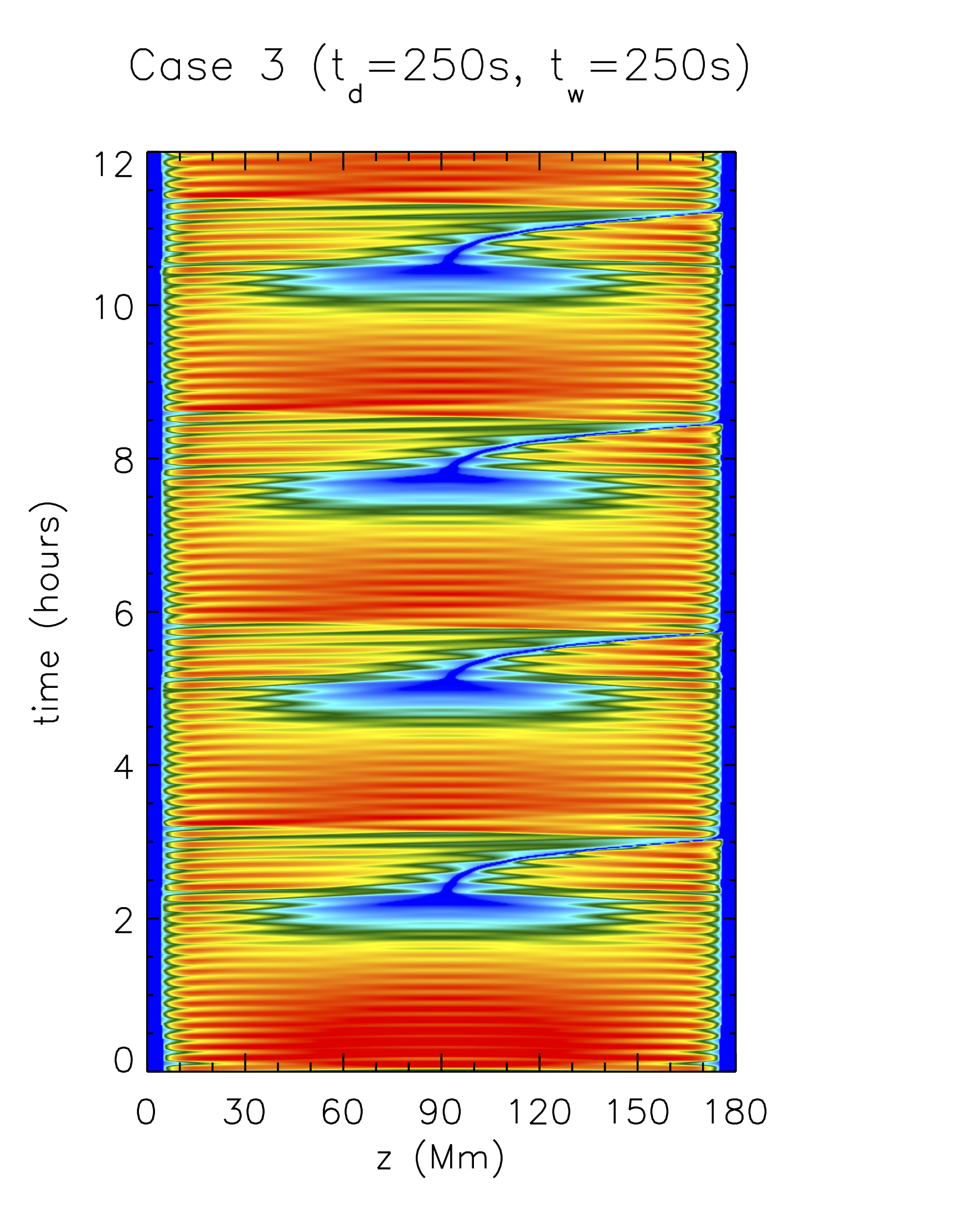}}
  \hspace*{-0.08\linewidth}
  \subfigure{  
  \includegraphics[width=0.35\linewidth]
  {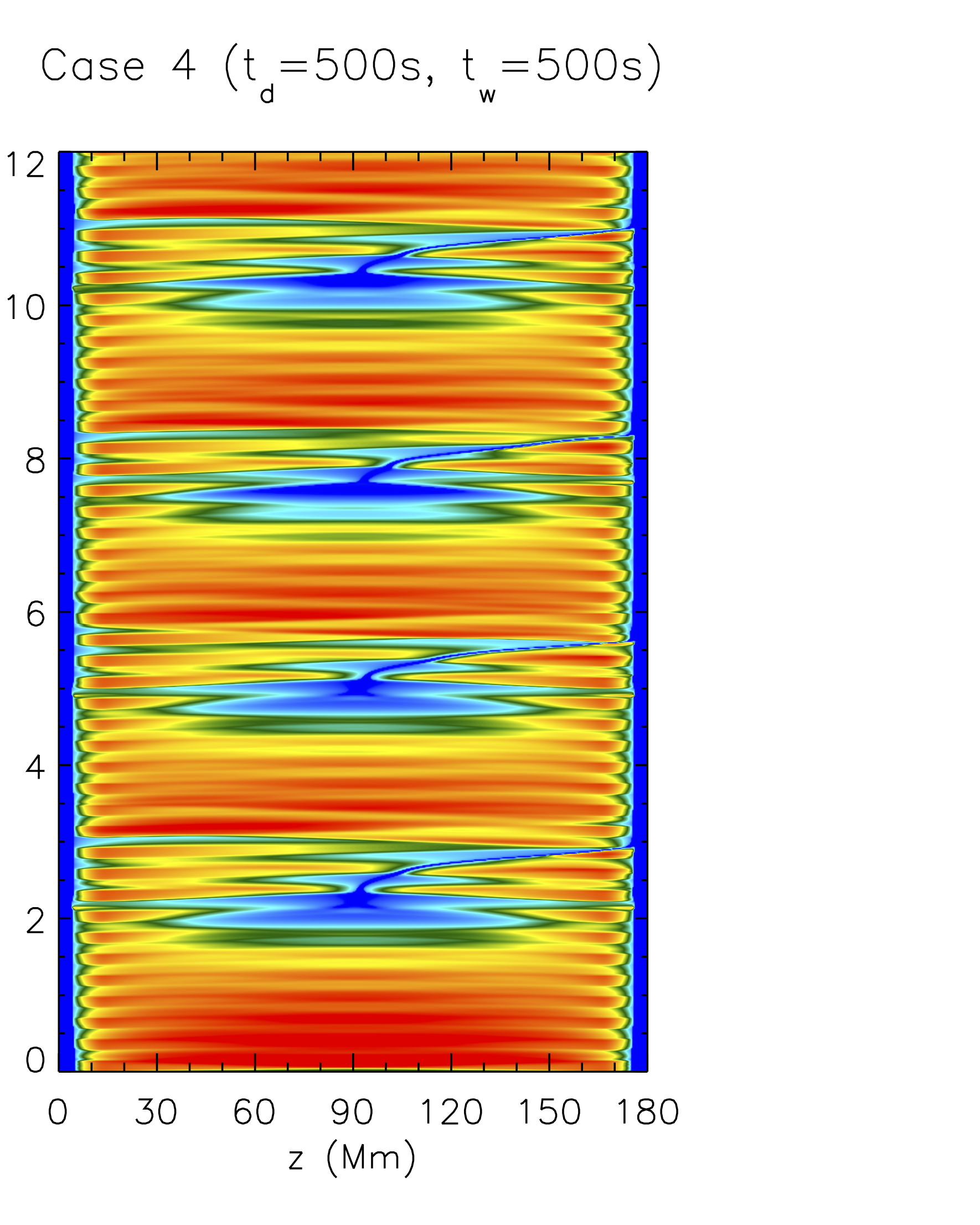}}
  \hspace*{-0.11\linewidth}
  \subfigure{  
  \includegraphics[width=0.35\linewidth]
  {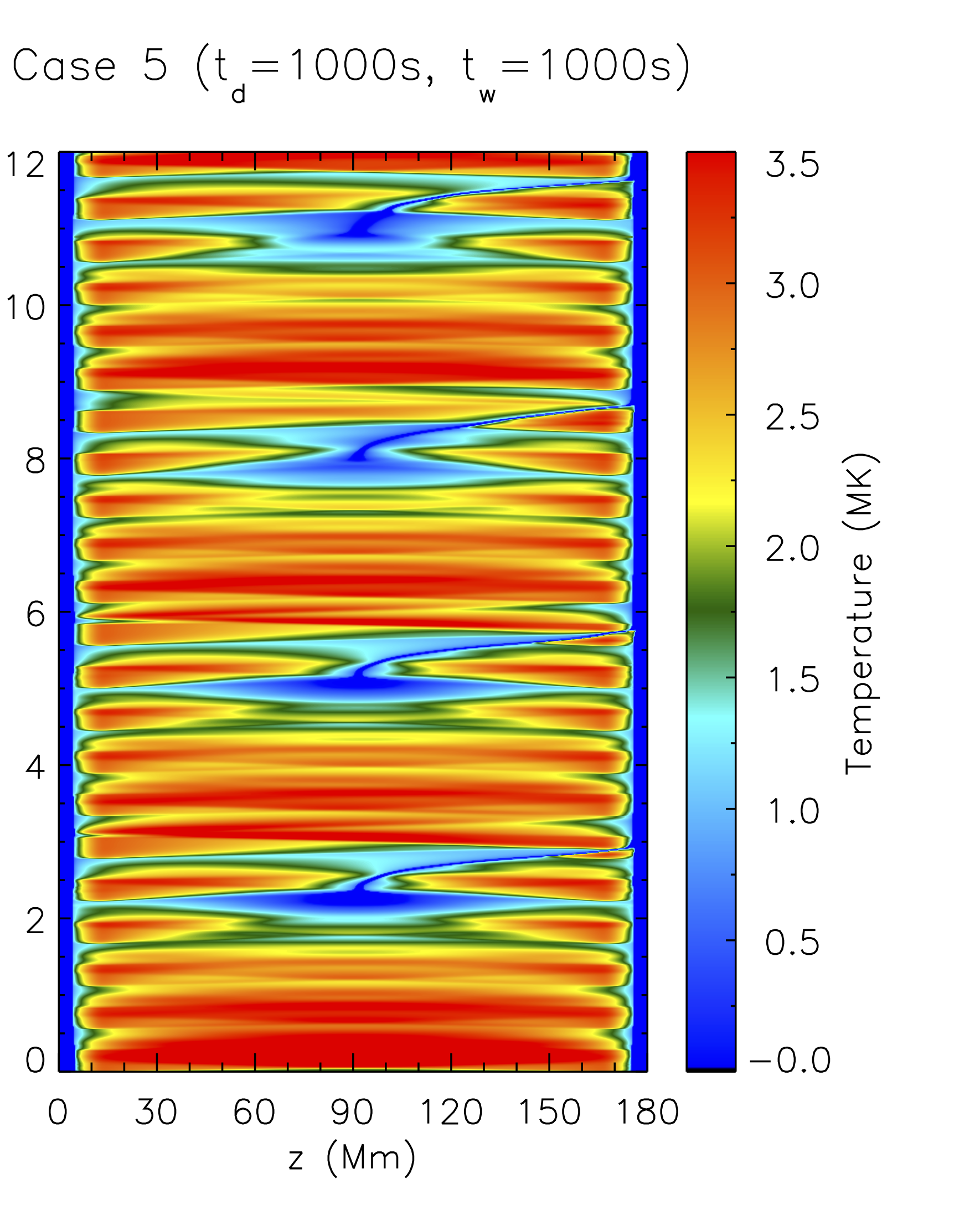}}
  \\[-0.5cm]
  \hspace*{0.05\linewidth}
  \subfigure{
  \includegraphics[width=0.35\linewidth]
  {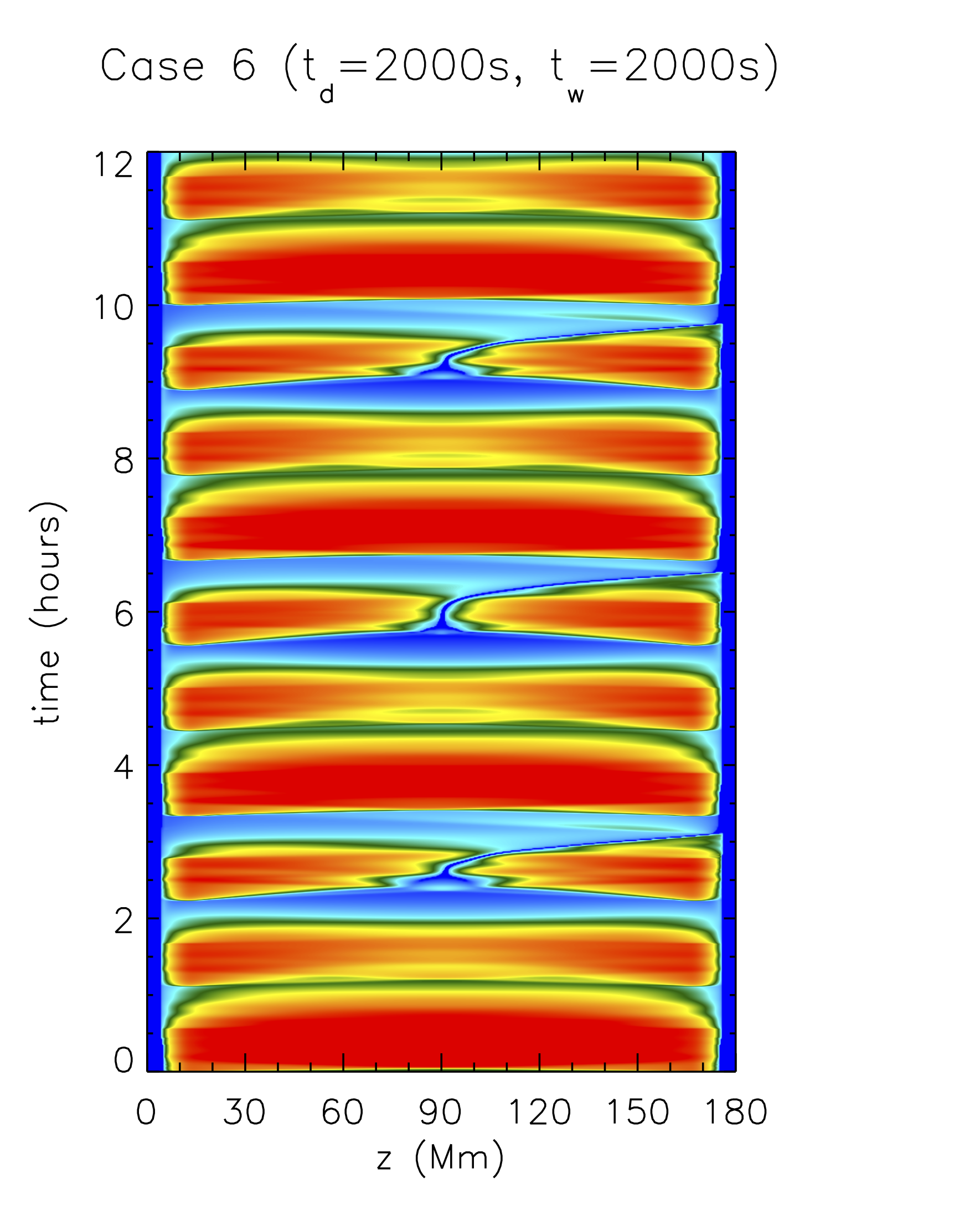}}
  \hspace*{-0.08\linewidth}
  \subfigure{  
  \includegraphics[width=0.35\linewidth]
  {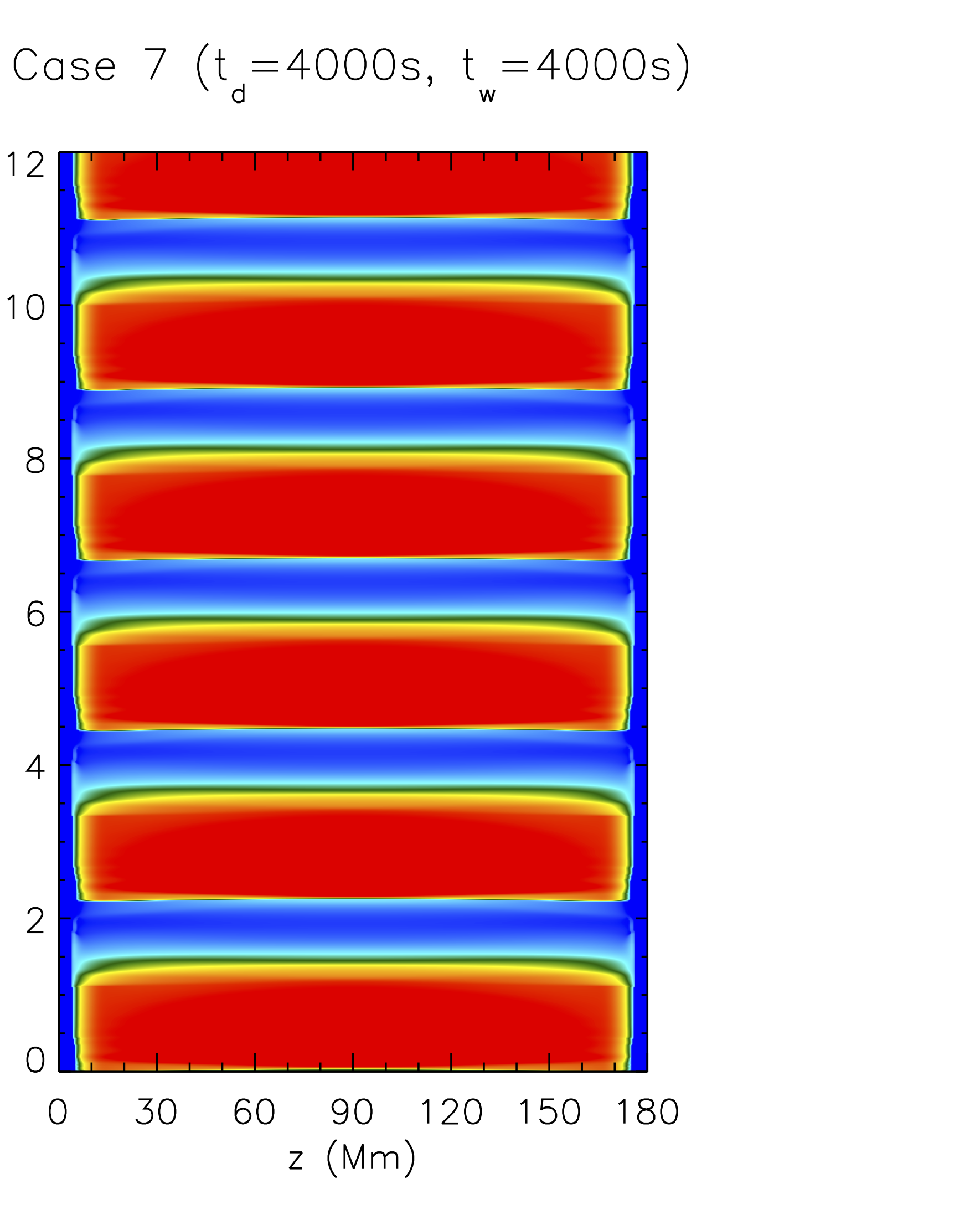}}
  \hspace*{-0.11\linewidth}
  \subfigure{ 
  \includegraphics[width=0.35\linewidth]
  {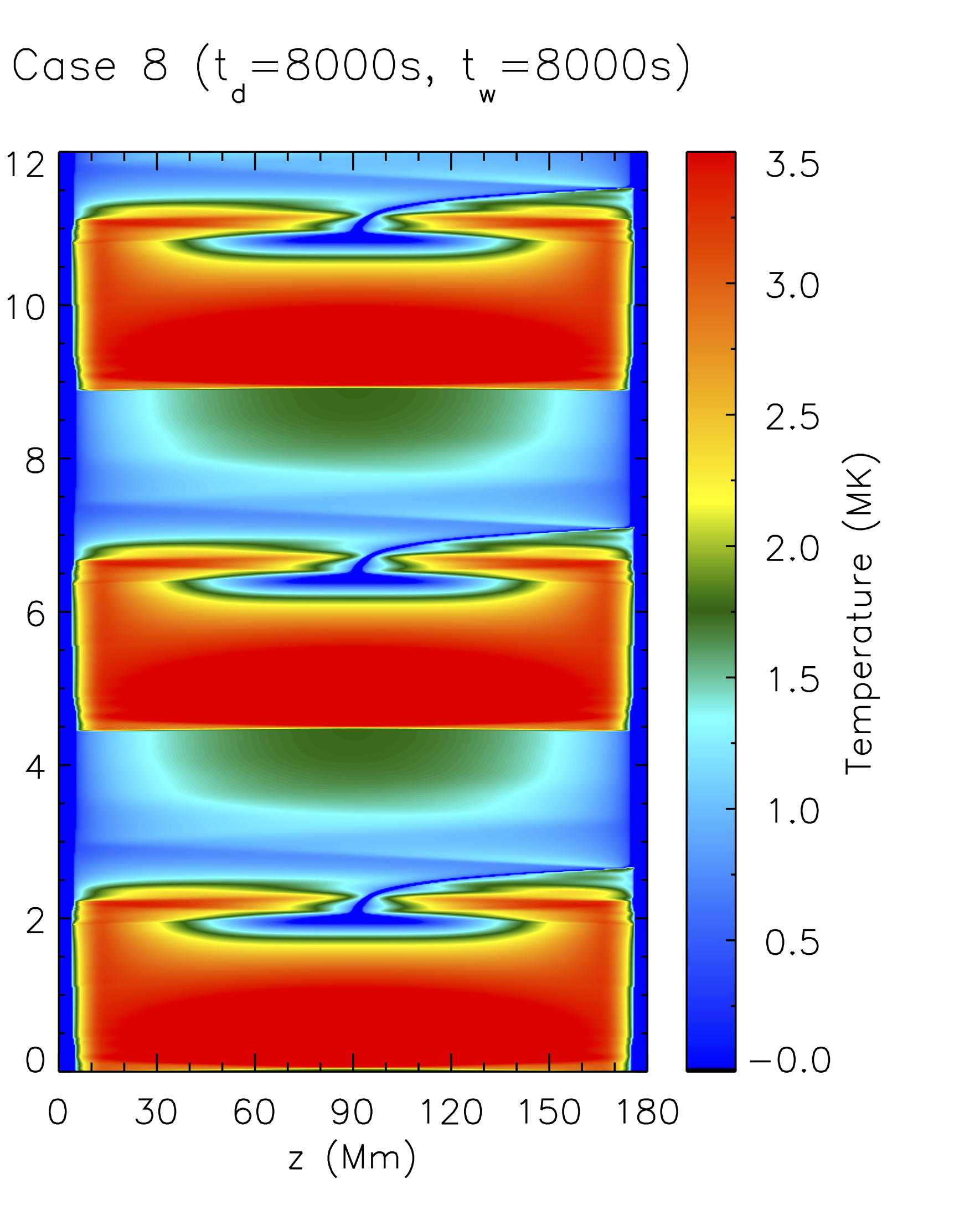}}
  \\[-0.5cm]
  \caption{
    Effect of heating timescales
    on TNE cycles.
    LareJ results for time dependent footpoint heating
    Cases 1--8 together with the steady 
    footpoint heating result.
    The panels show the time evolution of the
    temperature as a function of position along the loop.
    \label{Fig:td_fph_LareJ_T}
    }
\end{figure*}
  %
  %
  %%%%%%%%%%%%%%%%%%%%%%%%%%%%%%%%%%%%%%%%%%%%%%%%%%%%%%%%%%%%%    
  %
  % Fig:td_fph_LareJ_Tn_ca
  %
  \begin{figure*}
  \hspace*{-0.025\linewidth}
  \subfigure{\includegraphics[width=0.35\linewidth]
  {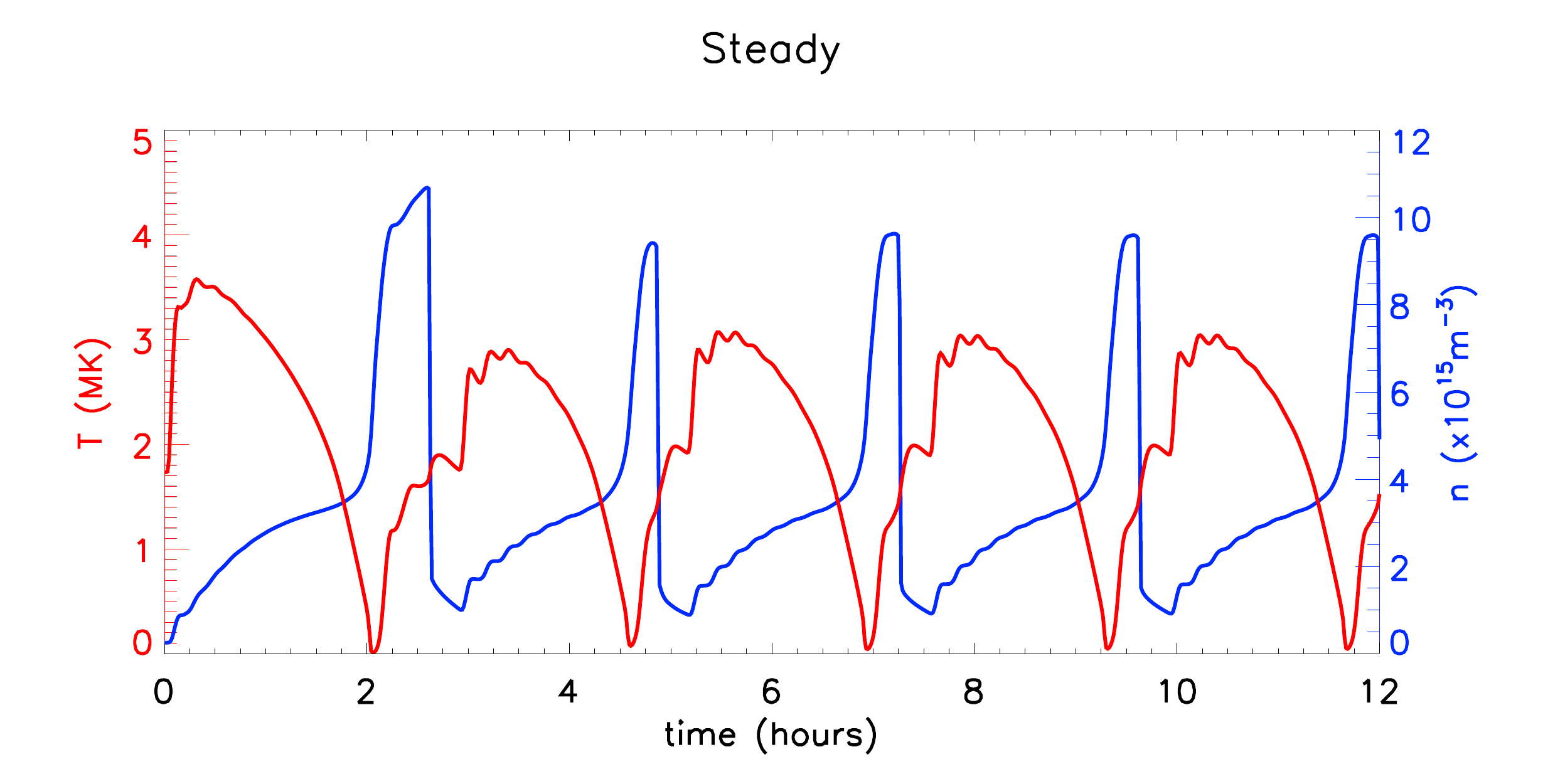}}
  \hspace*{-0.025\linewidth}
  \subfigure{\includegraphics[width=0.35\linewidth]
  {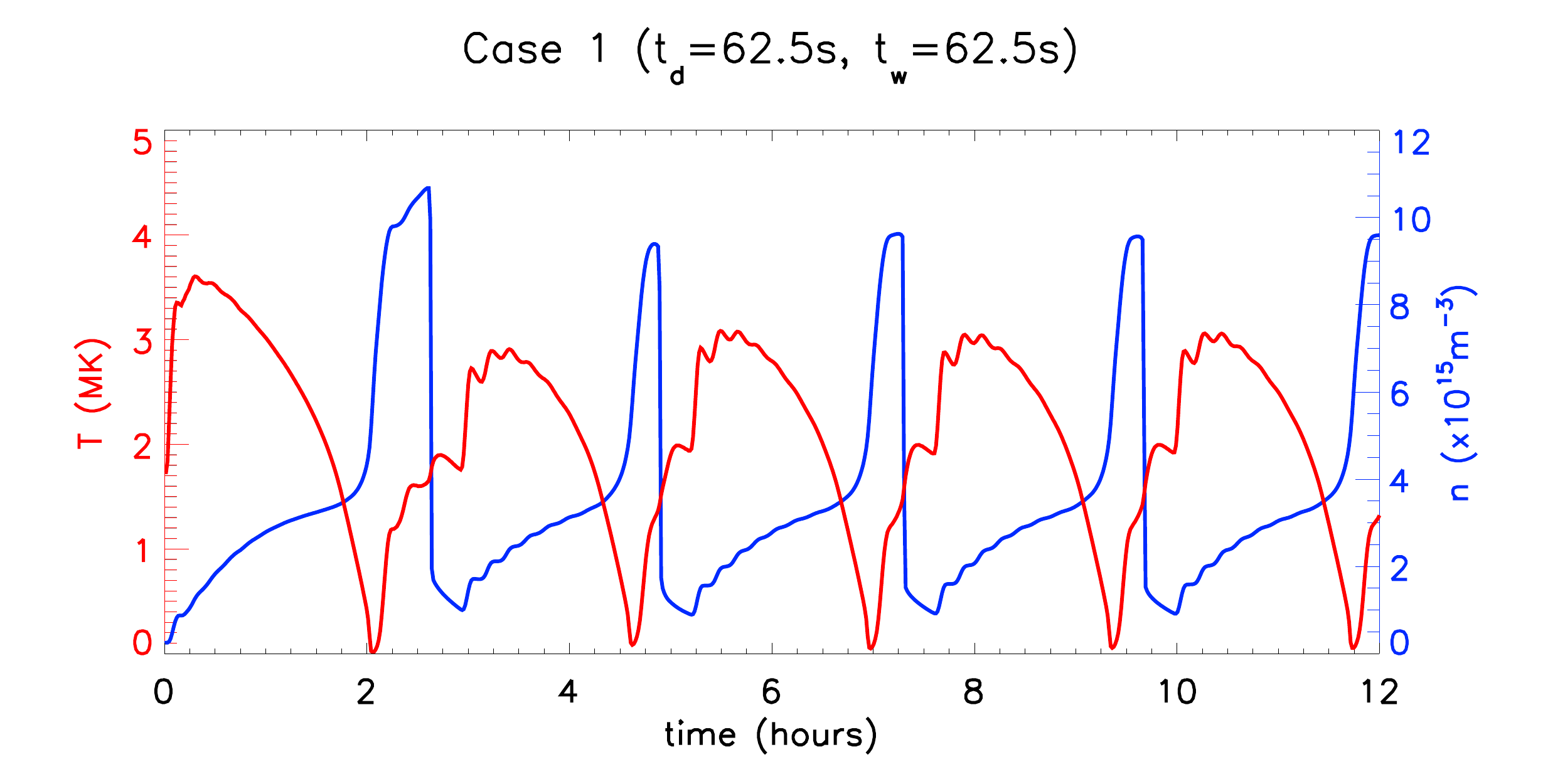}}
  \hspace*{-0.025\linewidth}
  \subfigure{\includegraphics[width=0.35\linewidth]
  {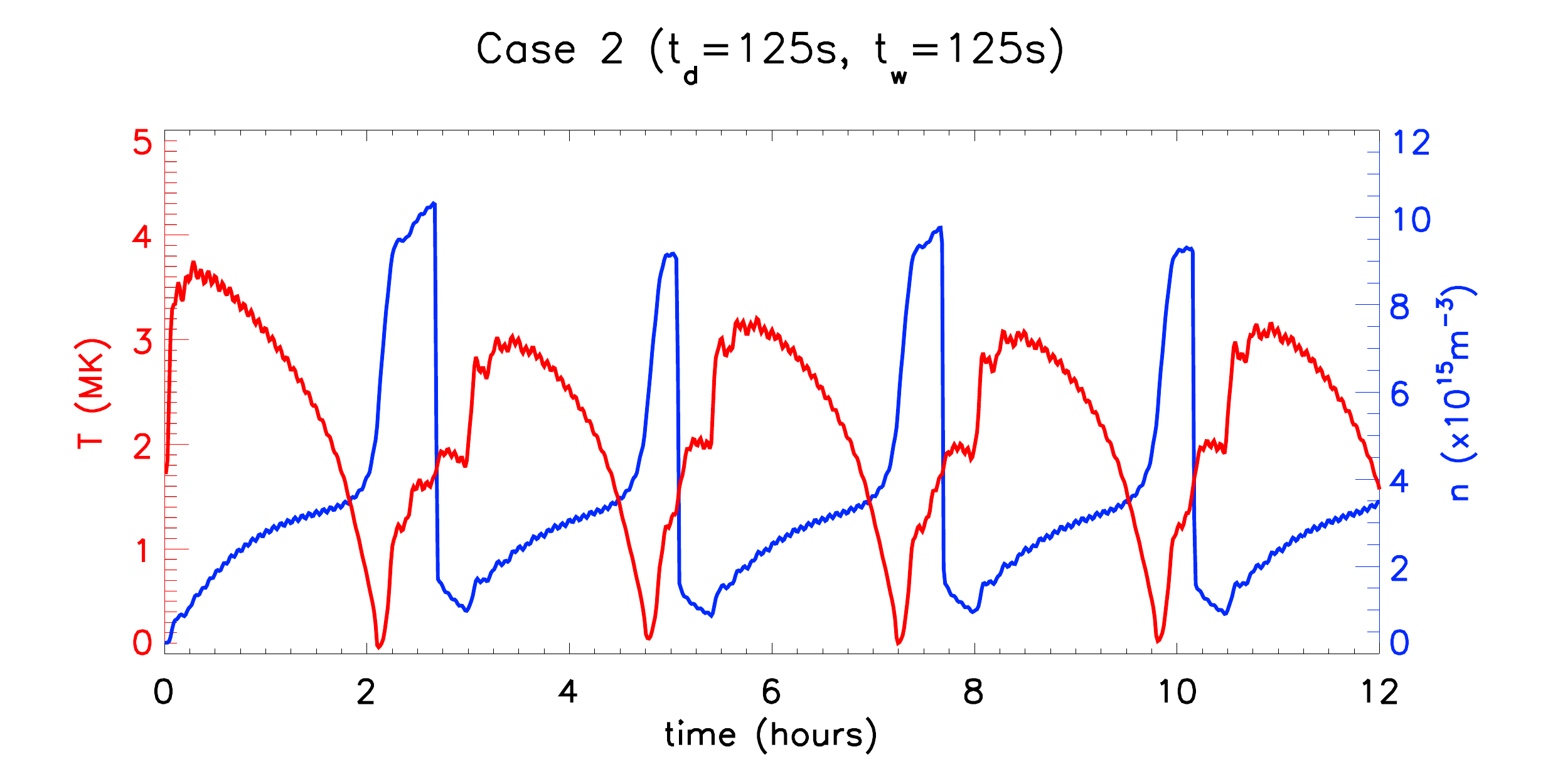}}
  \\
  \hspace*{-0.025\linewidth}
  \subfigure{\includegraphics[width=0.35\linewidth]
  {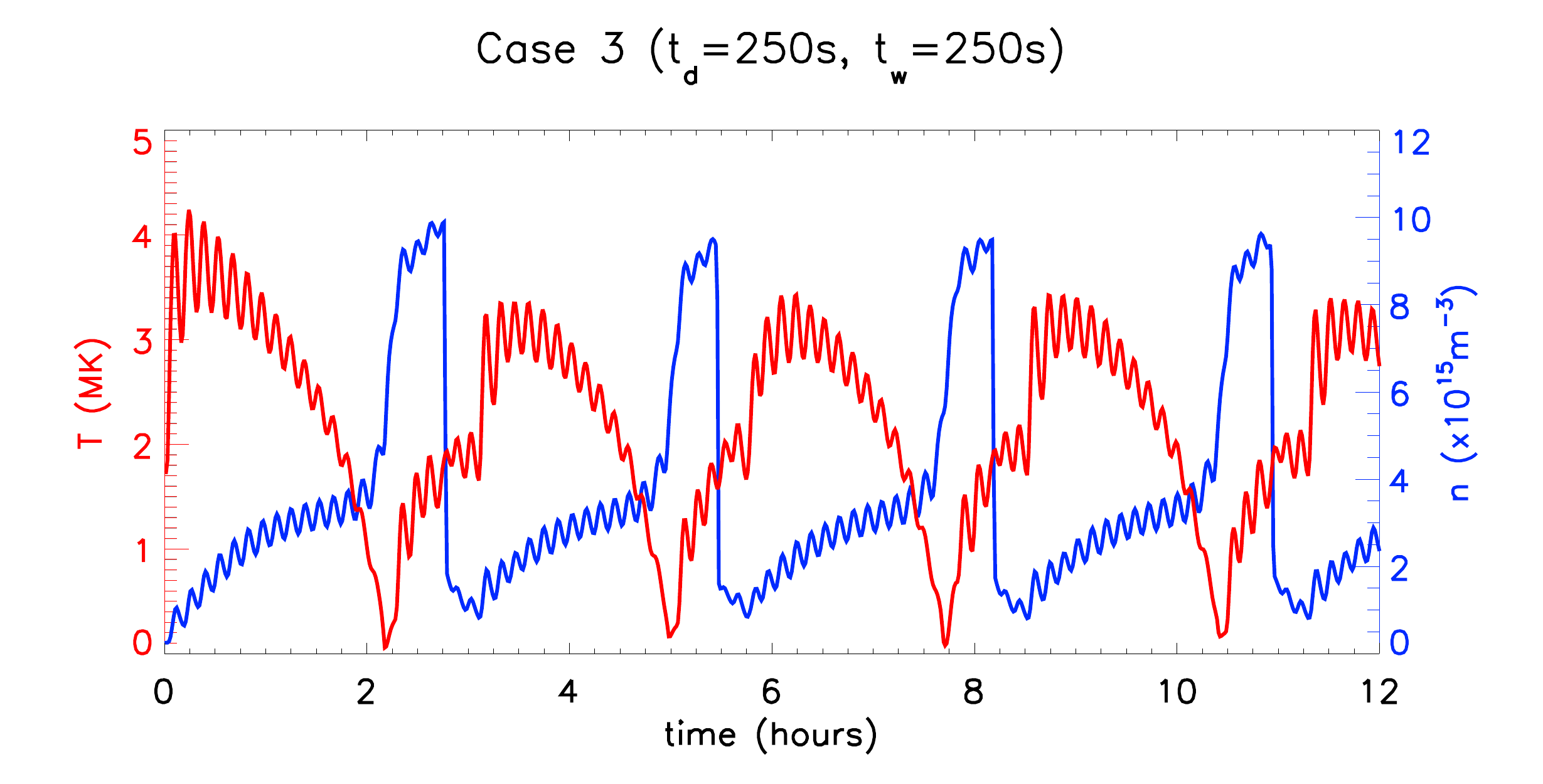}}
  \hspace*{-0.025\linewidth}
  \subfigure{\includegraphics[width=0.35\linewidth]
  {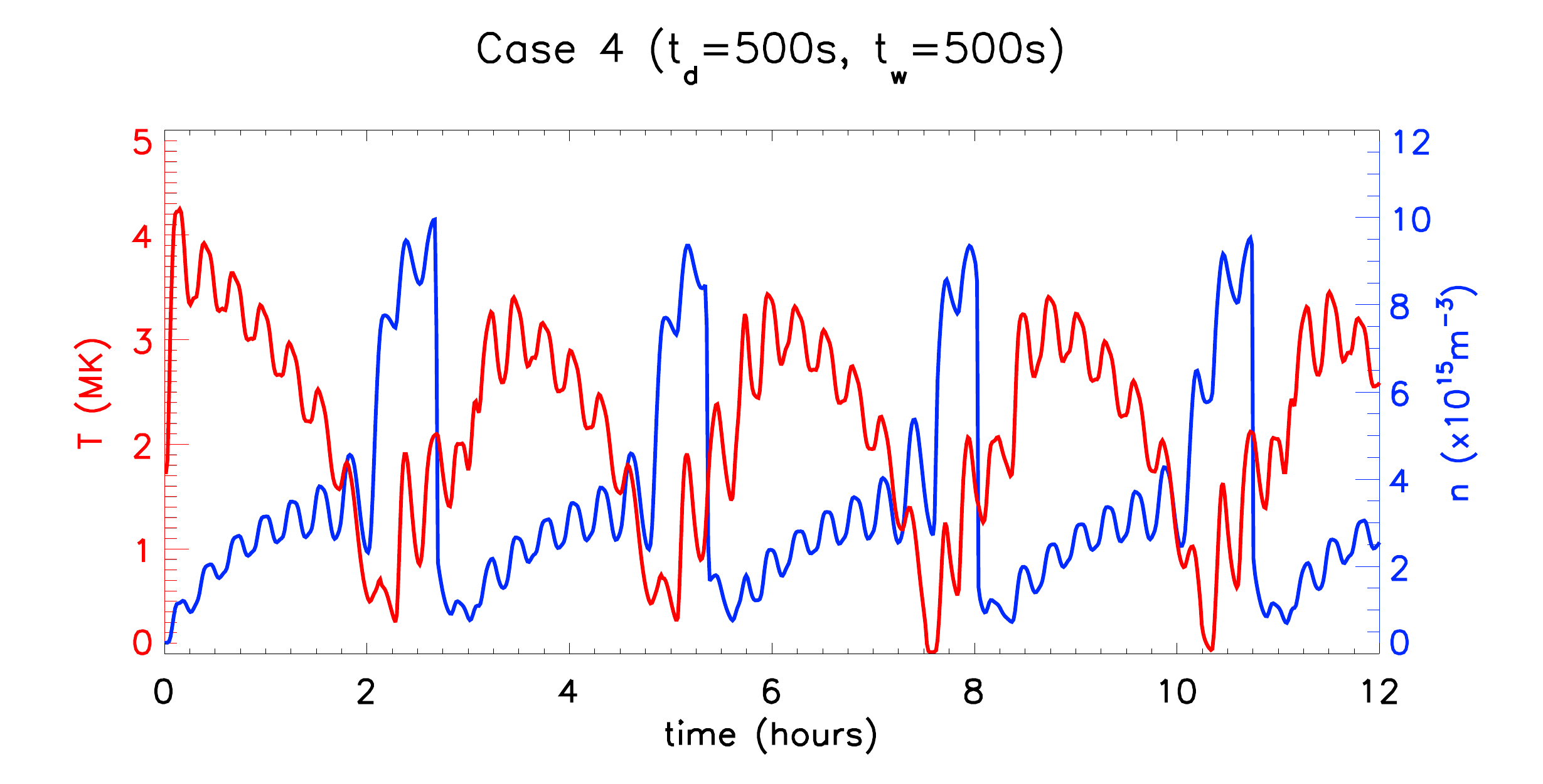}}
  \hspace*{-0.025\linewidth}
  \subfigure{\includegraphics[width=0.35\linewidth]
  {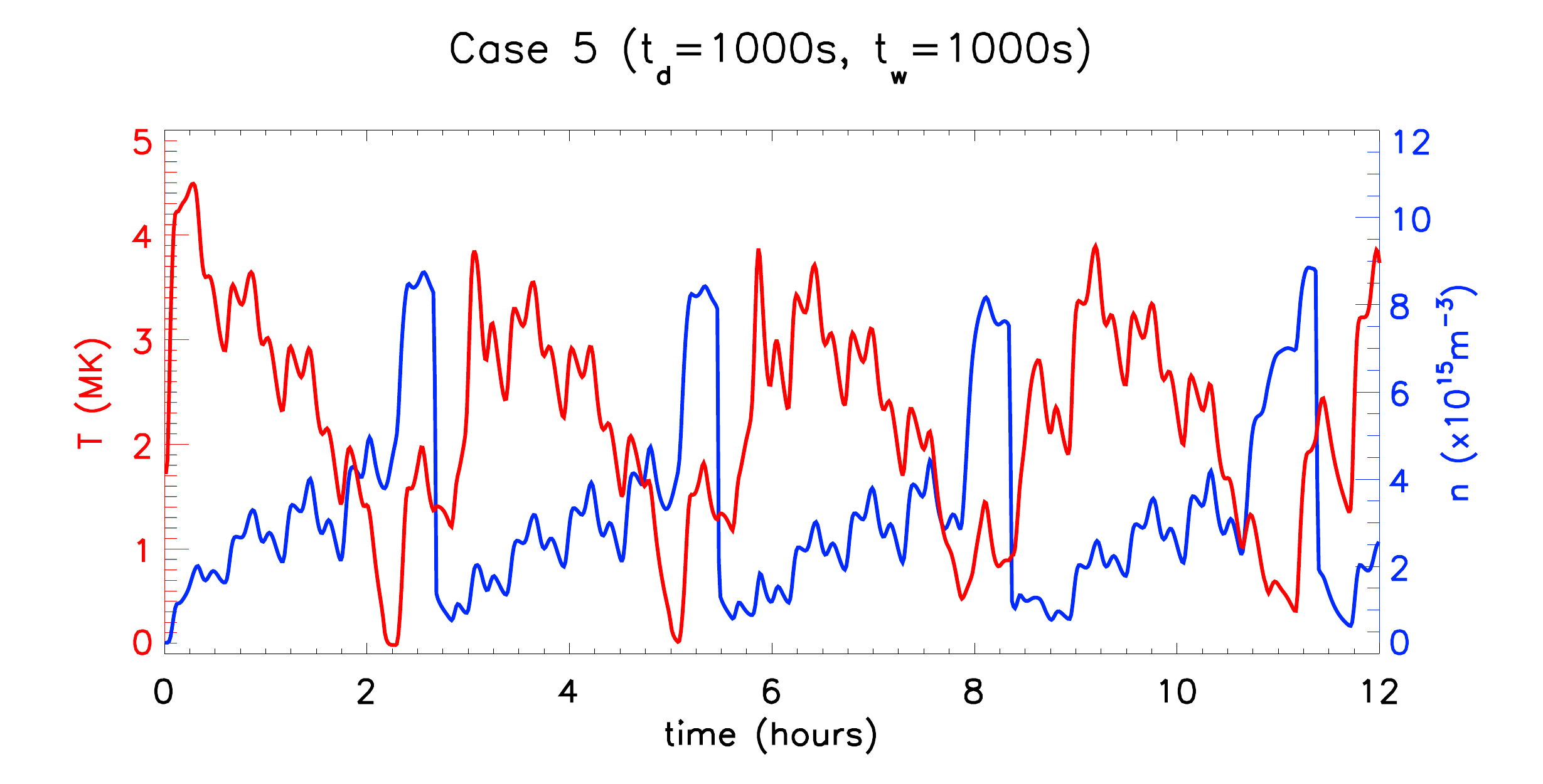}}
  \\
  \hspace*{-0.025\linewidth}
  \subfigure{\includegraphics[width=0.35\linewidth]
  {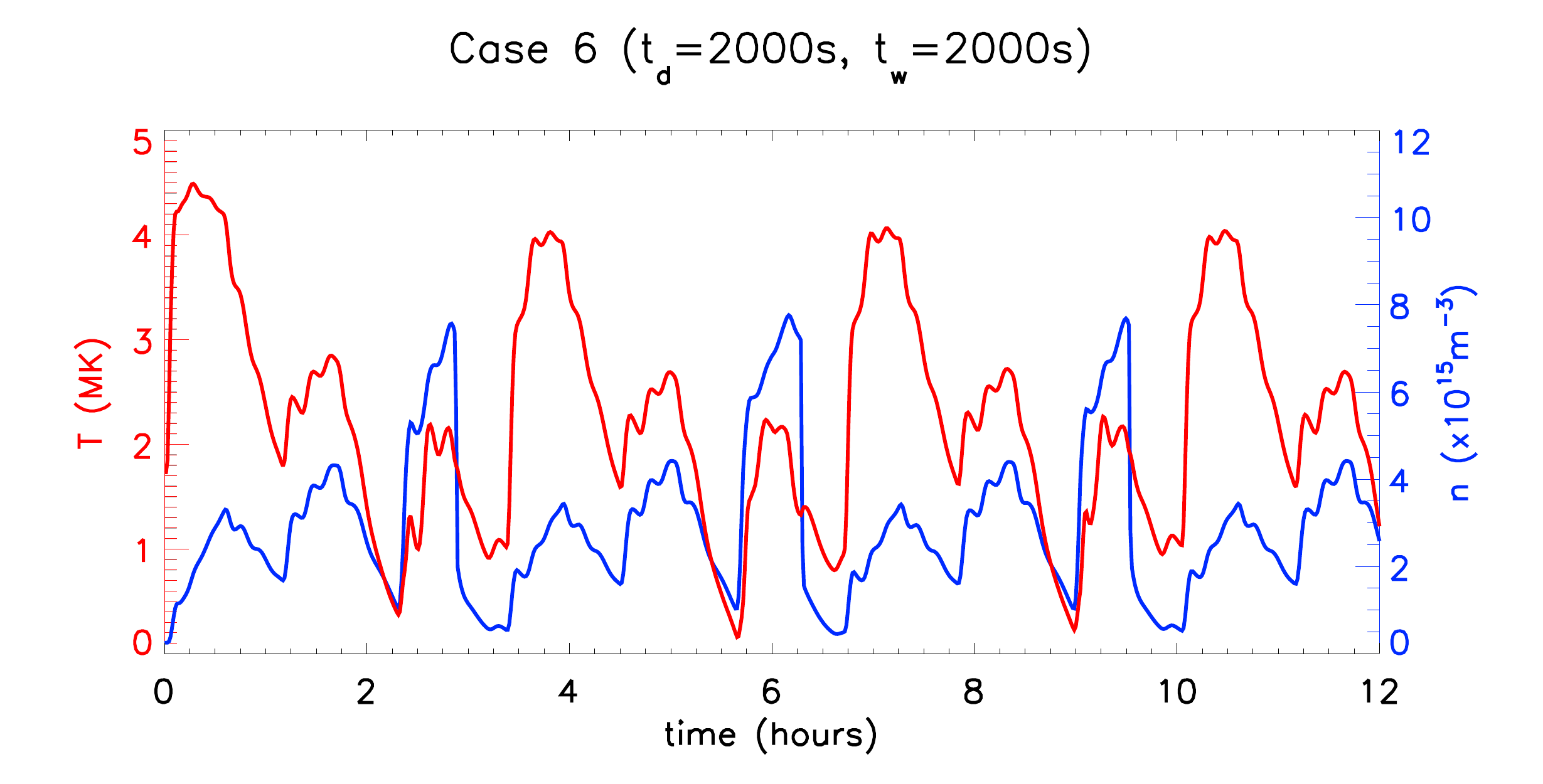}}
  \hspace*{-0.025\linewidth}
  \subfigure{\includegraphics[width=0.35\linewidth]
  {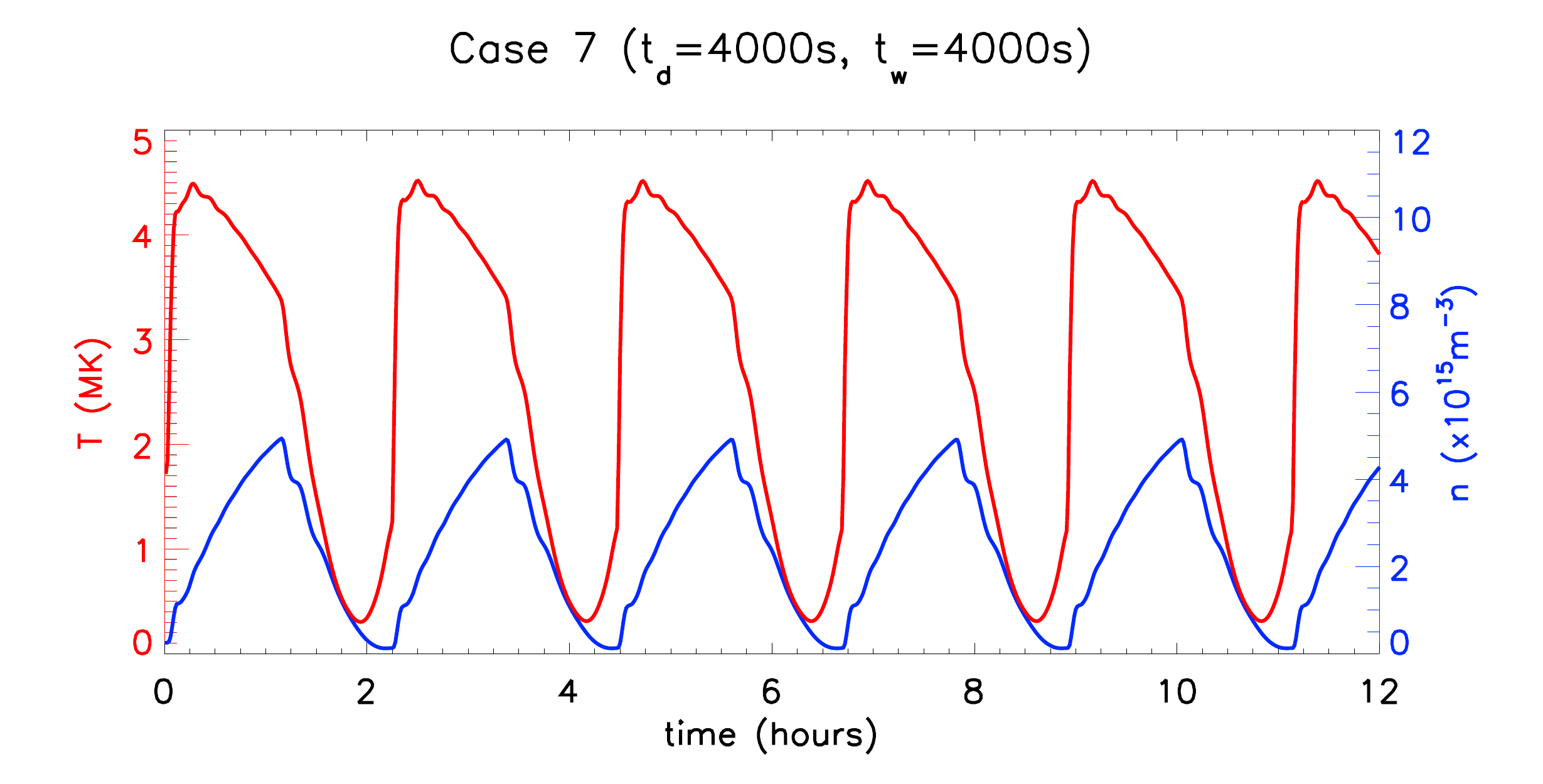}}
  \hspace*{-0.025\linewidth}
  \subfigure{\includegraphics[width=0.35\linewidth]
  {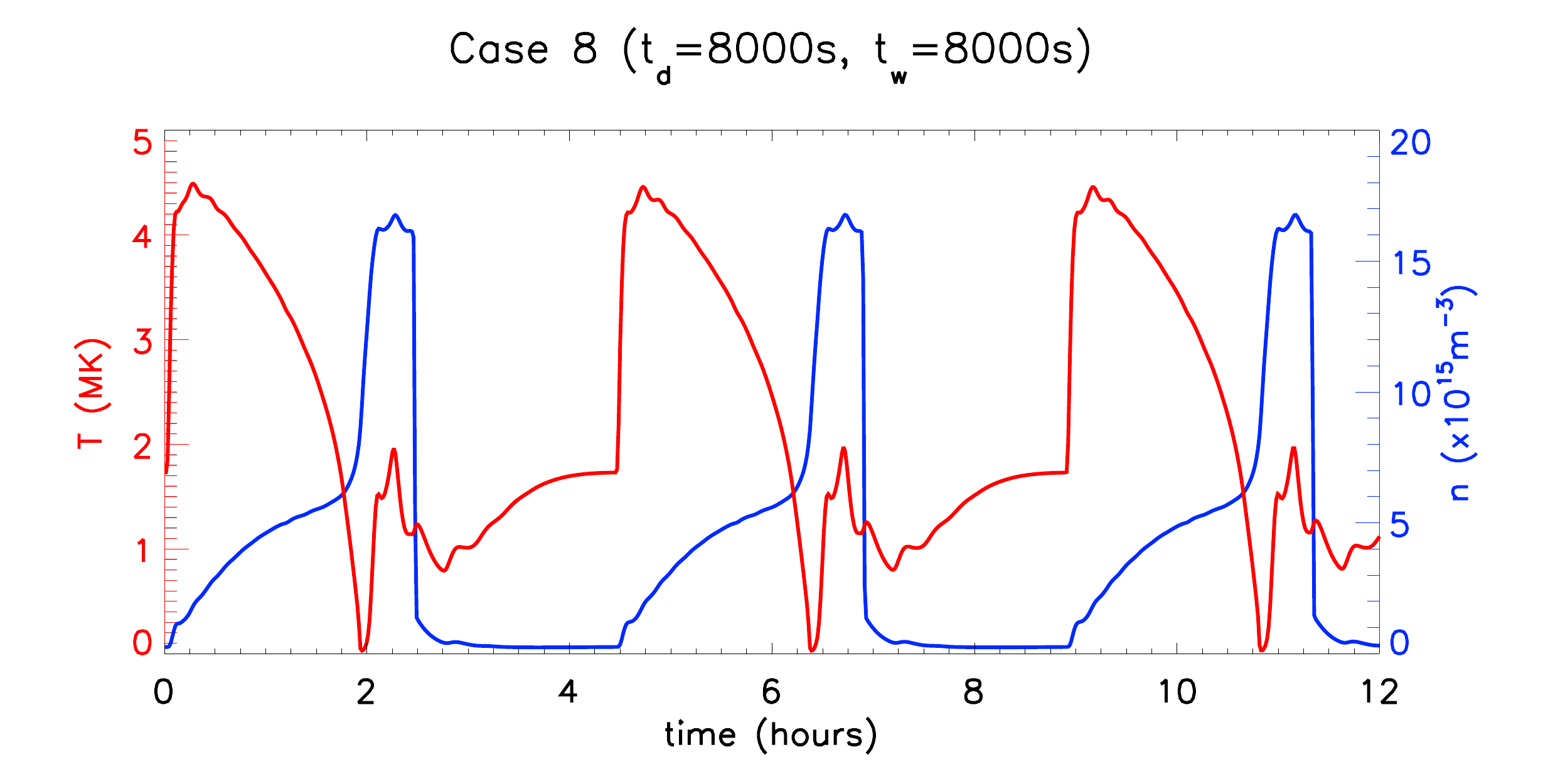}}
  \caption{
    Effect of heating timescales
    on TNE cycles.
    LareJ results for time dependent footpoint heating
    Cases 1--8 together with the steady 
    footpoint heating result.
    The panels show the time evolution of the coronal averaged
    temperature (red line, left hand axis) and
    density (blue line, right hand axis).
    \label{Fig:td_fph_LareJ_Tn_ca}
    }
\end{figure*}
  %
  %
  %%%%%%%%%%%%%%%%%%%%%%%%%%%%%%%%%%%%%%%%%%%%%%%%%%%%%%%%%%%%%    
  %
  % Fig:td_fph_LareJ_Tpnv
  %
  \begin{figure*}
  \hspace*{-0.02\linewidth}
  \subfigure{\includegraphics[width=0.5\linewidth]
  {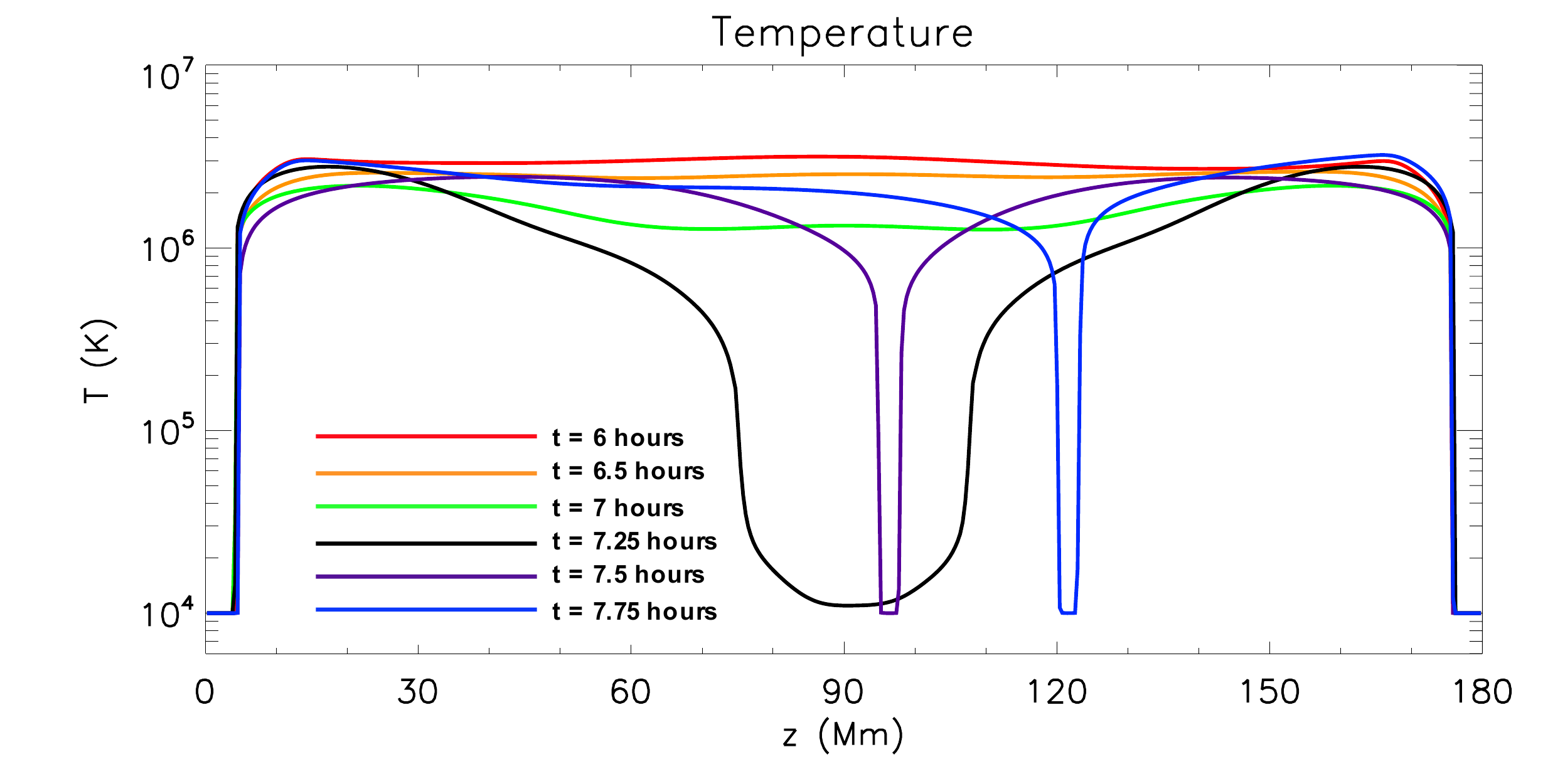}}
  \hspace*{-0.0125\linewidth}
  \subfigure{\includegraphics[width=0.5\linewidth]
  {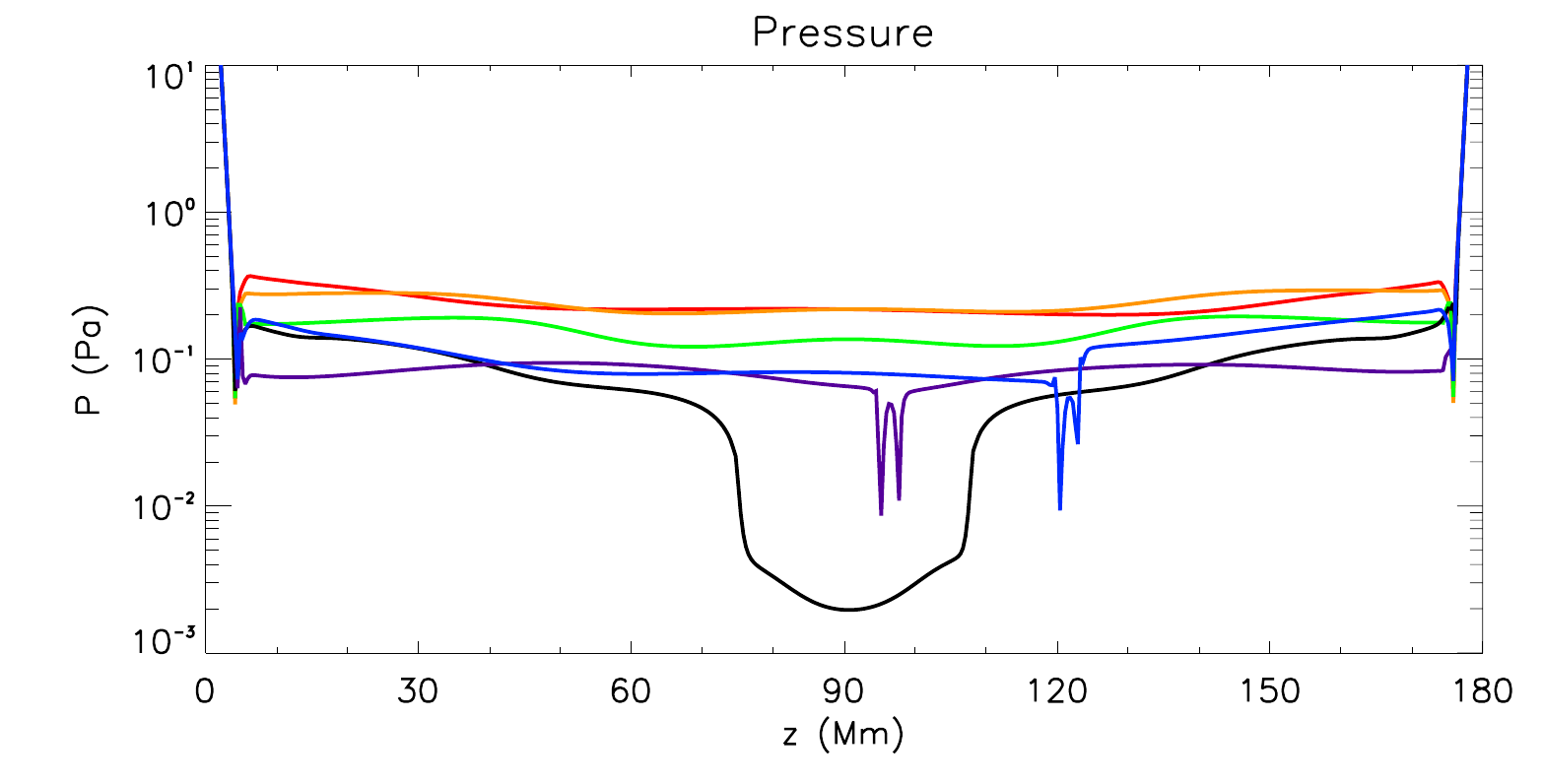}}
  \\
  \hspace*{-0.02\linewidth}
  \subfigure{\includegraphics[width=0.5\linewidth]
  {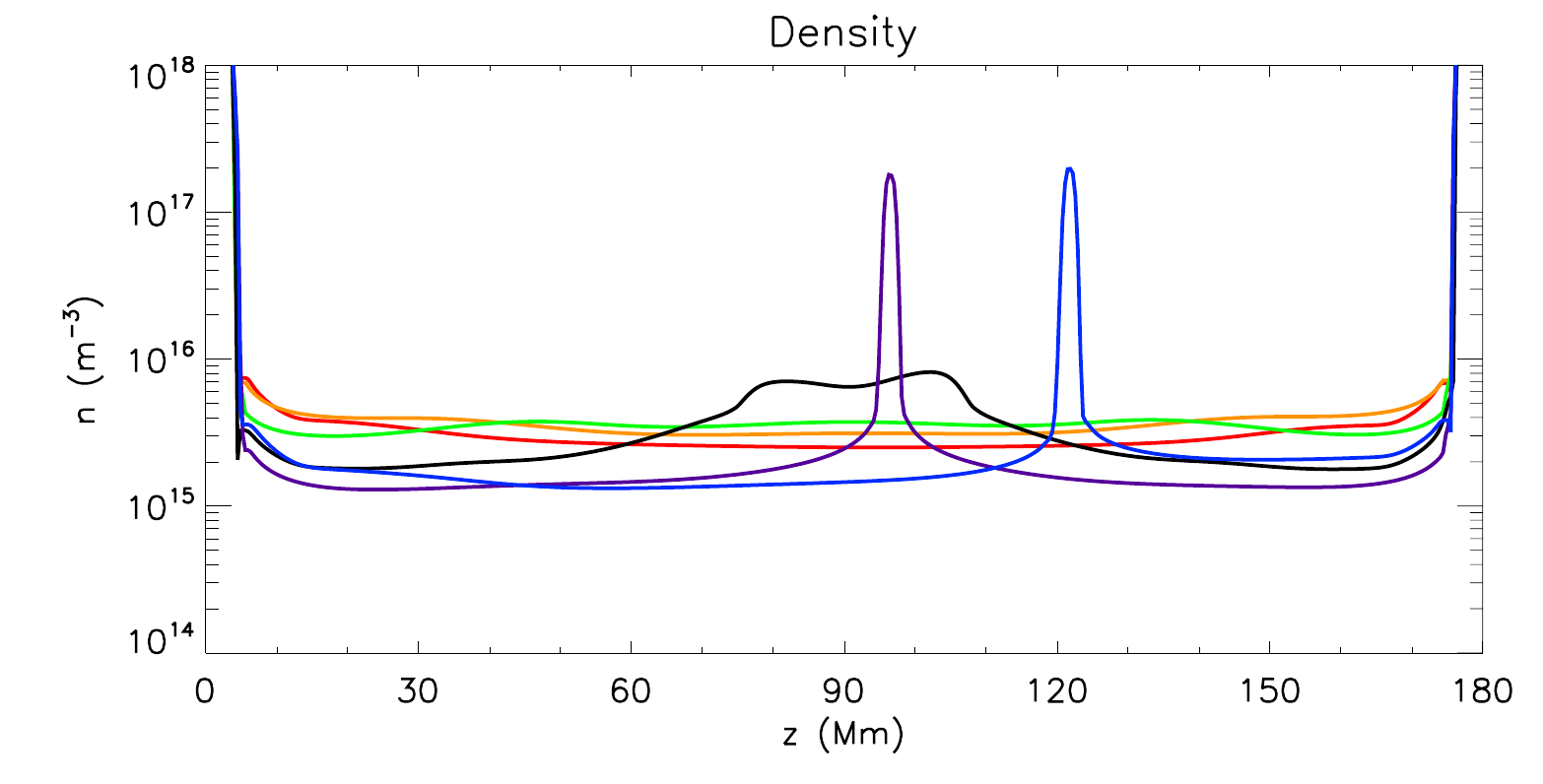}}
  \hspace*{-0.0125\linewidth}
  \subfigure{\includegraphics[width=0.5\linewidth]
  {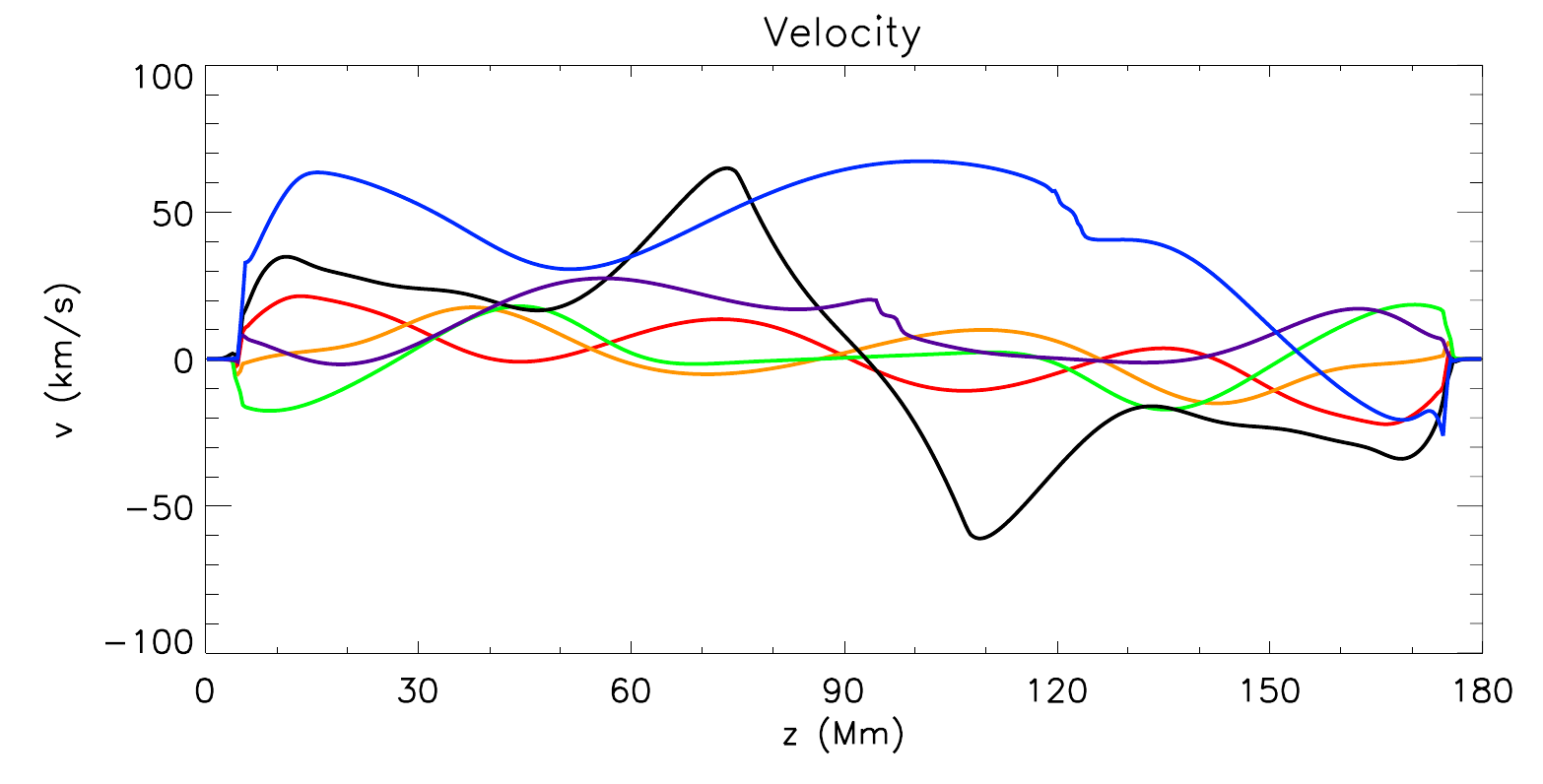}}
  \\
  \hspace*{-0.02\linewidth}
  \subfigure{\includegraphics[width=0.5\linewidth]
  {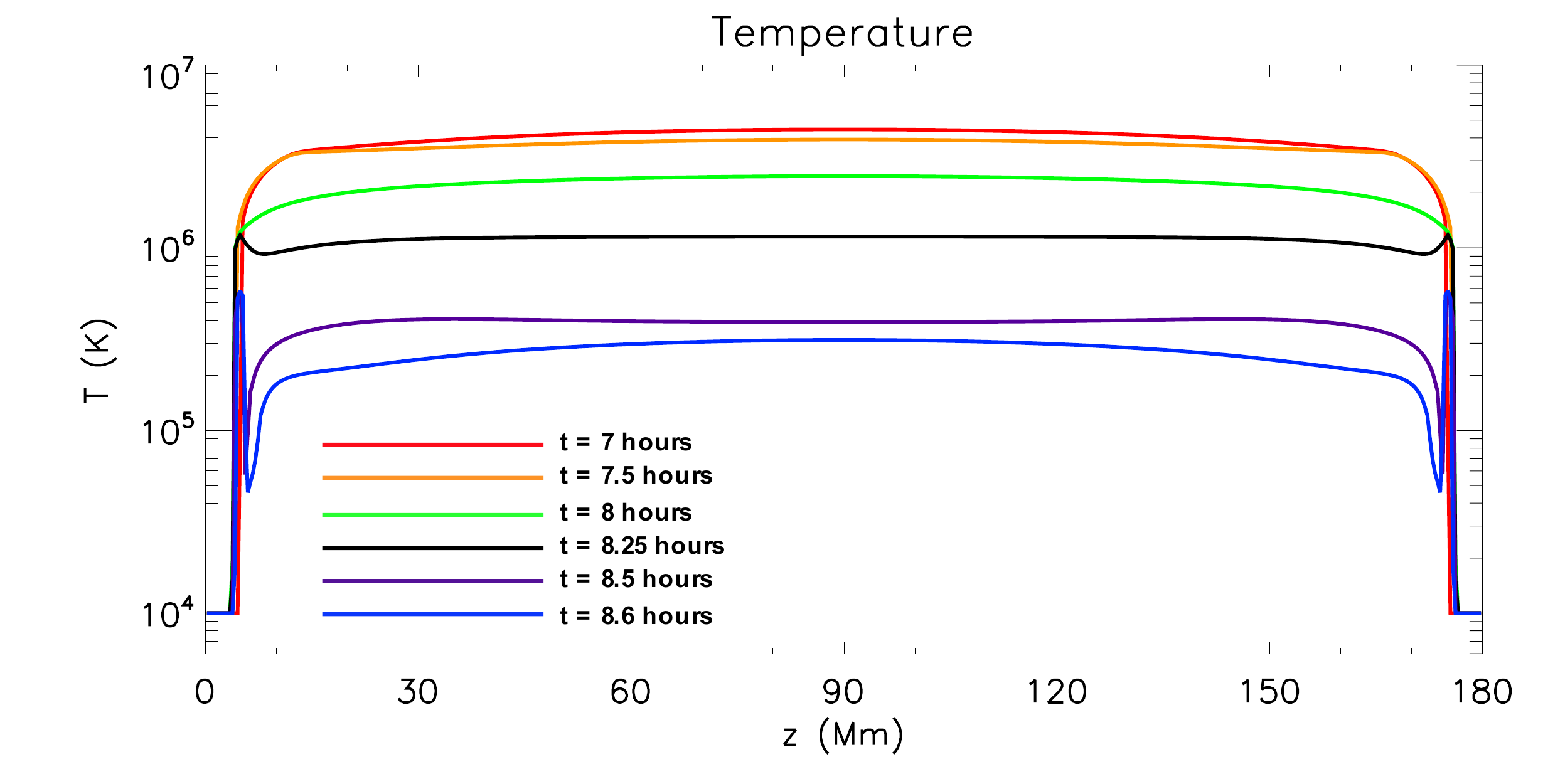}}
  \hspace*{-0.0125\linewidth}
  \subfigure{\includegraphics[width=0.5\linewidth]
  {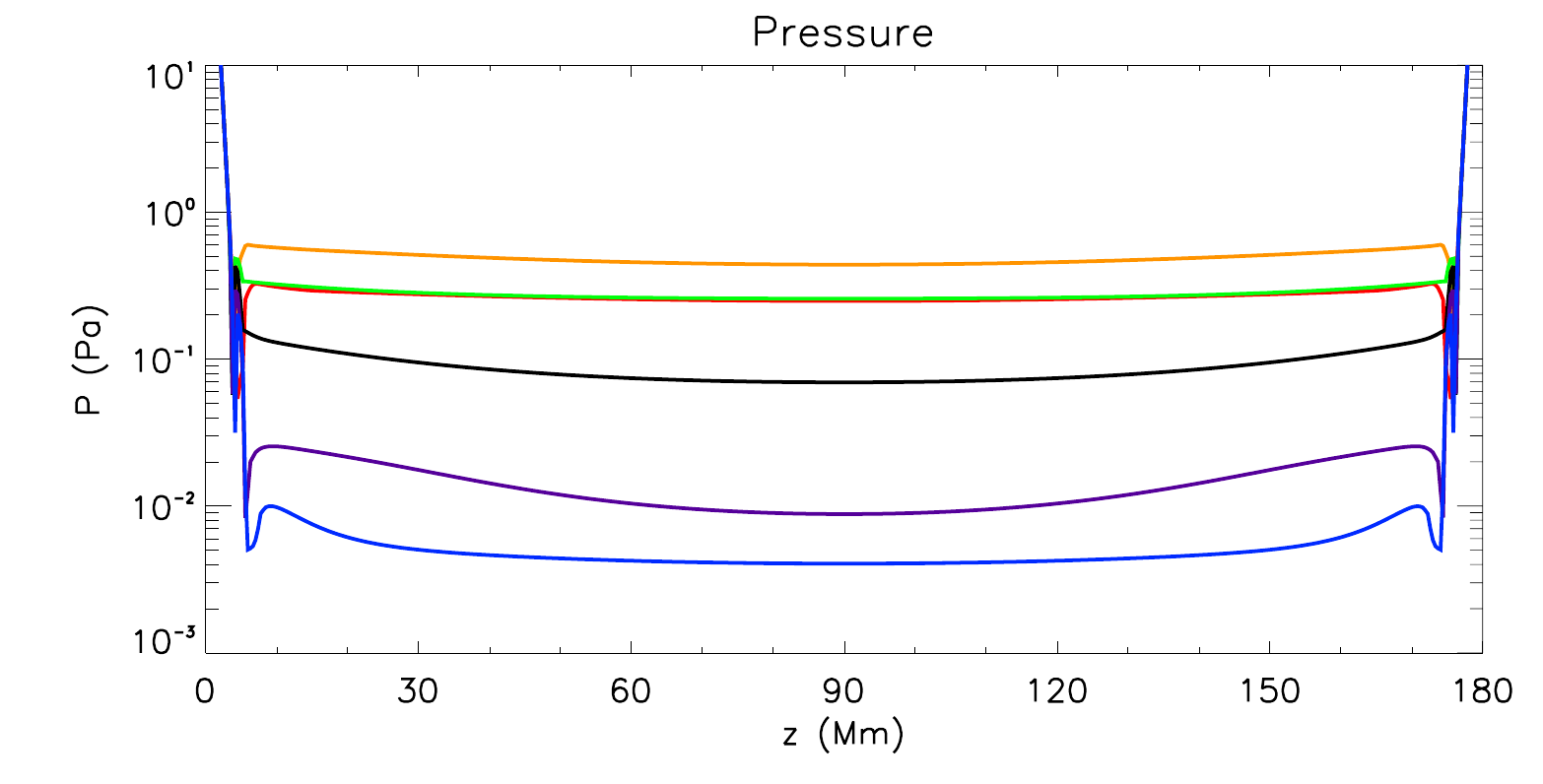}}
  \\
  \hspace*{-0.02\linewidth}
  \subfigure{\includegraphics[width=0.5\linewidth]
  {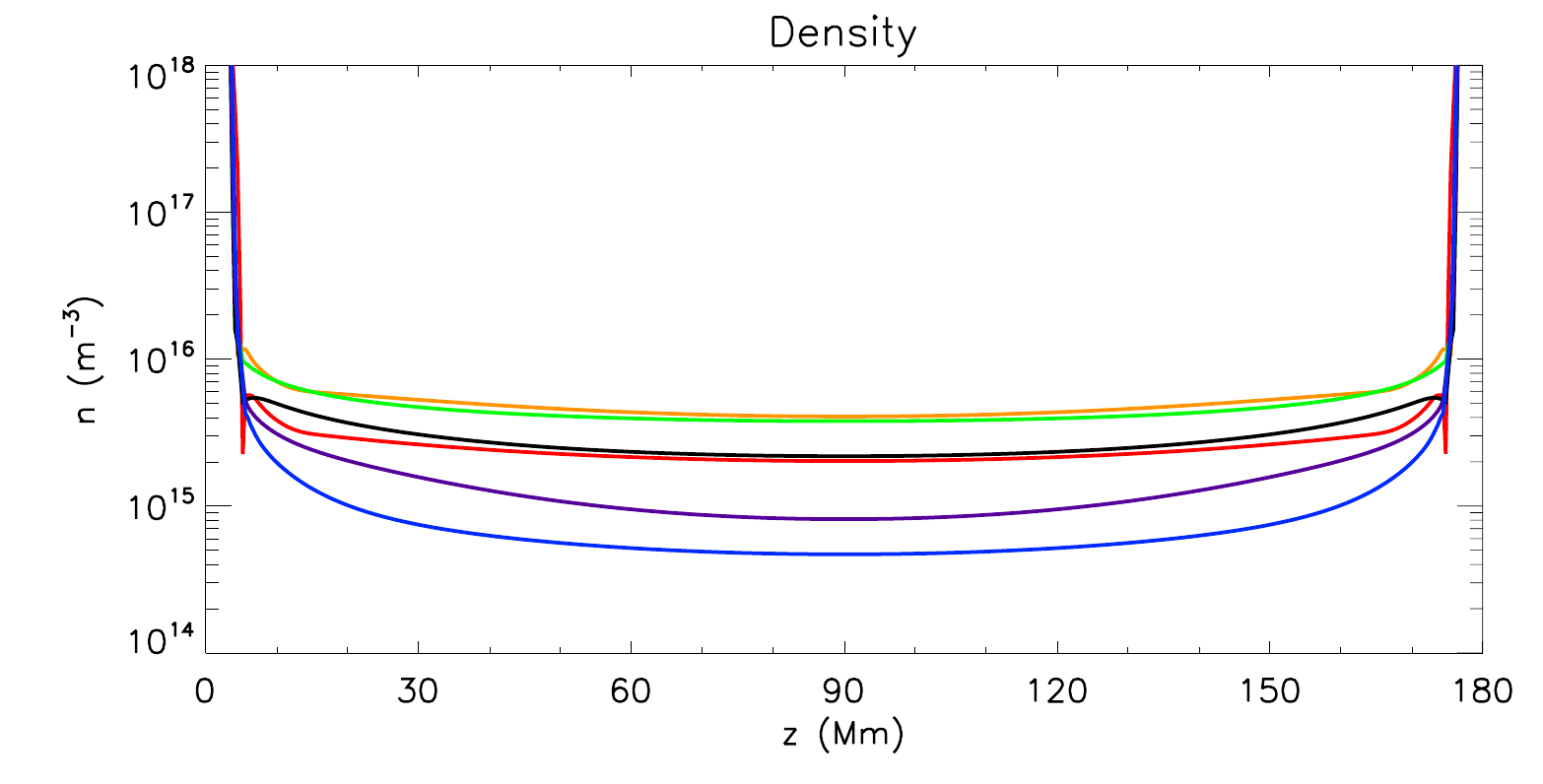}}
  \hspace*{-0.0125\linewidth}
  \subfigure{\includegraphics[width=0.5\linewidth]
  {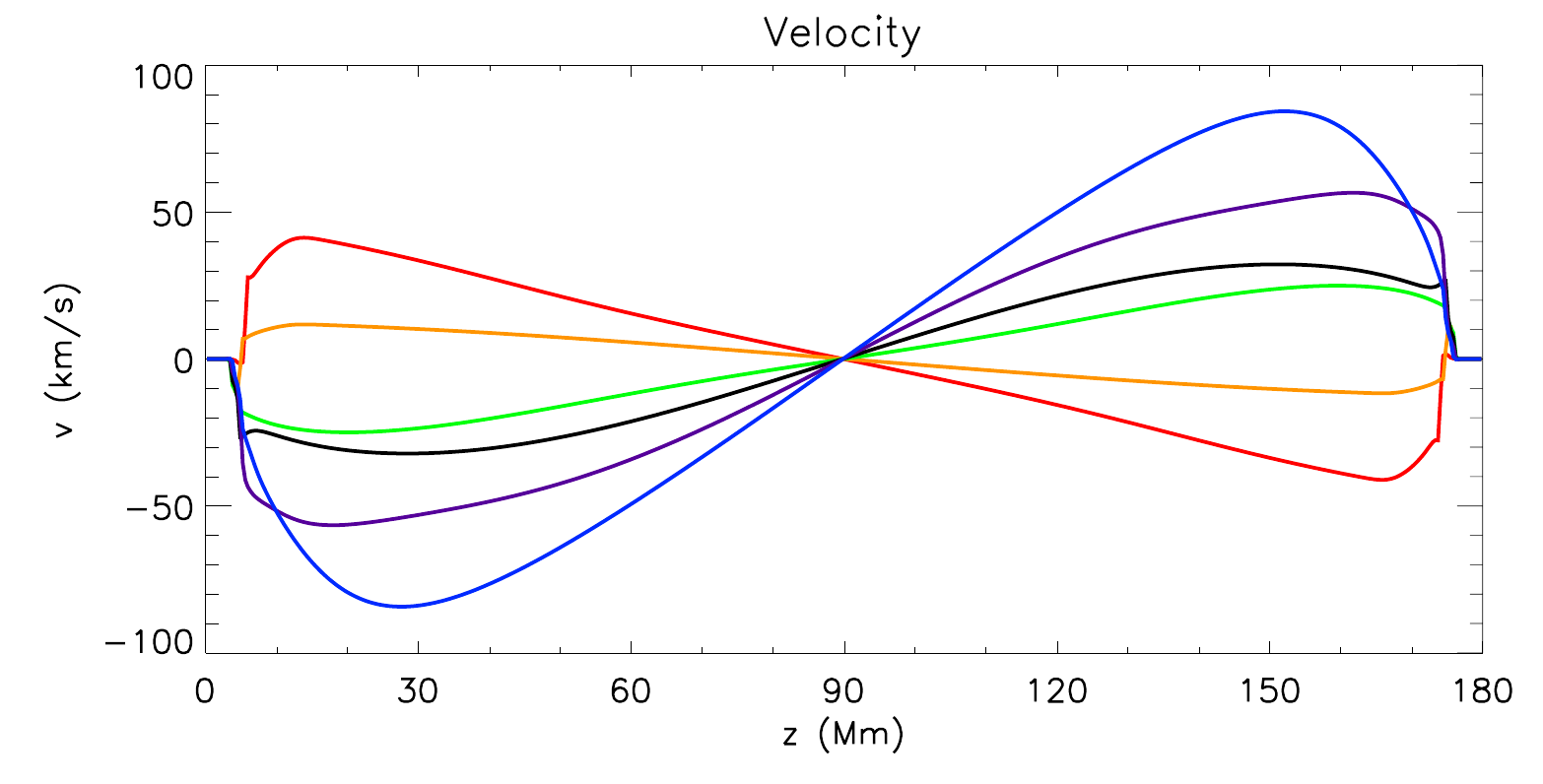}}
  \caption{
    Effect of heating timescales
    on TNE cycles.
    LareJ results for time dependent footpoint heating
    Cases 2 (upper four panels) \& 7 (lower four panels). 
    The panels show
    time ordered snapshots of the 
    temperature, pressure, density and velocity 
    as functions of position along the loop for times
    during the third TNE cycle of Case 2 and 
    fourth heating cycle of Case 7.
  \label{Fig:td_fph_LareJ_Tpnv}}
\end{figure*}
  %
  %
  %%%%%%%%%%%%%%%%%%%%%%%%%%%%%%%%%%%%%%%%%%%%%%%%%%%%%%%%%%%%%    
  %
  % Fig:IoQbg_LareJ
  %
  \begin{figure*}
  \hspace*{0.15\linewidth}
  \subfigure{\includegraphics[width=0.32\linewidth]
  {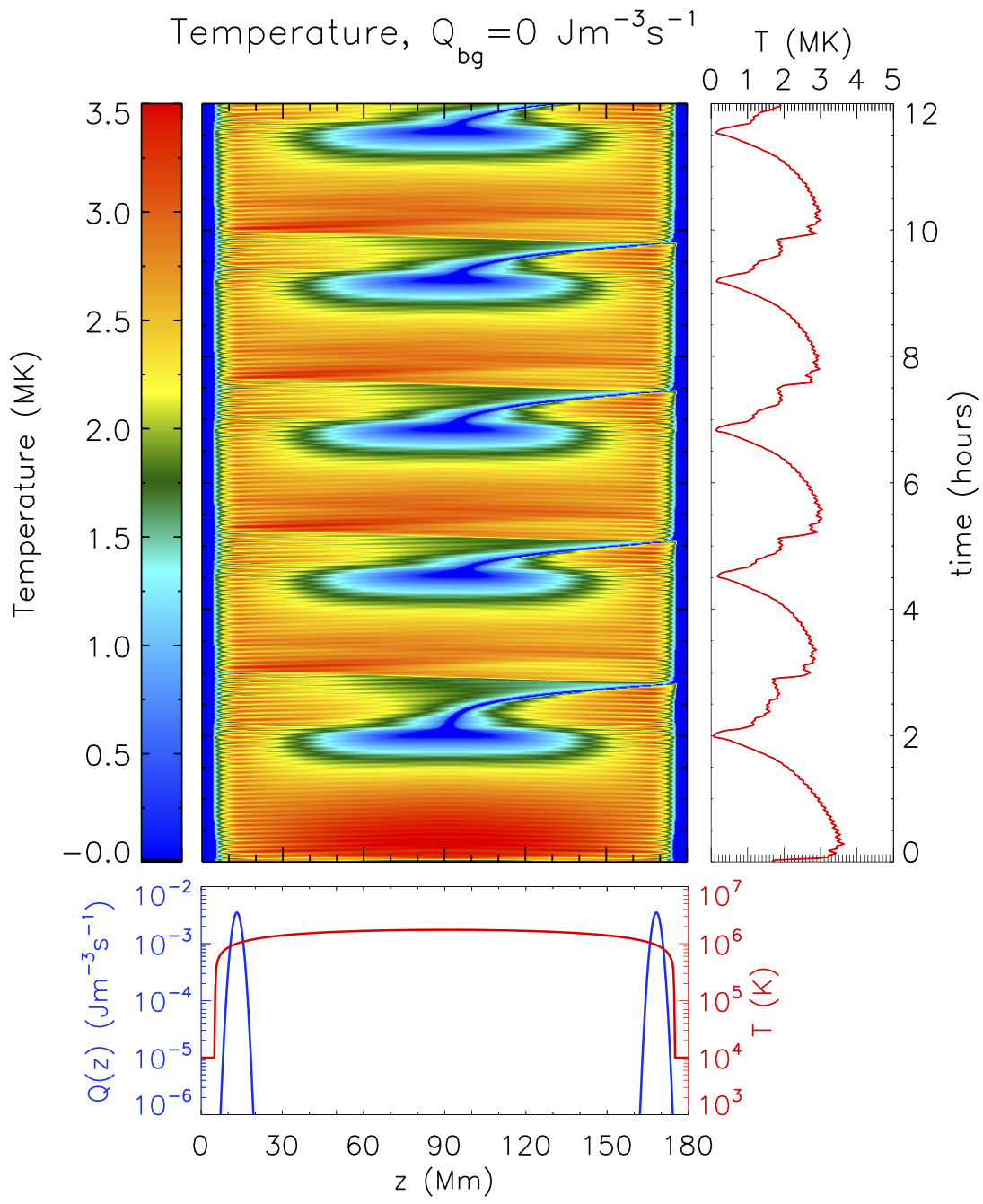}}
  \hspace*{0.05\linewidth}
  \subfigure{\includegraphics[width=0.32\linewidth]
  {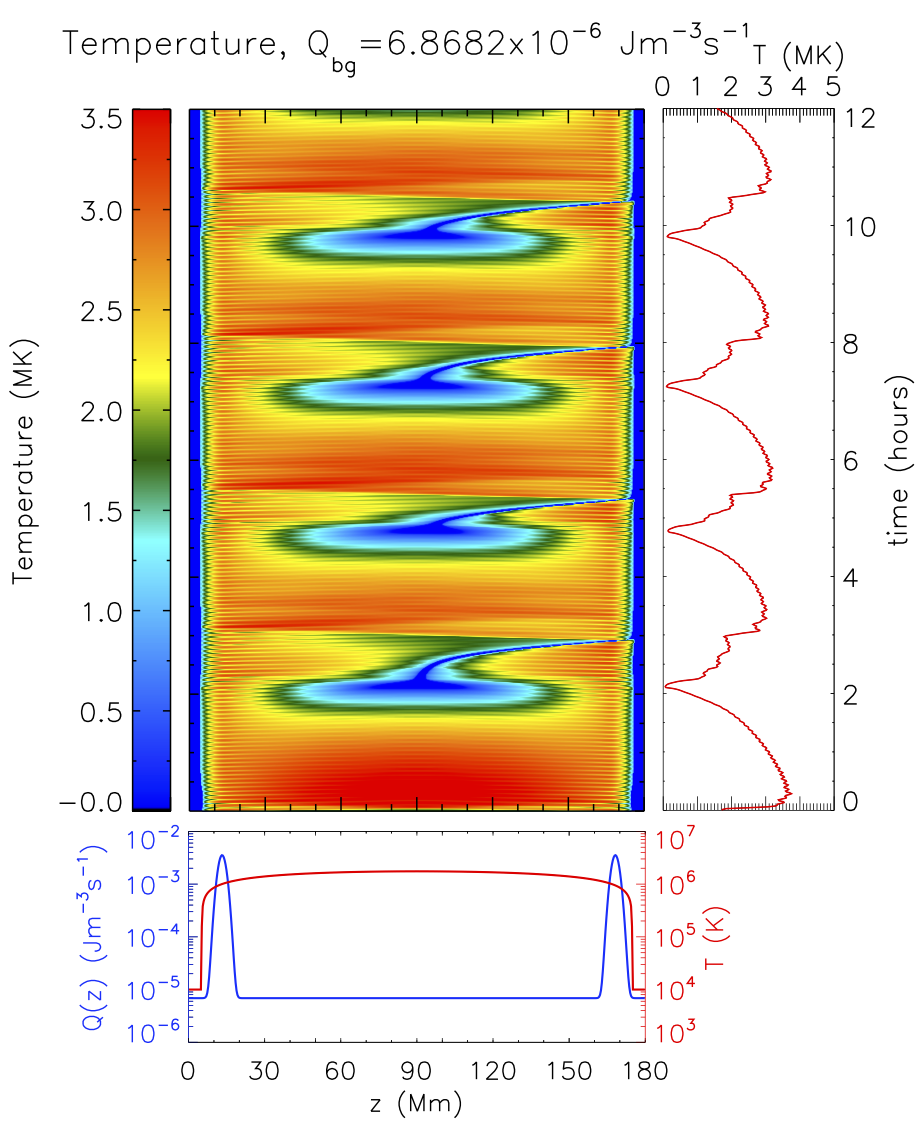}}
  \\
  \hspace*{0.15\linewidth}
  \subfigure{\includegraphics[width=0.32\linewidth]
  {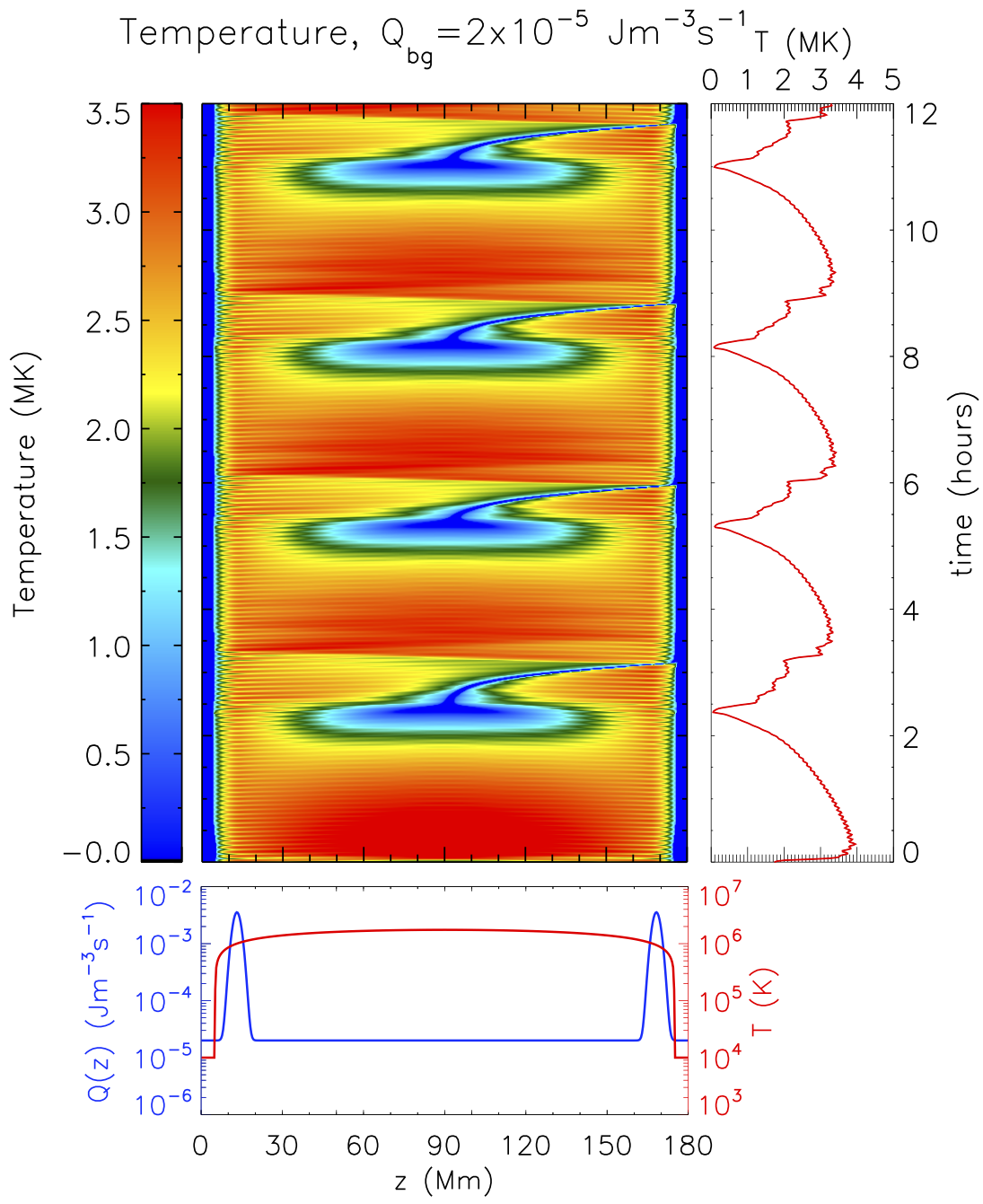}}
  \hspace*{0.05\linewidth}
  \subfigure{\includegraphics[width=0.32\linewidth]
  {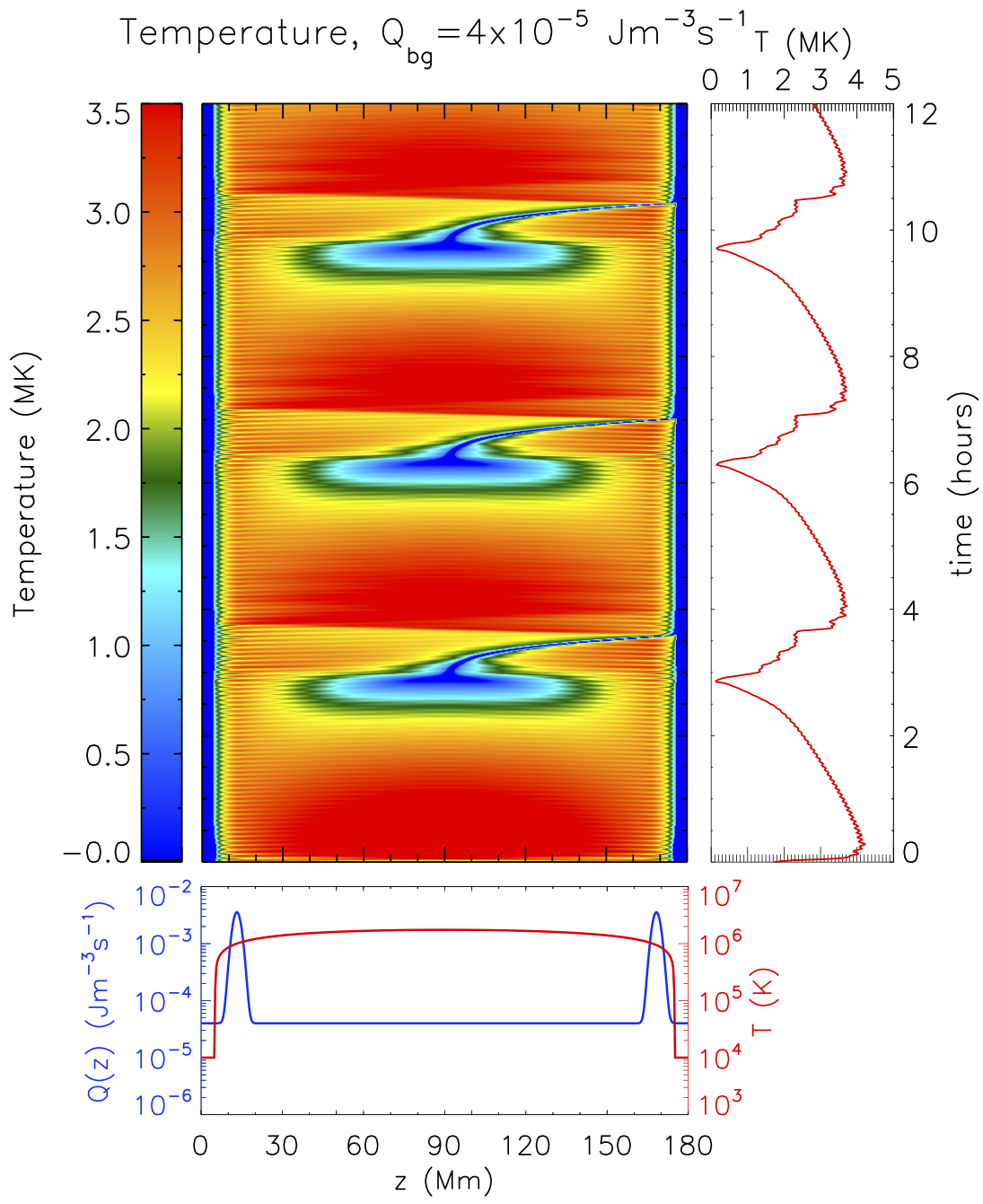}}
  \\
  \hspace*{0.15\linewidth}
  \subfigure{\includegraphics[width=0.32\linewidth]
  {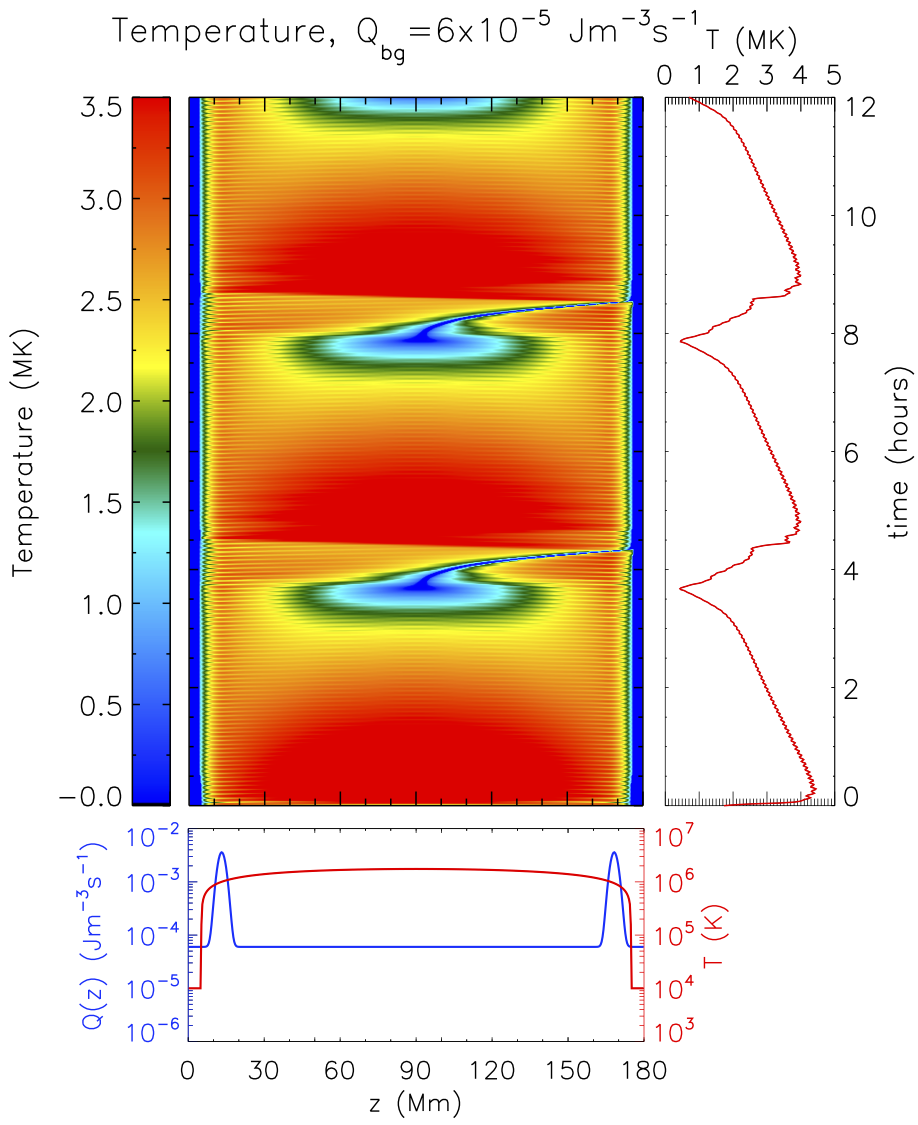}}
  \hspace*{0.05\linewidth}
  \subfigure{\includegraphics[width=0.32\linewidth]
  {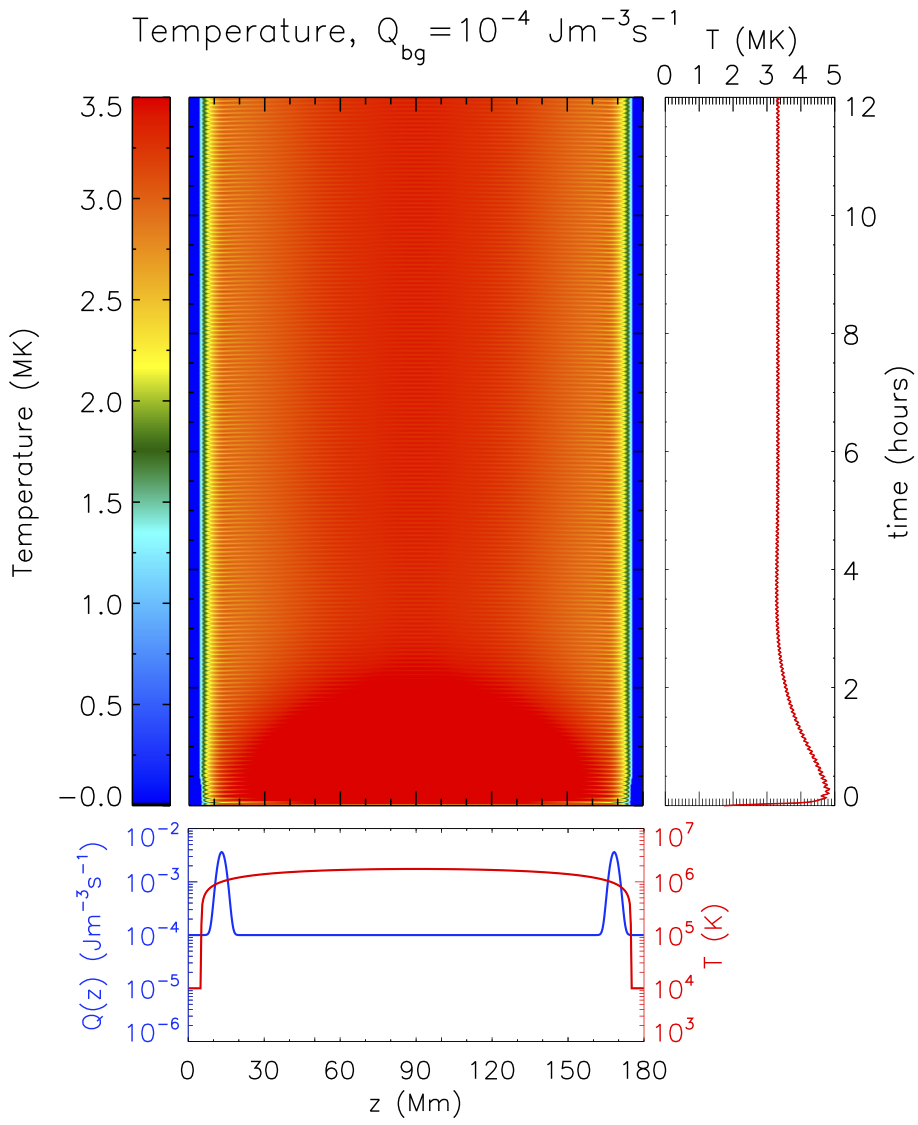}}
  \caption{Effect of the background heating on TNE  
    cycles. LareJ results for time dependent footpoint heating. 
    The 2D plots show the time evolution of
    the temperature as a function of position along the loop.
    The various panels represent different values of
    background heating ($Q_{bg}$).
    On the right of the 2D plots, we display the
    evolution of the coronal averaged temperature (computed by
    spatially averaging over the uppermost 25\% of the loop).
    At the bottom of the 2D plots, we show the time
    averaged footpoint
    heating profile (blue line, left-hand axis) imposed
    on top of the temperature initial condition (red line, 
    right-hand axis). 
    \label{Fig:IoQbg_LareJ}  
    }
\end{figure*}
  %
  %
  %%%%%%%%%%%%%%%%%%%%%%%%%%%%%%%%%%%%%%%%%%%%%%%%%%%%%%%%%%%%%    
  %
  % Fig:f_vs_Qbg_LareJ
  %
  \begin{figure*}
  \hspace*{1.5cm}
  \subfigure{\includegraphics[width=0.8\linewidth]  
  {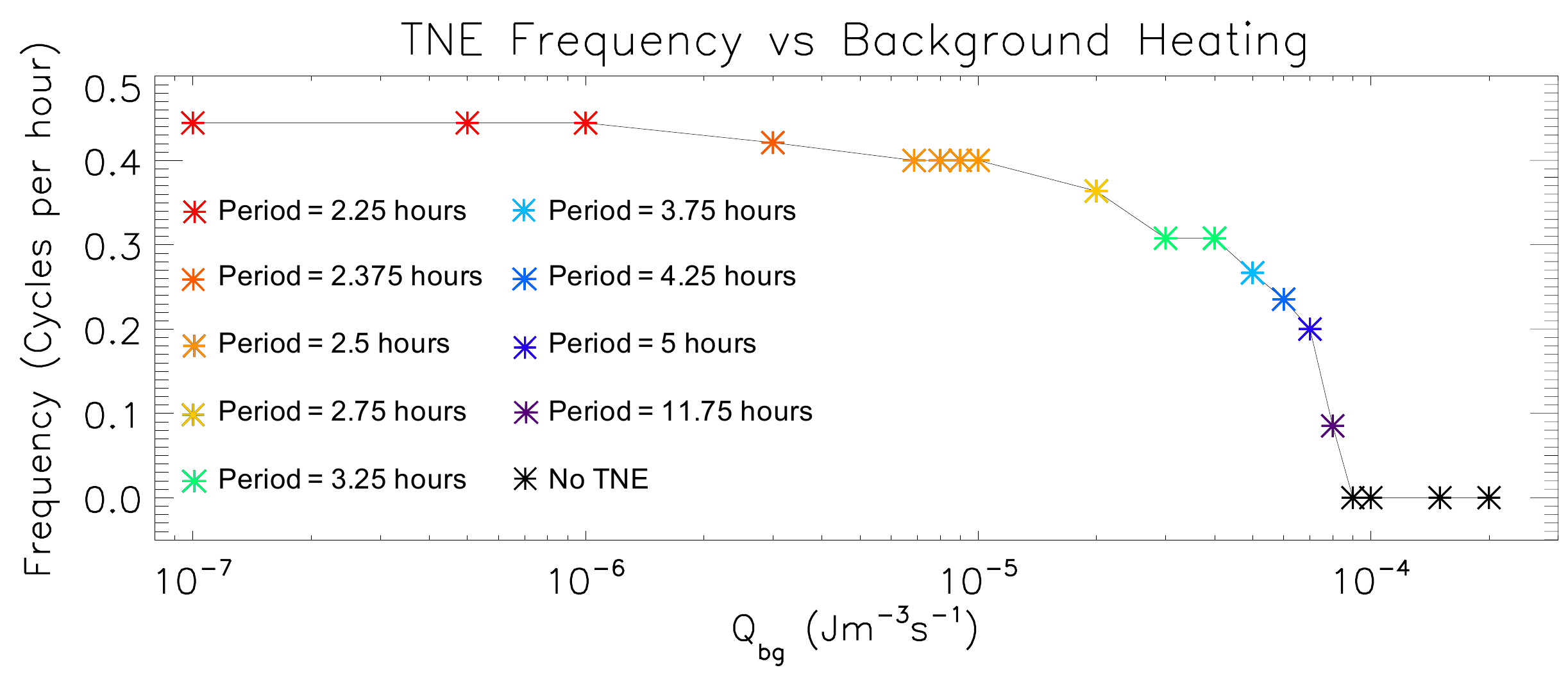}}
  \vspace*{-0.5cm}
  \caption{Effect of the background heating on TNE 
    cycles.
    The panel shows the relation between TNE cycle
    frequency and the background heating value. 
    Note from Figure \ref{Fig:IoQbg_LareJ} that
    the loop computed with no background heating has a
    cycle period of 2.25 hours, which is
    consistent with the asterisk at 
    $Q_{bg}=10^{-7}$ Jm$^{-3}$s$^{-1}$.
    \label{Fig:f_vs_Qbg_LareJ}  
    }
\end{figure*}
  %
  %
  %
  %
  %%%%%%%%%%%%%%%%%%%%%%%%%%%%%%%%%%%%%%%%%%%%%%%%%%%%%%%%%%%%%
  %
  %
  \subsubsection{Steady footpoint heating: Lare simulations
  \label{Sect:Resolution_LareJ}}
  \indent
  We have now shown that with the HYDRAD code adequate TR 
  resolution is required for the correct modelling of footpoint 
  heating and associated TNE, thus extending the result of BC13 
  which was limited to spatially uniform heating. 
  This suggests that the correct modelling of TNE is 
  unlikely to be a practical proposition in multi-stranded
  (thousands or more) models of a single observed loop
  or an entire active region, due to the excessive CPU 
  requirements.
  Instead, other approaches are required, of 
  which the UTR jump condition method 
  (see Section \ref{Sect:methods}) 
  is a well-documented example. 
  We have performed simulations using this approach (referred 
  to as LareJ) for a loop with 500 uniformly spaced grid points 
  and so a resolution of 
  360 km along the 180 Mm loop. This UTR LareJ approach is
  compared with the results obtained using the Lare code 
  (referred to as Lare1D) with a coarse spatial of 360 km 
  everywhere. LareJ is considered the benchmark solution
  because of our prior demonstration of good agreement with 
  HYDRAD 
  (e.g. \cite{paper:Johnstonetal2017a,paper:Johnstonetal2017b})  
  whereas Lare1D is considered to be 
  representative of a typical simulation with an under-resolved
  TR.
  \\
  \indent
  Figure \ref{Fig:IoR_Lare1d_vs_LareJ}
  shows
  the temporal evolution 
  and spatial variation  
  of the temperature in response to steady footpoint heating,
  for loops computed with and without 
  the UTR method (LareJ right and Lare1D left respectively).
  The LareJ approach shows the development of TNE and 
  associated condensations while the Lare1D simulation settles 
  down to a 
  static equilibrium after an initial adjustment
  to the energy
  deposition. 
  \\
  \indent
  The temporal evolution of the coronal averaged temperature
  from the Lare1D and LareJ loops 
  is shown on the right of each panel in Figure 
  \ref{Fig:IoR_Lare1d_vs_LareJ}. 
  For the Lare1D loop, it is clear that the 
  coronal temperature settles and remains at 3 MK from around
  1 hour onwards.
  On the other hand, the LareJ loop is
  initially heated to 3.6 MK before
  locally
  cooling to
  $10^4$K after 2 hours,
  in response to an increased coronal density. 
  The evolution then repeats and the loop follows a
  regular TNE cycle. 
  The period of the cycle is estimated from the troughs in the
  coronal averaged
  temperature as about 2.25 hours (three cycles in 
  seven hours from
  $t=4.5$ hours onwards). 
  \\
  \indent
  Thus, Figure \ref{Fig:IoR_Lare1d_vs_LareJ}
  again demonstrates that the existence of TNE cycles in
  coronal loop models is strongly dependent on obtaining the 
  \lq correct\rq\ plasma response.
  This is achieved in LareJ through the UTR approach whereas
  the Lare1D result is similar to the under-resolved results in 
  Section \ref{Sect:Resolution_HYDRAD}.
  \\
  \indent
  We do
  note that the TNE cycle period of the LareJ solution
  is slightly shorter than the
  fully resolved HYDRAD result and
  this discrepancy can be attributed to the sources of
  over-evaporation that are introduced when using the jump
  condition method (see \cite{paper:Johnstonetal2017b}). 
  However, if we focus on a comparison between Figures 
  \ref{Fig:IoR_HYDRAD} and \ref{Fig:IoR_Lare1d_vs_LareJ},
  LareJ agrees 
  qualitatively with the converged HYDRAD runs with
  only minor differences quantitatively,
  despite under-resolving the TR.
  The errors in the averaged 
  density and temperature 
  are just 3\% and 4\% respectively (over a TNE cycle). 
  This has the potential to be of great importance in 
  (a) 
  surveying the large parameter space associated with
  TNE
  \citep[e.g.][]
  {paper:Fromentetal2018}
  and (b)
  modelling TNE in active region simulations with multiple
  loop strands.
  \\
  \indent
  We note also that 
  the UTR method operates only at the 
  footpoints, not at the hot-cool transition at the edge of the 
  condensation. The simulations remain accurate at these 
  locations because the plasma cools largely in situ with no 
  flow through the interface.
  \\
  \indent
  Figure \ref{Fig:IoR_HYDRAD_ct} summarises the CPU
  requirements 
  of HYDRAD for all of the values of RL,
  and demonstrates the large decrease in CPU time of the
  UTR method (LareJ) over
  HYDRAD in the simulations where convergence of the 
  TNE cycle period is observed 
  ($\textrm{RL}\geq 9$). 
  In particular, LareJ 
  required at least one order of magnitude less computational
  time than HYDRAD run with 
  nine levels of refinement (i.e. 
  $\tau(\textrm{HYDRAD[RL=9])}/\tau(\textrm{LareJ})=18$). 
  This 
  is comparable to the improvements in run time described in 
  \cite{paper:Johnstonetal2017a}.
  Therefore,
  in the remainder
  of this paper to
  exploit the short computation time and because the general
  trends remain the same (i.e. the results are not
  method dependent),
  we use LareJ as a reference solution to 
  explore the effects of heating timescales and
  background heating.  
  %
  %
%%%%%%%%%%%%%%%%%%%%%%%%%%%%%%%%%%%%%%%%%%%%%%%%%%%%%%%%%%%%%%
%
% Effect of heating timescales
%
%%%%%%%%%%%%%%%%%%%%%%%%%%%%%%%%%%%%%%%%%%%%%%%%%%%%%%%%%%%%%%
\subsection{Effect of heating timescales
  \label{Sect:Time_dependent_fp_heating}}
  \indent
  In this Section we 
  explore the changes to TNE brought about by 
  unsteady footpoint heating.
  The heating ($Q_H$)
  is modified by assuming a
  time-dependent cycle
  comprising of a series of energy releases each lasting 
  $t_d$ seconds, 
  throughout which the maximum footpoint heating rate
  ($Q_{H_0}$) is constant, with a waiting time between these 
  heating events lasting $t_w$ seconds 
  when there is no footpoint heating (i.e. $Q_H=0$). The 
  background heating remains turned on during the waiting time. 
  Thus:
  \begin{align}
    Q(z,t) &= Q_{bg} + Q_H(z),
    \quad
    0 < t < t_d;
    \\[2mm]
    Q(z,t) &= Q_{bg},
    \quad
    t_d < t < t_d + t_w,
  \end{align}
  and so the cycle repeats over $t_d + t_w$ seconds.
  \\
  \indent
  The spatial footpoint heating profile 
  is the same as in Section \ref{Sect:Resolution}.
  However, we require that the
  total energy released is the same in all simulations when 
  averaged over
  a heating cycle, and 
  is equivalent to the steady 
  footpoint 
  heating simulations described previously. Thus the peak 
  heating rate ($Q_{H_0}$) in each simulation is
  increased by a factor $(t_d+t_w)/t_d$. 
  \\
  \indent
  The time-dependent footpoint heating cases are
  summarised in Table 
  \ref{Table:LareJ_td_fph_simulations}
  and include short and long heating pulses as well as a range 
  of the ratios $t_w/t_d$. The former ranges between 62.5 and 
  8000 seconds and the latter between one and seven. 
  We selected the values used for the 
  unsteady footpoint heating timescales, $t_d$ and $t_w$,
  based on current 
  coronal nanoflare models.   
  An upper limit of a few thousand seconds has been suggested
  for the waiting time 
  \citep[e.g.][]{paper:Cargill2014,
  Klimchuk_2015RSPTA.37340256K,paper:Marshetal2018}. 
  Thus this range is encompassed for $t_w$ but
  the nanoflare duration ($t_d$) is more problematic. 
  Some authors believe that $t_d$ is short (tens of seconds)
  while others 
  that it is long
  \citep[hundreds of 
  seconds e.g.][]{paper:Klimchuketal2008}.
  Hence, 
  we consider a large range for $t_d$ 
  due to the uncertainty on the real value. 
  \\
  \indent
  Figures 
  \ref{Fig:td_fph_LareJ_T} and
  \ref{Fig:td_fph_LareJ_Tn_ca}
  show the results for Cases 1 -- 8 in Table 
  \ref{Table:LareJ_td_fph_simulations}, 
  with steady footpoint heating shown for reference in the 
  upper left panel. 
  Figure \ref{Fig:td_fph_LareJ_T} shows the loop temperature as 
  a function of position (horizontal axis) and time (vertical 
  axis) and 
  Figure
  \ref{Fig:td_fph_LareJ_Tn_ca} the coronal averaged temperature 
  (red) and density (blue) as a function of time.
  The former 
  provides both spatial and temporal information to correlate
  while the latter allows a comparison between the coronal 
  properties
  around the loop apex (between $z=67.5$~Mm and $z=112.5$~Mm)
  as a function of time. 
  In particular,
  the phasing between the coronal averaged temperature and
  density is used to 
  identify the characteristic behaviour
  of the simulations (e.g. TNE has 
  peak density at the time of the temperature minimum).
  \\
  \indent
  There are two regimes evident, one exists for short 
  {\it and} long heating pulses and the other for intermediate 
  values, although there is also overlap between them. The 
  fifth column of Table \ref{Table:LareJ_td_fph_simulations} 
  provides a concise summary. For short $t_d$ and $t_w$ (up to 
  $t_d = 500 s$) the properties are similar to steady heating, 
  with TNE occuring and determining the cyclic 
  behaviour of the loop evolution. However, superposed on top 
  of this 
  behaviour is a jaggedness in both $T$ and $n$ 
  associated with the impulsive heating.
  \\
  \indent
  For longer $t_d$ and $t_w$ the 
  behaviour changes, with the TNE cycle becoming less evident 
  and the loop cyclic evolution being essentially the 
  same as the heating cycle
  and, by $t_d = 4000$ s, there is a transition to a loop that 
  undergoes a heating and cooling cycle, but without evidence 
  of TNE and catastrophic cooling
  (we refer to this 
  as 
  \lq global cooling\rq\ to indicate 
  the absence of the very localised cool and dense
  regions characteristic 
  of TNE). 
  The loop cyclic evolution in these cases is 
  entirely determined by the heating cycle.
  Such global cooling is the behaviour seen with
  \lq intermediate\rq\ and \lq low-frequency\rq\
  coronal nanoflares 
  \citep[e.g.][]{paper:Cargill2014,
  paper:Cargilletal2015}. 
  \\
  \indent
  However, for very large 
  $t_d$, 
  catastrophic cooling
  from the thermal instability 
  returns and occurs prior 
  to global cooling, without the cyclic character of TNE but 
  with the 
  loop evolution determined by the heating 
  cycle. The last two panels of Figure
  \ref{Fig:td_fph_LareJ_Tn_ca} make this point well: panel 8 
  shows global cooling (or nanoflare-like response) and panel 9 
  catastrophic cooling with
  global cooling.
  \\
  \indent
  The different regimes are highlighted further in Figure 
  \ref{Fig:td_fph_LareJ_Tpnv} with the upper and lower panels 
  showing a series of snapshots of $T$, $P$, $n$ and $v$ for 
  Cases 2 ($t_d = 125$ s) and 7 ($t_d = 4000$ s) 
  respectively. For Case 2, the evolution is, despite the 
  bursty nature of the heating, representative of TNE as 
  discussed in the literature. Thus the 250~s cycle of the 
  heating, being shorter than the characteristic time for TNE 
  to evolve, plays no significant role. 
  On the other hand, Case 
  7 shows evolution characteristic of an impulsively-heated 
  loop 
  \citep[e.g.][]{paper:Klimchuk2006}
  with a rise in 
  temperature, 
  followed by the density increase due to evaporation, then, 
  after the time of maximum density, an enthalpy and radiative 
  global cooling phase \citep{paper:Bradshaw&Cargill2010b}. 
  In this 
  case, the heating is turned on for just over one hour and 
  thermal instability 
  does not have time to develop before the heating declines. In 
  other words, the density due to the evaporation is limited to 
  a value below that needed for thermal instability. 
  \\
  \indent
  Figure \ref{Fig:td_fph_LareJ_Tpnv} also
  demonstrates that TNE in the corona (upper four panels)
  can be characterised by quantities such as the 
  skew and flatness. This is clear from the top left-hand
  panel where the temperature 
  evolution between 6.5 and 7.25 hours shows both developing.
  In contrast, 
  the low frequency nanoflare-like response
  (lower four panels)
  does not show this type of behaviour.
  \\
  \indent  
  It is also 
  interesting to note that for Case 8, there is a return to 
  thermal instability, and the TNE cyclic 
  evolution would return if the heating was kept for longer 
  times. Here the heating pulse is long enough for the loop to 
  see it as being \lq steady\rq, 
  so that the density in the corona 
  is large enough for thermal instability (and 
  the corresponding catastrophic cooling) to set in.
  \\
  \indent
  For cases 1 - 8, the transition between the TNE 
  cycle and the heating (and global cooling) cycle 
  occurs 
  roughly when $t_d = t_w = 2000$ s with the 
  4000~s 
  cycle roughly being equal to the characteristic time 
  for 
  thermal instability onset (and subsequent TNE 
  cycle), with a partial transition back 
  to catastrophic
  cooling for Case 8. Cases 9 - 16 reinforce our conclusions 
  with TNE occurring 
  only for short pulse cycles. 
  \\
  \indent
  However, within this general 
  classification, there are some  subtleties when we switch 
  between the two types of
  characteristic behaviour.
  This transition takes place when
  the waiting time ($t_w$)
  between heating periods becomes comparable 
  to the loop cooling time. 
  The outcome is a mixed regime which is
  characterised by properties that
  incorporate
  both types of behaviour.
  For example, Cases 6 and 11 
  exhibit catastrophic cooling from
  the triggering of the thermal instability
  and global cooling from the ending of the
  heating pulse.
  These cases have waiting times of 2000~s and 1500~s,
  respectively. 
  \\
  \indent
  We stress that these examples are limited in that strong 
  symmetry is assumed both with the intensity and time 
  variability of the heating at both footpoints. While the 
  breaking of this symmetry could lead to new forms of 
  behaviour 
  such as siphon flows 
  \citep[e.g.][]{paper:Cargill&Priest1980,
  paper:Mikicetal2013,
  paper:Fromentetal2018}, the 
  dependence of the loop evolution on time-variability of the 
  heating can be expected to persist.
  %
  %
%%%%%%%%%%%%%%%%%%%%%%%%%%%%%%%%%%%%%%%%%%%%%%%%%%%%%%%%%%%%%%
%
% Effect of the background heating
%
%%%%%%%%%%%%%%%%%%%%%%%%%%%%%%%%%%%%%%%%%%%%%%%%%%%%%%%%%%%%%%
\subsection{Effect of the background heating
  \label{Sect:Q_bg}}
  \indent
  Next we investigate the effect of the background heating
  on the TNE cycle evolution. 
  The background heating ($Q_{bg}$) is varied over several 
  orders of
  magnitude, ranging from no background heating up to 
  $10^{-4}$ Jm$^{-3}$s$^{-1}$. All other parameters are as in 
  previous Sections, the footpoint heating uses the values of 
  Case 2 in Section \ref{Sect:Time_dependent_fp_heating} and 
  LareJ results are shown.
  \\
  \indent
  Figure \ref{Fig:IoQbg_LareJ}
  shows the
  evolution of the temperature,
  as a function of time and
  position
  along the loop
  with
  $Q_{bg}=$ [0, 6.8682$\times 10^{-6}$, 2$\times 10^{-5}$, 
  4$\times 10^{-5}$, 6$\times 10^{-5}$, $10^{-4}$]  
  Jm$^{-3}$s$^{-1}$ in the six panels. The 
  sum of the spatial distribution 
  of the footpoint heating at the peak of the heating cycle and 
  the background heating (blue) are shown below each panel.
  For reference
  we note that 
  $Q_{bg}=6.8682\times 10^{-6}$ Jm$^{-3}$s$^{-1}$
  corresponds to
  the minimum value
  required to achieve thermal balance in the hydrostatic
  initial condition and will be referred to as the 
  equilibrium background heating value.
  Furthermore, 
  $Q_{bg}=6\times 10^{-5}$ Jm$^{-3}$s$^{-1}$ was
  previously used as the background heating value in the
  Model 1 heating profile
  that was considered by
  \cite{paper:Mikicetal2013}. 
  \\
  \indent
  Figure \ref{Fig:IoQbg_LareJ} shows that five of the loops 
  experience TNE 
  with condensations but
  the cycle periods (and thermodynamic evolution during a 
  cycle) are
  significantly different.
  For example, the case with
  \lq equilibrium\rq\ background heating value
  has a TNE cycle period of 2.5 hours, while increasing 
  $Q_{bg}$ to $6\times 10^{-5}$ Jm$^{-3}$s$^{-1}$
  increases the period to 4.25 hours.  Moreover, for no 
  background heating the 
  period is 2.25 hours.
  We also note that the
  case with the largest background
  heating 
  ($Q_{bg}=10^{-4}$ Jm$^{-3}$s$^{-1}$), 
  is stable to the thermal instability and instead
  settles to a static equilibrium.
  \\
  \indent
  Figure \ref{Fig:f_vs_Qbg_LareJ} shows the range of
  TNE cycle periods obtained
  for all of the background heating runs.
  Convergence to a period of
  2.25 hours is only observed for small background
  heating values ($Q_{bg} \leq 10^{-6}$ Jm$^{-3}$s$^{-1}$)
  while loops computed with 
  $Q_{bg} \geq 9\times 10^{-5}$ Jm$^{-3}$s$^{-1}$ do not 
  experience TNE.  
  \\
  \indent
  The suppression of TNE arises as $Q_{bg}$ increases
  because for it to occur, 
  the radiative losses must exceed the 
  heating in the corona. Obviously as $Q_{bg}$ increases, this 
  becomes more difficult.  
  Therefore,
  triggering the thermal 
  instability when an increased background heating value
  is used requires
  either (i) an extended heating duration 
  or (ii) an increase in the 
  magnitude of the maximum footpoint heating rate
  ($Q_{H_0}$) 
  in order to  
  accumulate a sufficient amount of mass in the 
  corona. 
  The influence of the 
  latter is observed as the suppression of TNE cycles
  when the footpoint heating rate
  is not increased relative to
  background heating rate while the
  effect of the former is seen as  
  an increase in the TNE cycle period.
  We can thus conclude that 
  the ratio of
  the maximum footpoint heating rate
  to the background heating value
  plays a key role in the onset criteria for TNE.
  %
  %
%%%%%%%%%%%%%%%%%%%%%%%%%%%%%%%%%%%%%%%%%%%%%%%%%%%%%%%%%%%%%%
%
% Discussion and Conclusions
%
%%%%%%%%%%%%%%%%%%%%%%%%%%%%%%%%%%%%%%%%%%%%%%%%%%%%%%%%%%%%%%
\section{Discussion and conclusions
  \label{Sect:dis_concl}}
  \indent
  The phenomenon of TNE is a very challenging one for numerical 
  models for several reasons, in particular the need to 
  correctly resolve the TR and so sustain precise periodicity 
  and thermodynamic characteristics 
  of the coronal condensations. We have shown that inadequate 
  TR resolution can lead to incorrect properties of 
  the TNE cycles and even 
  the suppression of TNE. An approximate method of handling the 
  TR is shown to eliminate this problem with the benefit of 
  significantly shorter computational times while introducing a 
  small discrepancy of order 15\% in the condensation 
  periodicity. Furthermore, when averaged over a TNE cycle,
  the error in the coronal density and temperature evolution
  is only 3\% and 4\% respectively.  
  \\
  \indent
  The approximate method is applied to models of TNE with 
  steady, uniformly distributed \lq background\rq\ 
  heating as 
  well as unsteady footpoint heating.
  In the former case we find a trend evident in previous work 
  \cite[e.g. see Models 1 and 2 in][]{paper:Mikicetal2013}, 
  namely that TNE in a loop 
  with footpoint 
  heating is suppressed unless
  the  background heating is 
  sufficiently small. Such a small steady background heating 
  is sometimes included in models for computational reasons 
  (e.g. Section \ref{Sect:res}) 
  and a physical motivation is currently unclear. 
  However, any 
  background heating need not be 
  steady for TNE to be suppressed. For example high 
  frequency coronal nanoflares
  \citep[][]{paper:Warrenetal2011, paper:Cargilletal2015} would 
  achieve this, as discussed in \citet{paper:Antolinetal2010}, 
  where the dissipation of torsional finite-amplitude
  Alfv\'en waves by shock heating 
  \citep{Antolin_2008ApJ...688..669A} 
  eliminates the TNE cycles present in a footpoint 
  heated loop and leads instead to a uniformly heated loop in 
  thermal 
  equilibrium.
  \\
  \indent  
  On the other hand, a background of low 
  frequency coronal nanoflares \citep[]{paper:Cargilletal2015} 
  are sufficiently infrequent that 
  TNE can 
  proceed superimposed on their behaviour (although 
  they may increase the resulting period of the loop's cyclic 
  evolution). Intermediate 
  frequency nanoflares, now widely believed to be responsible 
  for active region heating 
  \citep[][]{paper:Cargill2014,
  paper:Barnesetal2016,
  paper:Viall&Klimchuk2017}
  repeat on 
  the loop cooling time, so that results similar to those shown 
  in Section \ref{Sect:Q_bg}
  will be found. 
  \\
  \indent
  Unsteady footpoint heating leads to further 
  complications. 
  For heating bursts of a specified duration 
  separated by a waiting time, 
  thermal instability arises for both 
  short and very long (> 6000 s) heating times.  
  Cyclic evolution is given by the 
  TNE cycle for the short heating times and by the heating 
  cycle itself for the longer.
  In addition, a new intermediate regime of mixed 
  catastrophic cooling from thermal instability and global 
  cooling from the cessation of heating arises for waiting
  times of order a few thousand seconds and is comparable to
  the behaviour of a corona heated by low-frequency 
  nanoflares. 
  \\
  \indent
  When the condensation dynamics are considered, 
  important 
  differences arise that can help distinguish this intermediate 
  regime in 
  observations. During a TNE cycle the speed of a 
  falling condensation is determined by its mass and by the 
  surrounding coronal gas pressure 
  \citep{paper:Antolinetal2010,Oliver_2014ApJ...784...21O}. 
  Instead, condensations occur along 
  with global cooling, so that the coronal pressure 
  drops as the condensations fall, thus rapidly increasing 
  their speeds with accelerations tending to the gravitational 
  value. Coronal rain is usually observed to fall with an 
  acceleration lower than due
  to gravity along a curved loop
  \citep{Schrijver_2001SoPh..198..325S,
  DeGroof_2004AA...415.1141D,
  Antolin_etal_2012SoPh..280..457A,
  Antolin_Rouppe_2012ApJ...745..152A}, 
  although the observed distribution of 
  speeds for the rain has a long tail towards higher values. 
  These results suggest
  a broad range of possible TNE evolution depending on the 
  waiting time but one that in an average sense does not
  correspond to the intermediate regime.
\subsection{Parametric dependence of TNE}
  \indent
  To conclude this paper, 
  we address two points that seem to be
  essential in advancing the topic of TNE in
  coronal loops. Clearly 
  there is a complex relation between the onset of 
  TNE as various footpoint heating parameters are varied. While 
  models of TNE predict
  generic global and 
  local features in spectral diagnostics of the solar corona 
  including
  loops undergoing a cyclic 
  heating and cooling evolution of evaporation and 
  condensation, the existence of these cycles 
  depends on several parameters  
  \citep[e.g.][]
  {
  paper:Antiochosetal2000,
  paper:Karpenetal2001,
  paper:Mulleretal2003,
  paper:Mulleretal2004,
  paper:Mulleretal2005, 
  Mendozabriceno_2005ApJ...624.1080M,
  paper:Moketal2008, 
  paper:Antolinetal2010, 
  paper:Susinoetal2010,
  paper:Lionelloetal2013,
  paper:Mikicetal2013,
  paper:Susinoetal2013, 
  paper:Moketal2016}
  which include (but need not be limited to) geometrical 
  factors 
  (such as the loop length and area expansion) and the nature 
  of the heating mechanism (its spatial and temporal 
  distribution, the degree of asymmetry between both 
  footpoints, its stochasticity and so on). As shown by 
  \citet{paper:Fromentetal2018} the existence of TNE cycles 
  seem to be very sensitive to some of these parameters, 
  suggesting that if they are approximately 
  uniformly distributed, the vast 
  majority of loops should not exhibit TNE.
  \\
  \indent
  Assuming that loops consist 
  of independently heated strands (and therefore undergo an 
  independent thermodynamic evolution) 
  \citet{Klimchuk_2010ApJ...714.1239K} 
  have shown that the TNE theory leads to inconsistencies with 
  observational constraints for the solar corona 
  \citep[see also][]{Klimchuk_2015RSPTA.37340256K}.  
  While TNE does seem to explain very well loops with 
  coronal rain 
  \citep{paper:Moketal2008,Kamio_etal_2011AA...532A..96K,
  Peter_2012AA...548A...1P,
  paper:Antolinetal2010,
  Antolin_Rouppe_2012ApJ...745..152A,
  paper:Antolinetal2015,
  Fang_2013ApJ...771L..29F,
  Fang_2015ApJ...807..142F,
  Fang_2016ApJ...833...36F} and loops exhibiting highly 
  periodic EUV intensity pulsations 
  \citep{paper:Auchereetal2014,paper:Fromentetal2015,paper:Fromentetal2018,paper:Auchereetal2018}, 
  it is unclear how much of the coronal volume is 
  involved in these phenomena. Recent 
  numerical work has shown that by varying some of the above 
  mentioned parameters 
  \citep{paper:Mikicetal2013,paper:Fromentetal2018,
  Winebarger_2018ApJ...865..111W} 
  and including multi-dimensional effects 
  \citep{Lionello_2013ApJ...773..134L,
  paper:Moketal2016,Winebarger_2016ApJ...831..172W,
  Xia_2017AA...603A..42X} some 
  of the difficulties may be resolved. 
  \\
  \indent
  Thus the large parameter space involved 
  in the existence of TNE constitutes a challenge for
  disentangling the 
  properties of the underlying heating. As a further example,
  in our parameter space investigation we have 
  produced loops with very similar global parameters (TNE cycle 
  period, average coronal density and temperature, siphon flow 
  velocities and so on) but with significant variation in the 
  amount of heating input (a factor of one third to one half).
  Caution must 
  therefore be placed on using only one of these proxies (such 
  as the TNE cycle period) rather than the ensemble of 
  observational constraints, both at a global and local level 
  (for instance, regarding the dynamics of condensations and 
  their observational signatures).	
  \subsection{Footpoint heating}
  \indent
  Many results have shown that the presence of TNE 
  requires 
  footpoint heating. It is thus pertinent to ask what the 
  direct evidence for such localised heating actually is. One 
  can dismiss the idea that the heating is strictly steady: 
  that is not how plasmas release energy, yet many models 
  impose just such a condition. However, as we have shown, high 
  frequency bursty heating does approximate steady heating. 
  \\
  \indent
  Direct observational evidence for footpoint heating is 
  limited
  \citep[e.g.][]{2008ApJ...678L..67H,2011ApJ...737L..43N,
  2014Sci...346B.315T}, the 
  latter being interpreted as the footpoint response to a 
  coronal acceleration of electrons. 
  Further, observations of spicules suggest 
  that associated
  heating to coronal temperatures along with nonthermal line
  broadenings and upflows all arise 
  in the lower solar atmosphere 
  \citep[e.g.][]{2011Sci...331...55D,2018ApJ...860..116M}.
  Whether this corresponds to
  adequate footpoint heating is unclear, but it has been
  suggested that minimal coronal heating in fact results 
  \citep[e.g.][]{2012JGRA..11712102K}. 
  Studies of TNE where both 
  its coronal manifestation and spectral observations at the 
  footpoints are thus an essential future requirement.
  \\
  \indent
  A number of theoretical models 
  also give results where
  heating is concentrated at the base of loops. These include
  \lq braiding\rq\ 
  type models
  \citep[e.g.][]{2005ApJ...618.1031G}
  and turbulent cascades due to interacting Alfven waves
  \citep[e.g.][]{2011ApJ...736....3V}. 
  However, 
  apart from limited numerical resolution, such models also 
  include unrealistically high dissipation coefficients, in 
  part to avoid numerical instabilities, which can lead to a 
  lack of ability to resolve fine-scale currents. A consequence 
  of this is an 
  effectively (temporally) constant coronal heating background 
  which may also be in the wrong place due to the lack of 
  ability to model the small-scale current structure throughout 
  the atmosphere. The problem confronting such models is 
  compounded by the difficulty of TR resolution. 
  \\
  \indent
  On the other 
  hand, models that do better in resolving such currents 
  \citep[e.g.][]{paper:Bareford&Hood2015} 
  do not attempt to model the 
  transition between chromosphere and corona, so cannot address 
  the question of footpoint heating. Thus it seems evident that 
  details of the \lq what, where or why\rq\
  of footpoint heating 
  are very unclear at this time.
  \\
  \indent
  A further important 
  result from this series of papers 
  \citep{paper:Johnstonetal2017a,paper:Johnstonetal2017b} and 
  previous work \citep{paper:Bradshaw&Cargill2013} is that 
  intensities of coronal lines may not be realistic 
  simply because of the wrong evaporative response of the 
  atmosphere to the heating input due to the limited numerical 
  resolution. 
  This is particularly likely to be a problem in 3D 
  MHD models.
  The present work also suggests that,
  besides coronal lines, TR and chromospheric lines may also be 
  affected since TNE cycles and the accompanying coronal rain 
  (which have strong signatures in chromospheric, transition 
  region and EUV lines) may be far more common than previously 
  thought in more realistic global 3D MHD models
  (and as suggested by observations).
  Thus approximate methods such as 
  presented here and by \cite{paper:Mikicetal2013} are 
  vital for large-scale contemporary MHD models since they can 
  \lq free up\rq\ grid points which can then be used to 
  resolve 
  better the currents responsible for the heating. 
  However,
  their extension to 3D
  MHD requires a more sophisticated treatment than in 1D, 
  in particular how the magnetic field and transverse velocity
  modify the jump relations. 
  %
  %
%%%%%%%%%%%%%%%%%%%%%%%%%%%%%%%%%%%%%%%%%%%%%%%%%%%%%%%%%%%%%%
%
% Acknowledgements
%
%%%%%%%%%%%%%%%%%%%%%%%%%%%%%%%%%%%%%%%%%%%%%%%%%%%%%%%%%%%%%%
\begin{acknowledgements} 
  This research has received funding from the 
  European Research Council (ERC) 
  under the European Union's Horizon 2020 research and 
  innovation programme (grant agreement No 647214) and the
  UK Science and 
  Technology Facilities Council through the consolidated 
  grant ST/N000609/1.
  P.A. has received funding from his 
  STFC Ernest Rutherford Fellowship 
  (grant agreement No. ST/R004285/1). 
  S.J.B. is grateful to the National Science Foundation 
  for supporting this work through CAREER award AGS-1450230.
  C.D.J. and P.A. 
  acknowledge support from the International Space Science 
  Institute (ISSI), Bern, Switzerland to the International Team 
  401 \lq\lq Observed Multi-Scale Variability of Coronal Loops 
  as a 
  Probe of Coronal Heating\rq\rq.
\end{acknowledgements}
  %
  %
%%%%%%%%%%%%%%%%%%%%%%%%%%%%%%%%%%%%%%%%%%%%%%%%%%%%%%%%%%%%%%
%
% References
%
%%%%%%%%%%%%%%%%%%%%%%%%%%%%%%%%%%%%%%%%%%%%%%%%%%%%%%%%%%%%%%
\bibliographystyle{aa}
\bibliography{tne_paper}

\begin{thebibliography}{81}
\expandafter\ifx\csname natexlab\endcsname\relax\def\natexlab#1{#1}\fi

\bibitem[{{Antiochos} \& {Klimchuk}(1991)}]{paper:Antiochos&Klimchuk1991}
{Antiochos}, S.~K. \& {Klimchuk}, J.~A. 1991, \apj, 378, 372

\bibitem[{{Antiochos} {et~al.}(2000){Antiochos}, {MacNeice}, \&
  {Spicer}}]{paper:Antiochosetal2000}
{Antiochos}, S.~K., {MacNeice}, P.~J., \& {Spicer}, D.~S. 2000, \apj, 536, 494

\bibitem[{{Antiochos} {et~al.}(1999){Antiochos}, {MacNeice}, {Spicer}, \&
  {Klimchuk}}]{paper:Antiochosetal1999}
{Antiochos}, S.~K., {MacNeice}, P.~J., {Spicer}, D.~S., \& {Klimchuk}, J.~A.
  1999, \apj, 512, 985

\bibitem[{{Antiochos} \& {Sturrock}(1978)}]{paper:Antiochos&Sturrock1978}
{Antiochos}, S.~K. \& {Sturrock}, P.~A. 1978, \apj, 220, 1137

\bibitem[{{Antolin} \& {Rouppe van der
  Voort}(2012)}]{Antolin_Rouppe_2012ApJ...745..152A}
{Antolin}, P. \& {Rouppe van der Voort}, L. 2012, \apj, 745, 152

\bibitem[{{Antolin} {et~al.}(2008){Antolin}, {Shibata}, {Kudoh}, {Shiota}, \&
  {Brooks}}]{Antolin_2008ApJ...688..669A}
{Antolin}, P., {Shibata}, K., {Kudoh}, T., {Shiota}, D., \& {Brooks}, D. 2008,
  \apj, 688, 669

\bibitem[{{Antolin} {et~al.}(2010){Antolin}, {Shibata}, \&
  {Vissers}}]{paper:Antolinetal2010}
{Antolin}, P., {Shibata}, K., \& {Vissers}, G. 2010, \apj, 716, 154

\bibitem[{{Antolin} {et~al.}(2015){Antolin}, {Vissers}, {Pereira}, {Rouppe van
  der Voort}, \& {Scullion}}]{paper:Antolinetal2015}
{Antolin}, P., {Vissers}, G., {Pereira}, T.~M.~D., {Rouppe van der Voort}, L.,
  \& {Scullion}, E. 2015, \apj, 806, 81

\bibitem[{{Antolin} {et~al.}(2012){Antolin}, {Vissers}, \& {Rouppe van der
  Voort}}]{Antolin_etal_2012SoPh..280..457A}
{Antolin}, P., {Vissers}, G., \& {Rouppe van der Voort}, L. 2012, \solphys,
  280, 457

\bibitem[{{Arber} {et~al.}(2001){Arber}, {Longbottom}, {Gerrard}, \&
  {Milne}}]{paper:Arber2001}
{Arber}, T.~D., {Longbottom}, A.~W., {Gerrard}, C.~L., \& {Milne}, A.~M. 2001,
  Journal of Computational Physics, 171, 151

\bibitem[{{Auch{\`e}re} {et~al.}(2014){Auch{\`e}re}, {Bocchialini}, {Solomon},
  \& {Tison}}]{paper:Auchereetal2014}
{Auch{\`e}re}, F., {Bocchialini}, K., {Solomon}, J., \& {Tison}, E. 2014, \aap,
  563, A8

\bibitem[{{Auch{\`e}re} {et~al.}(2018){Auch{\`e}re}, {Froment}, {Soubri{\'e}},
  {Antolin}, {Oliver}, \& {Pelouze}}]{paper:Auchereetal2018}
{Auch{\`e}re}, F., {Froment}, C., {Soubri{\'e}}, E., {et~al.} 2018, \apj, 853,
  176

\bibitem[{{Bareford} \& {Hood}(2015)}]{paper:Bareford&Hood2015}
{Bareford}, M.~R. \& {Hood}, A.~W. 2015, Philosophical Transactions of the
  Royal Society of London Series A, 373, 20140266

\bibitem[{{Barnes} {et~al.}(2016){Barnes}, {Cargill}, \&
  {Bradshaw}}]{paper:Barnesetal2016}
{Barnes}, W.~T., {Cargill}, P.~J., \& {Bradshaw}, S.~J. 2016, \apj, 833, 217

\bibitem[{{Betta} {et~al.}(1997){Betta}, {Peres}, {Reale}, \&
  {Serio}}]{paper:Bettaetal1997}
{Betta}, R., {Peres}, G., {Reale}, F., \& {Serio}, S. 1997, \aaps, 122

\bibitem[{{Bradshaw} \& {Cargill}(2006)}]{paper:Bradshaw&Cargill2006}
{Bradshaw}, S.~J. \& {Cargill}, P.~J. 2006, \aap, 458, 987

\bibitem[{{Bradshaw} \& {Cargill}(2010b)}]{paper:Bradshaw&Cargill2010b}
{Bradshaw}, S.~J. \& {Cargill}, P.~J. 2010b, \apj, 717, 163

\bibitem[{{Bradshaw} \& {Cargill}(2013)}]{paper:Bradshaw&Cargill2013}
{Bradshaw}, S.~J. \& {Cargill}, P.~J. 2013, \apj, 770, 12

\bibitem[{{Bradshaw} \& {Mason}(2003)}]{paper:Bradshaw&Mason2003}
{Bradshaw}, S.~J. \& {Mason}, H.~E. 2003, \aap, 407, 1127

\bibitem[{{Cargill}(2014)}]{paper:Cargill2014}
{Cargill}, P.~J. 2014, \apj, 784, 49

\bibitem[{{Cargill} {et~al.}(2012a){Cargill}, {Bradshaw}, \&
  {Klimchuk}}]{paper:Cargilletal2012a}
{Cargill}, P.~J., {Bradshaw}, S.~J., \& {Klimchuk}, J.~A. 2012a, \apj, 752, 161

\bibitem[{{Cargill} \& {Priest}(1980)}]{paper:Cargill&Priest1980}
{Cargill}, P.~J. \& {Priest}, E.~R. 1980, \solphys, 65, 251

\bibitem[{{Cargill} {et~al.}(2015){Cargill}, {Warren}, \&
  {Bradshaw}}]{paper:Cargilletal2015}
{Cargill}, P.~J., {Warren}, H.~P., \& {Bradshaw}, S.~J. 2015, Philosophical
  Transactions of the Royal Society of London Series A, 373, 20140260

\bibitem[{{De Groof} {et~al.}(2005){De Groof}, {Bastiaensen}, {M{\"u}ller},
  {Berghmans}, \& {Poedts}}]{DeGroof05}
{De Groof}, A., {Bastiaensen}, C., {M{\"u}ller}, D.~A.~N., {Berghmans}, D., \&
  {Poedts}, S. 2005, \aap, 443, 319

\bibitem[{{De Groof} {et~al.}(2004){De Groof}, {Berghmans}, {van
  Driel-Gesztelyi}, \& {Poedts}}]{DeGroof_2004AA...415.1141D}
{De Groof}, A., {Berghmans}, D., {van Driel-Gesztelyi}, L., \& {Poedts}, S.
  2004, \aap, 415, 1141

\bibitem[{{De Pontieu} {et~al.}(2011){De Pontieu}, {McIntosh}, {Carlsson},
  {Hansteen}, {Tarbell}, {Boerner}, {Martinez-Sykora}, {Schrijver}, \&
  {Title}}]{2011Sci...331...55D}
{De Pontieu}, B., {McIntosh}, S.~W., {Carlsson}, M., {et~al.} 2011, Science,
  331, 55

\bibitem[{{Fang} {et~al.}(2013){Fang}, {Xia}, \&
  {Keppens}}]{Fang_2013ApJ...771L..29F}
{Fang}, X., {Xia}, C., \& {Keppens}, R. 2013, \apjl, 771, L29

\bibitem[{{Fang} {et~al.}(2015){Fang}, {Xia}, {Keppens}, \& {Van
  Doorsselaere}}]{Fang_2015ApJ...807..142F}
{Fang}, X., {Xia}, C., {Keppens}, R., \& {Van Doorsselaere}, T. 2015, \apj,
  807, 142

\bibitem[{{Fang} {et~al.}(2016){Fang}, {Yuan}, {Xia}, {Van Doorsselaere}, \&
  {Keppens}}]{Fang_2016ApJ...833...36F}
{Fang}, X., {Yuan}, D., {Xia}, C., {Van Doorsselaere}, T., \& {Keppens}, R.
  2016, \apj, 833, 36

\bibitem[{{Field}(1965)}]{paper:Field1965}
{Field}, G.~B. 1965, \apj, 142, 531

\bibitem[{{Froment} {et~al.}(2017){Froment}, {Auch{\`e}re}, {Aulanier},
  {Miki{\'c}}, {Bocchialini}, {Buchlin}, \& {Solomon}}]{paper:Fromentetal2017}
{Froment}, C., {Auch{\`e}re}, F., {Aulanier}, G., {et~al.} 2017, \apj, 835, 272

\bibitem[{{Froment} {et~al.}(2015){Froment}, {Auch{\`e}re}, {Bocchialini},
  {Buchlin}, {Guennou}, \& {Solomon}}]{paper:Fromentetal2015}
{Froment}, C., {Auch{\`e}re}, F., {Bocchialini}, K., {et~al.} 2015, \apj, 807,
  158

\bibitem[{{Froment} {et~al.}(2018){Froment}, {Auch{\`e}re}, {Miki{\'c}},
  {Aulanier}, {Bocchialini}, {Buchlin}, {Solomon}, \&
  {Soubri{\'e}}}]{paper:Fromentetal2018}
{Froment}, C., {Auch{\`e}re}, F., {Miki{\'c}}, Z., {et~al.} 2018, \apj, 855, 52

\bibitem[{{Gudiksen} \& {Nordlund}(2005)}]{2005ApJ...618.1031G}
{Gudiksen}, B.~V. \& {Nordlund}, {\AA}. 2005, \apj, 618, 1031

\bibitem[{{Hara} {et~al.}(2008){Hara}, {Watanabe}, {Harra}, {Culhane}, {Young},
  {Mariska}, \& {Doschek}}]{2008ApJ...678L..67H}
{Hara}, H., {Watanabe}, T., {Harra}, L.~K., {et~al.} 2008, \apjl, 678, L67

\bibitem[{{Hildner}(1974)}]{paper:Hildner1974}
{Hildner}, E. 1974, \solphys, 35, 123

\bibitem[{{Johnston} {et~al.}(2017a){Johnston}, {Hood}, {Cargill}, \& {De
  Moortel}}]{paper:Johnstonetal2017a}
{Johnston}, C.~D., {Hood}, A.~W., {Cargill}, P.~J., \& {De Moortel}, I. 2017a,
  \aap, 597, A81

\bibitem[{{Johnston} {et~al.}(2017b){Johnston}, {Hood}, {Cargill}, \& {De
  Moortel}}]{paper:Johnstonetal2017b}
{Johnston}, C.~D., {Hood}, A.~W., {Cargill}, P.~J., \& {De Moortel}, I. 2017b,
  \aap, 605, A8

\bibitem[{{Kamio} {et~al.}(2011){Kamio}, {Peter}, {Curdt}, \&
  {Solanki}}]{Kamio_etal_2011AA...532A..96K}
{Kamio}, S., {Peter}, H., {Curdt}, W., \& {Solanki}, S.~K. 2011, \aap, 532, A96

\bibitem[{{Karpen} {et~al.}(2001){Karpen}, {Antiochos}, {Hohensee}, {Klimchuk},
  \& {MacNeice}}]{paper:Karpenetal2001}
{Karpen}, J.~T., {Antiochos}, S.~K., {Hohensee}, M., {Klimchuk}, J.~A., \&
  {MacNeice}, P.~J. 2001, \apjl, 553, L85

\bibitem[{{Kawaguchi}(1970)}]{Kawaguchi_1970PASJ...22..405K}
{Kawaguchi}, I. 1970, \pasj, 22, 405

\bibitem[{{Kjeldseth-Moe} \&
  {Brekke}(1998)}]{Kjeldseth_Brekke_1998SoPh..182...73K}
{Kjeldseth-Moe}, O. \& {Brekke}, P. 1998, \solphys, 182, 73

\bibitem[{{Klimchuk}(2006)}]{paper:Klimchuk2006}
{Klimchuk}, J.~A. 2006, \solphys, 234, 41

\bibitem[{{Klimchuk}(2012)}]{2012JGRA..11712102K}
{Klimchuk}, J.~A. 2012, Journal of Geophysical Research (Space Physics), 117,
  A12102

\bibitem[{{Klimchuk}(2015)}]{Klimchuk_2015RSPTA.37340256K}
{Klimchuk}, J.~A. 2015, Philosophical Transactions of the Royal Society of
  London Series A, 373, 20140256

\bibitem[{{Klimchuk} {et~al.}(2010){Klimchuk}, {Karpen}, \&
  {Antiochos}}]{Klimchuk_2010ApJ...714.1239K}
{Klimchuk}, J.~A., {Karpen}, J.~T., \& {Antiochos}, S.~K. 2010, \apj, 714, 1239

\bibitem[{{Klimchuk} {et~al.}(2008){Klimchuk}, {Patsourakos}, \&
  {Cargill}}]{paper:Klimchuketal2008}
{Klimchuk}, J.~A., {Patsourakos}, S., \& {Cargill}, P.~J. 2008, \apj, 682, 1351

\bibitem[{{Kuin} \& {Martens}(1982)}]{paper:Kuin&Martens1982}
{Kuin}, N.~P.~M. \& {Martens}, P.~C.~H. 1982, \aap, 108, L1

\bibitem[{{Leroy}(1972)}]{Leroy_1972SoPh...25..413L}
{Leroy}, J. 1972, \solphys, 25, 413

\bibitem[{{Levine} \& {Withbroe}(1977)}]{Levine_Withbroe_1977SoPh...51...83L}
{Levine}, R.~H. \& {Withbroe}, G.~L. 1977, \solphys, 51, 83

\bibitem[{{Lionello} {et~al.}(2016){Lionello}, {Alexander}, {Winebarger},
  {Linker}, \& {Miki{\'c}}}]{Lionello_2016ApJ...818..129L}
{Lionello}, R., {Alexander}, C.~E., {Winebarger}, A.~R., {Linker}, J.~A., \&
  {Miki{\'c}}, Z. 2016, \apj, 818, 129

\bibitem[{{Lionello} {et~al.}(2013{\natexlab{a}}){Lionello}, {Winebarger},
  {Mok}, {Linker}, \& {Miki{\'c}}}]{paper:Lionelloetal2013}
{Lionello}, R., {Winebarger}, A.~R., {Mok}, Y., {Linker}, J.~A., \&
  {Miki{\'c}}, Z. 2013{\natexlab{a}}, \apj, 773, 134

\bibitem[{{Lionello} {et~al.}(2013{\natexlab{b}}){Lionello}, {Winebarger},
  {Mok}, {Linker}, \& {Miki{\'c}}}]{Lionello_2013ApJ...773..134L}
{Lionello}, R., {Winebarger}, A.~R., {Mok}, Y., {Linker}, J.~A., \&
  {Miki{\'c}}, Z. 2013{\natexlab{b}}, \apj, 773, 134

\bibitem[{{Marsh} {et~al.}(2018){Marsh}, {Smith}, {Glesener}, {Klimchuk},
  {Bradshaw}, {Vievering}, {Hannah}, {Christe}, {Ishikawa}, \&
  {Krucker}}]{paper:Marshetal2018}
{Marsh}, A.~J., {Smith}, D.~M., {Glesener}, L., {et~al.} 2018, \apj, 864, 5

\bibitem[{{Mart{\'{\i}}nez-Sykora} {et~al.}(2018){Mart{\'{\i}}nez-Sykora}, {De
  Pontieu}, {De Moortel}, {Hansteen}, \& {Carlsson}}]{2018ApJ...860..116M}
{Mart{\'{\i}}nez-Sykora}, J., {De Pontieu}, B., {De Moortel}, I., {Hansteen},
  V.~H., \& {Carlsson}, M. 2018, \apj, 860, 116

\bibitem[{{Mendoza-Brice{\~n}o} {et~al.}(2005){Mendoza-Brice{\~n}o},
  {Sigalotti}, \& {Erd{\'e}lyi}}]{Mendozabriceno_2005ApJ...624.1080M}
{Mendoza-Brice{\~n}o}, C.~A., {Sigalotti}, L.~D.~G., \& {Erd{\'e}lyi}, R. 2005,
  \apj, 624, 1080

\bibitem[{{Miki{\'c}} {et~al.}(2013){Miki{\'c}}, {Lionello}, {Mok}, {Linker},
  \& {Winebarger}}]{paper:Mikicetal2013}
{Miki{\'c}}, Z., {Lionello}, R., {Mok}, Y., {Linker}, J.~A., \& {Winebarger},
  A.~R. 2013, \apj, 773, 94

\bibitem[{{Mok} {et~al.}(1990){Mok}, {Drake}, {Schnack}, \& {van
  Hoven}}]{paper:Moketal1990}
{Mok}, Y., {Drake}, J.~F., {Schnack}, D.~D., \& {van Hoven}, G. 1990, \apj,
  359, 228

\bibitem[{{Mok} {et~al.}(2016){Mok}, {Miki{\'c}}, {Lionello}, {Downs}, \&
  {Linker}}]{paper:Moketal2016}
{Mok}, Y., {Miki{\'c}}, Z., {Lionello}, R., {Downs}, C., \& {Linker}, J.~A.
  2016, \apj, 817, 15

\bibitem[{{Mok} {et~al.}(2008){Mok}, {Miki{\'c}}, {Lionello}, \&
  {Linker}}]{paper:Moketal2008}
{Mok}, Y., {Miki{\'c}}, Z., {Lionello}, R., \& {Linker}, J.~A. 2008, \apjl,
  679, L161

\bibitem[{{M{\"u}ller} {et~al.}(2005){M{\"u}ller}, {De Groof}, {Hansteen}, \&
  {Peter}}]{paper:Mulleretal2005}
{M{\"u}ller}, D.~A.~N., {De Groof}, A., {Hansteen}, V.~H., \& {Peter}, H. 2005,
  \aap, 436, 1067

\bibitem[{{M{\"u}ller} {et~al.}(2003){M{\"u}ller}, {Hansteen}, \&
  {Peter}}]{paper:Mulleretal2003}
{M{\"u}ller}, D.~A.~N., {Hansteen}, V.~H., \& {Peter}, H. 2003, \aap, 411, 605

\bibitem[{{M{\"u}ller} {et~al.}(2004){M{\"u}ller}, {Peter}, \&
  {Hansteen}}]{paper:Mulleretal2004}
{M{\"u}ller}, D.~A.~N., {Peter}, H., \& {Hansteen}, V.~H. 2004, \aap, 424, 289

\bibitem[{{Nishizuka} \& {Hara}(2011)}]{2011ApJ...737L..43N}
{Nishizuka}, N. \& {Hara}, H. 2011, \apjl, 737, L43

\bibitem[{{Oliver} {et~al.}(2014){Oliver}, {Soler}, {Terradas}, {Zaqarashvili},
  \& {Khodachenko}}]{Oliver_2014ApJ...784...21O}
{Oliver}, R., {Soler}, R., {Terradas}, J., {Zaqarashvili}, T.~V., \&
  {Khodachenko}, M.~L. 2014, \apj, 784, 21

\bibitem[{{O'Shea} {et~al.}(2007){O'Shea}, {Banerjee}, \&
  {Doyle}}]{Oshea_etal_2007AA...475L..25O}
{O'Shea}, E., {Banerjee}, D., \& {Doyle}, J.~G. 2007, \aap, 475, L25

\bibitem[{{Parker}(1953)}]{paper:Parker1953}
{Parker}, E.~N. 1953, \apj, 117, 431

\bibitem[{{Peter} \& {Bingert}(2012)}]{Peter_2012AA...548A...1P}
{Peter}, H. \& {Bingert}, S. 2012, \aap, 548, A1

\bibitem[{{Peter} {et~al.}(2012){Peter}, {Bingert}, \&
  {Kamio}}]{paper:Peteretal2012}
{Peter}, H., {Bingert}, S., \& {Kamio}, S. 2012, \aap, 537, A152

\bibitem[{{Schrijver}(2001)}]{Schrijver_2001SoPh..198..325S}
{Schrijver}, C.~J. 2001, \solphys, 198, 325

\bibitem[{{Serio} {et~al.}(1981){Serio}, {Peres}, {Vaiana}, {Golub}, \&
  {Rosner}}]{Serio_1981ApJ...243..288S}
{Serio}, S., {Peres}, G., {Vaiana}, G.~S., {Golub}, L., \& {Rosner}, R. 1981,
  \apj, 243, 288

\bibitem[{{Susino} {et~al.}(2010){Susino}, {Lanzafame}, {Lanza}, \&
  {Spadaro}}]{paper:Susinoetal2010}
{Susino}, R., {Lanzafame}, A.~C., {Lanza}, A.~F., \& {Spadaro}, D. 2010, \apj,
  709, 499

\bibitem[{{Susino} {et~al.}(2013){Susino}, {Spadaro}, {Lanzafame}, \&
  {Lanza}}]{paper:Susinoetal2013}
{Susino}, R., {Spadaro}, D., {Lanzafame}, A.~C., \& {Lanza}, A.~F. 2013, \aap,
  552, A17

\bibitem[{{Testa} {et~al.}(2014){Testa}, {De Pontieu}, {Allred}, {Carlsson},
  {Reale}, {Daw}, {Hansteen}, {Martinez-Sykora}, {Liu}, {DeLuca}, {Golub},
  {McKillop}, {Reeves}, {Saar}, {Tian}, {Lemen}, {Title}, {Boerner},
  {Hurlburt}, {Tarbell}, {Wuelser}, {Kleint}, {Kankelborg}, \&
  {Jaeggli}}]{2014Sci...346B.315T}
{Testa}, P., {De Pontieu}, B., {Allred}, J., {et~al.} 2014, Science, 346,
  1255724

\bibitem[{{Tripathi} {et~al.}(2009){Tripathi}, {Mason}, {Dwivedi}, {del Zanna},
  \& {Young}}]{Tripathi_etal_2009ApJ...694.1256T}
{Tripathi}, D., {Mason}, H.~E., {Dwivedi}, B.~N., {del Zanna}, G., \& {Young},
  P.~R. 2009, \apj, 694, 1256

\bibitem[{{van Ballegooijen} {et~al.}(2011){van Ballegooijen}, {Asgari-Targhi},
  {Cranmer}, \& {DeLuca}}]{2011ApJ...736....3V}
{van Ballegooijen}, A.~A., {Asgari-Targhi}, M., {Cranmer}, S.~R., \& {DeLuca},
  E.~E. 2011, \apj, 736, 3

\bibitem[{{Viall} \& {Klimchuk}(2017)}]{paper:Viall&Klimchuk2017}
{Viall}, N.~M. \& {Klimchuk}, J.~A. 2017, \apj, 842, 108

\bibitem[{{Warren} {et~al.}(2011){Warren}, {Brooks}, \&
  {Winebarger}}]{paper:Warrenetal2011}
{Warren}, H.~P., {Brooks}, D.~H., \& {Winebarger}, A.~R. 2011, \apj, 734, 90

\bibitem[{{Winebarger} {et~al.}(2018){Winebarger}, {Lionello}, {Downs},
  {Miki{\'c}}, \& {Linker}}]{Winebarger_2018ApJ...865..111W}
{Winebarger}, A.~R., {Lionello}, R., {Downs}, C., {Miki{\'c}}, Z., \& {Linker},
  J. 2018, \apj, 865, 111

\bibitem[{{Winebarger} {et~al.}(2016){Winebarger}, {Lionello}, {Downs},
  {Miki{\'c}}, {Linker}, \& {Mok}}]{Winebarger_2016ApJ...831..172W}
{Winebarger}, A.~R., {Lionello}, R., {Downs}, C., {et~al.} 2016, \apj, 831, 172

\bibitem[{{Xia} {et~al.}(2017){Xia}, {Keppens}, \&
  {Fang}}]{Xia_2017AA...603A..42X}
{Xia}, C., {Keppens}, R., \& {Fang}, X. 2017, \aap, 603, A42

\end{thebibliography}
\end{document}